\documentstyle[aps,eqsecnum,preprint]{revtex}
\begin{document}
\draft
\title{Scaling Theory of the Integer Quantum Hall Effect}
\author{Bodo Huckestein\footnote{Present address: Institut f\"ur
    Theoretische Physik, Universit\"at zu K\"oln, 50937 K\"oln, Germany}}
\address{Max-Planck-Institut f\"ur Kernphysik, 69029 Heidelberg,
  Germany}
\date{\today}
\maketitle
\begin{abstract}
  The scaling theory of the transitions between plateaus of the Hall
  conductivity in the integer Quantum Hall effect is reviewed. In the
  model of two-dimensional noninteracting electrons in strong magnetic
  fields the transitions are disorder-induced
  localization-delocalization transitions. While experimental and
  analytical approaches are surveyed, the main emphasis is on
  numerical studies, which successfully describe the experiments. The
  theoretical models for disordered systems are described in detail.
  An overview of the finite-size scaling theory and its relation to
  Anderson localization is given. The field-theoretical approach to
  the localization problem is outlined.  Numerical methods for the
  calculation of scaling quantities, in particular the localization
  length, are detailed. The properties of local observables at the
  localization-delocalization transition are discussed in terms of
  multifractal measures. Finally, the results of extensive numerical
  investigations are compared with experimental findings.
\end{abstract}
\tableofcontents

\section{Introduction}
\label{cha:intro}

As announced in a paper by von Klitzing, Dorda, and Pepper (1980),
Klaus von Klitzing discovered that under certain experimental
conditions the Hall conductivity $\sigma_{xy}$ of a
quasi-two-dimensional electron gas is quantized to a high precision in
integer multiples of $e^2/h$ while at the same time the longitudinal
conductivity $\sigma_{xx}$ vanishes. A remarkable aspect
of this integer quantum Hall effect (QHE) is that the quantization
exists over a finite range of physical parameters, like the magnetic
field $B$ or the carrier concentration $n$. This observation cannot be
explained by the semiclassical Drude theory of conductivity. If the
longitudinal conductivity $\sigma_{xx}$ vanishes the Drude result for
the Hall conductivity is given by $\sigma_{xy}= \nu e^2/h$, where
$\nu=nh/eB=n2\pi{}l_c^2$ is the filling factor. Only if $\nu$ is an
integer, i.e., if the Fermi level lies exactly between two Landau
levels, is the Hall conductivity quantized.  Laughlin
(1981), Aoki and Ando (1981), and Halperin
(1982) showed that the Hall conductivity takes on its
quantized value $ie^2/h$ over a finite range in filling factor around
$\nu=i$ if in that range the states {\em at\/} the Fermi energy are
localized. By the same argument, when the Hall conductivity is
quantized there are states {\em below\/} the Fermi energy that are
{\em not\/} localized. The width of the quantized plateaus then
depends on the ratio of localized vs.\ extended states.

The concept of localized states in disordered systems was developed by
Anderson (1958). He showed that if a quantum-mechanical
system is sufficiently disordered, states have a finite probability to
return to a given site in the long-time limit. This absence of
diffusion implies that these states are localized in a finite region
of space. The transmission probability decays exponentially on a
length scale, which is called the localization length. For localized
states the static conductivity vanishes at zero temperature. On the
other hand, if the disorder is weak enough extended states might exist
that do not decay exponentially and fill the whole system. Their
contribution to the conductivity is finite even at zero temperature.
The energy that separates extended from localized states is called the
mobility edge. At the mobility edge the character of eigenstates is
different from both extended and localized states and we will call
these states critical states.

The observation that an explanation of the QHE involves both extended
and localized states was the more unanticipated as the scaling theory
of localization ({Wegner, }1976; {Abrahams {\em et~al.\/},
  }1979; {Wegner, }1979) predicted the absence of
extended states in two-dimensional systems. It is the presence of a
strong magnetic field that leads to the emergence of nonlocalized
states in two dimensions ({Ono, }1982b). Chalker (1987)
showed that these states exist only at a single energy in the limit
that scattering between Landau levels can be neglected. At zero
temperature the Hall conductivity is thus expected to exhibit sharp
steps whenever the Fermi energy passes the critical energy. The
longitudinal conductivity vanishes for all energies except at the
critical energy. The
quantum Hall plateau transition is thus a special case of a
metal-insulator transition. The insulating phases correspond to the
plateau regions, in which $\sigma_{xx}$ vanishes and $\sigma_{xy}$
is quantized. These insulating phases are separated by mobility edges
with a finite $\sigma_{xx}$. Truly metallic phases are absent in the
quantum Hall system. The states at the mobility edge show critical
fluctuations ({Wegner, }1980; {Aoki, }1983; {Aoki, }1986).

The description of this localization--delocalization transition in
terms of a zero-temperature phase transition is the topic of this
review. We will focus on the theory of noninteracting electrons. Thus
we do not consider the fractional QHE ({Tsui {\em et~al.\/}, }1982),
where the
quantization of the Hall conductivity in rational fractions of $e^2/h$
is due to electron-electron interactions. One notes, however, that
experimentally the fractional quantum Hall transitions show a
remarkable similarity to the integer transitions (see
Sec.~\ref{cha:exp}). In the absence of an analytical theory that is
capable of providing quantitative results we will mostly rely on
numerical simulations.

The review is organized as follows: In Sec.~\ref{cha:exp} we
shortly survey the experimental investigations of the quantum Hall
transitions. Section \ref{cha:model} discusses the different models of
disorder used in theoretical calculations. Section \ref{cha:scaling}
provides an overview of the finite-size scaling theory.
Section \ref{cha:ft} outlines the description of the Anderson transition
in terms of a field theory and its extension to the quantum Hall
system. Sections \ref{cha:loclen} and \ref{cha:others} explain which
physical quantities are used in numerical finite-size scaling studies
and how they are computed. In Sec.~\ref{cha:num} the results of
numerical calculations are presented and analyzed in terms of the
finite-size scaling theory. In Sec.~\ref{cha:multi} we characterize
the critical states using a multifractal analysis. In
Sec.~\ref{cha:conc} we compare the experimental and theoretical
results presented in the previous sections and discuss their
implications for an understanding of the quantum Hall transition.
Except where noted otherwise we try to provide enough detail to allow
the reader to use this review as a starting point for own
investigations of the subject.

A comprehensive introduction to the QHE is provided in the book edited
by Prange and Girvin (1987). For reviews about disordered
electronic systems and localization see Thouless (1974),
Lee and Ramakrishnan (1985), and Kramer and MacKinnon
(1993).

\section{Experimental Results}
\label{cha:exp}

In this section we want to review experimental investigations of the
transition between plateaus in the Hall conductivity.  We will focus
on those results that are uniquely related to these phase transitions.
In later sections we will show how a simple scaling theory for the
integer quantum Hall transitions accounts for most of the observed
features. We will therefore postpone the discussion of the
experimental results until we are familiar with the necessary
theoretical framework (Sec.~\ref{cha:conc}).

The sections \ref{sec:temp} to \ref{sec:freq} are ordered roughly
chronologically. We will start with experiments on the temperature
dependence of the transition. Although these experiments give the most
indirect information about the scaling behavior they were the first
to be performed and instrumental in developing the idea that the plateau
transitions are continuous phase transitions. Much more direct
contact with the concept of finite-size scaling is made in experiments
with varying sample sizes. Finally, the dynamic aspects of the
transition are elucidated in high-frequency experiments.

\subsection{Temperature-Dependent Scaling}
\label{sec:temp}

Soon after the discovery of the QHE ({von Klitzing {\em et~al.\/}, }1980) it
became apparent that the quantized plateaus can become extremely broad
and the transition between them extremely sharp as the temperature was
lowered ({Kawaji and Wakabayashi, }1981; {Paalanen {\em et~al.\/},
  }1982). At the same time the longitudinal conductivity
exhibits a series of sharp spikes at the position of the plateau
transitions. Since the Fermi energy is in a region of localized
states when the Hall conductivity is quantized and the longitudinal
conductivity vanishes, it can be concluded from the wide plateaus that
most of the electronic states are localized at low temperatures.
Paalanen, Tsui, and Gossard (1982) estimated that in an
AlGaAs/GaAs  heterostructure at 50 mK 95\% of the states in each
Landau level are localized.

The temperature dependences of the longitudinal conductivity
$\sigma_{xx}$ and the Hall conductivity $\sigma_{xy}$ of a
Si-MOSFET are shown in Fig.~\ref{fig:temp_dep}. Even in the absence
of localization the temperature dependence of the Fermi-Dirac
distribution leads to a temperature dependence of the conductivity,
\begin{equation}
  \sigma_{\mu\nu} (T) = \int dE \left( - \frac{\partial f(T)}{\partial
      E} \right) \sigma_{\mu\nu}(T=0).
  \label{sigma_fd}
\end{equation}
If $\sigma_{xx}(T=0)$ is finite only in an interval of width
$\Delta$ near the center of each Landau band, Eq.~(\ref{sigma_fd})
predicts an increase of $\sigma_{xx}$ in the band center with
decreasing temperature as long as $k_B T \gtrsim \Delta$. This
increase is seen in the peaks of Fig.~\ref{fig:temp_dep}.

When plotting $\sigma_{xx}$ versus $\sigma_{xy}$ as a function of
temperature and filling factor striking features appear. The data for
a Si-MOSFET for temperatures between 0.35 K and about 1 K in
Fig.~\ref{fig:two_scale_jpn} lie on a single curve indicating that
$\sigma_{xx}$ and $\sigma_{xy}$ are not independent but dependent
on a single parameter. At higher temperatures the effect of the Fermi
distribution is still visible in the increase of $\sigma_{xx}$. By
contrast, the conductivities seem to be independent and follow a
two-parameter scaling flow in the temperature driven flow diagram for
an InGaAs-InP heterostructure in Fig.~\ref{fig:two_scale_pu}. In
both figures a flow towards
$(\sigma_{xx},\sigma_{xy})=(0,{\text{integer}})e^2/h$, corresponding to
the widening plateaus, is obvious. Kravchenko and Pudalov
(1989) and Kravchenko {\em et al.\/} (1990)
studied Si-MOSFET and AlGaAs/GaAs heterostructures and observed both
types of behavior depending on the filling factor.  A single curve was
also obtained by McEuen {\em et al.\/} (1990) in a study of
the non-local transport properties in high-mobility GaAs-AlGaAs
samples.

It is experimentally extremely difficult to obtain reliable
flow-diagrams close to the transitions at half-integer Hall
conductivity. As the temperature approaches zero the transition
regions become very narrow and inhomogeneities in the electron density
will let the transition happen at slightly different values of the
magnetic field in different parts of the sample. Since $\sigma_{xx}$
and $\sigma_{xy}$ are measured on different parts of the sample this
will lead to irregular behavior near the transition. Wei, Tsui,
Paalanen, and Pruisken (1988) avoided this problem by
studying the critical behavior of each component of the resistivity
tensor separately. Fig.~\ref{fig:wei_fig1} shows the transport
coefficients $\rho_{xy}$, $\rho_{xx}$, and the derivative
$d\rho_{xy}/dB$ for an InGaAs-InP heterostructure. Note the
similarity between $\rho_{xx}$ and $d\rho_{xy}/dB$. However, while
the maximum values of the former decrease, the maximum values of the
latter increase with decreasing temperature. In
Fig.~\ref{fig:wei_fig2} the temperature dependence of the maximum
$(d\rho_{xy}/dB)^{\text{max}}$ and the inverse width
$(\Delta{}B)^{-1}$ of the $\rho_{xx}$ peaks is plotted for several
Landau levels. Over the range of temperatures from 0.1 K to 4.2 K they
show power-law behavior $(d\rho_{xy}/dB)^{\text{max}}\propto
T^{-\kappa}$ and $\Delta{}B\propto T^{\kappa}$ with
$\kappa=0.42\pm0.04$. Further measurements of higher-order
derivatives showed that the extrema of $d^2\rho_{xx}/dB^2$ and
$d^3\rho_{xy}/dB^3$ diverge like $T^{-2\kappa}$ and
$T^{-3\kappa}$, respectively ({Wei {\em et~al.\/}, }1990). In experiments on
low-mobility AlGaAs/GaAs samples the same exponent was observed but
the characteristic temperature below which this scaling was observed
was much lower (200 mK) than in the InGaAs/InP samples (4.2 K)
({Wei {\em et~al.\/}, }1992). Huckestein {\em et al.\/} (1991) studied the
temperature dependence of the plateau width in AlGaAs/GaAs heterostructures and
found agreement with $\kappa=0.42$ for temperatures below 4.2 K. For
Landau levels where the spin splitting was not resolved
$(d\rho_{xy}/dB)^{\text{max}}$ and $(\Delta{}B)^{-1}$ diverge like
$T^{-\kappa/2}$ ({Wei {\em et~al.\/}, }1990; {Hwang {\em et~al.\/},
  }1993). The same exponent $\kappa$ was
measured in the fractional QHE regime of high-mobility
AlGaAs/GaAs heterostructures for the scaling between filling factors
$2/5$ and $1/3$ ({Engel {\em et~al.\/}, }1990).

The universality of the exponent $\kappa$ was questioned by
Wakabayashi {\em et al.\/} (1989; 1990) and by Koch {\em et
  al.\/} (1991a). Wakabayashi {\em et al.\/} studied two of
the valley- and spin-split subbands of Landau levels $n=0$ and $n=1$
in a Si-MOSFET. For the rather limited temperature range between 0.35
K and 1 K they observe $\kappa=0.29\pm0.10$ and $\kappa=0.16\pm0.02$
for subbands $(0\downarrow-)$ and $(1\uparrow-)$, respectively. Below
0.2 K the temperature dependence saturates, presumably due to Joule
heating ({Wakabayashi {\em et~al.\/}, }1990). Koch {\em et al.\/}
found a dependence of
$\kappa$ on the mobility of AlGaAs/GaAs heterostructures
deliberately doped to decrease the mobility. They observed power-law
scaling with temperature over the range 40 mK to 1.1 K. For spin-split
Landau levels the measured values for $\kappa$ increased from 0.28
to 0.81 with decreasing mobility. If the spin splitting is not
resolved, $\kappa$ is considerably smaller but not by a universal
factor of two as found by Wei {\em et al.\/} ({Koch {\em et~al.\/}, }1991a).

In the context of the finite-size scaling theory (see
Sec.~\ref{cha:scaling}) the power law behavior of the transport
coefficients reflects the single-parameter scaling of the conductivity
tensor ({Pruisken, }1988)
\begin{equation}
  \sigma_{\mu\nu}(B) =
  \frac{e^2}{h}S_{\mu\nu}\left(L_{\text{eff}}^{1/\nu}(B-B^*)\right),
  \label{sigma_single_scale}
\end{equation}
where $B^*$ is the critical magnetic field of the plateau transition,
$L_{\text{eff}}$ is the effective system size, and $\nu$ is the
critical exponent of the localization length [cf.\ Eq.~(\ref{xi_t})].
The strongest support for this assertion comes from the peculiar
temperature dependence of the magnetic field derivatives of the
conductivity tensor.  At $T=0$, $L_{\text{eff}}$ is equal to the
system size. At finite temperatures the effective system size is given
by the phase coherence length $L_{\Phi}$ ({Thouless, }1977). If $L_{\Phi}$
diverges as $T^{-p/2}$ as the temperature approaches zero then
$L_{\text{eff}}^{1/\nu}\propto T^{-\kappa}$ with $\kappa=p/2\nu$. With
Eq.~(\ref{sigma_single_scale}) the $n$-th derivative of the
conductivity tensor at the critical point is
\begin{equation}
  \frac{d^n\sigma_{\mu\nu}(B^*)}{dB^n} \propto
  L_{\text{eff}}^{n/\nu} \propto T^{-n\kappa},
  \label{sigma_deriv}
\end{equation}
unless it vanishes due to symmetry.

\subsection{Size-Dependent Scaling}
\label{sec:size}

Koch {\em et al.\/} (1991b) measured the exponent $\nu$
directly by studying samples of the same shape but different
sizes. For sufficiently small samples $(d\rho_{xy}/dB)^{\text{max}}$
and $\Delta{}B$ saturate at low temperatures (see
Fig.~\ref{fig:koch_exp}). The saturation temperature decreases with
increasing system size. This is interpreted as the temperature where
the phase coherence length $L_{\Phi}$ becomes comparable to the
system size and the temperature-dependent scaling at higher
temperature crosses over to size-dependent scaling. The saturation
value of $\Delta{}B$ at low temperatures is then given by the
condition $L/\xi(\Delta{}B)\approx1$, i.e., $\Delta{}B\propto
L^{-1/\nu}$.  By fitting their data to this relation Koch {\em et
  al.\/} obtain $\nu=2.3\pm0.1$ for the three lowest Landau levels.
{}From the measured values of $\kappa$ they deduce that $p$ is not
universal and varies between $2.7$ and $3.4$. For a Landau level in
which the spin splitting was not resolved $\nu=6.5\pm0.6$ was
measured while the value of $p$ was comparable to its value in other
Landau levels.

\subsection{Frequency-Dependent Scaling}
\label{sec:freq}

The dynamical conductivity $\sigma_{xx}(f)$ of a low-mobility AlGaAs/GaAs
heterostructure was investigated by Engel {\em et al.\/}
(1993) by measuring the attenuation of a coplanar
transmission line on the sample surface. The losses in the
transmission line are due to the finite conductivity of the
two-dimensional electron gas below the surface. In contrast to
previous measurements using crossed rectangular waveguides
({Kuchar {\em et~al.\/}, }1986; {Huckestein {\em et~al.\/}, }1991) that
were limited to a single frequency $f=35$ GHz
Engel {\em et al.\/} were able to sweep the frequency from 0.2 to 14
GHz. As can be seen from Fig.~\ref{fig:engels_fig} the effects of
raising the temperature and increasing the frequency are quite
similar. With decreasing frequency the width $\Delta{}B$ of the
peaks in Re$(\sigma_{xx})$ decreases and saturates below a
temperature-dependent frequency. Above this frequency $\Delta{}B$
scales like
\begin{equation}
  \Delta{}B \propto f^{\gamma},
  \label{freq_scale}
\end{equation}
with $\gamma=0.41\pm0.04$ for spin-split Landau levels and
$\gamma=0.21$ for non-split Landau levels. The saturation
temperature corresponds to $hf\approx k_B T$ in agreement with the
results of Huckestein {\em et al.\/} (1991).

Similar to Eq.~(\ref{sigma_single_scale}) a dynamical scaling ansatz
for $\sigma_{xx}(L,f)$ is
\begin{equation}
  \sigma_{xx}(L,f) = \frac{e^2}{h}
  S_{xx}(L/\xi(B),f\tau_0(B)),
  \label{sigma_dyn_scale}
\end{equation}
where $\tau_0(B)\propto \xi(B)^z\propto |B-B^*|^{-\nu{}z}$ is the
relaxation time of the system that diverges as the critical point is
approached. For sufficiently high frequencies $f\tau_0(B)$ dominates
the scaling behavior of $S_{xx}$ and $\Delta{}B$ scales like
$f^{1/\nu{}z}$. Equation (\ref{freq_scale}) thus implies
$\gamma=1/\nu{}z$, and using the result $\nu=2.3$ from the
size-dependent scaling experiments, $z=1$ is obtained ({Engel {\em
    et~al.\/}, }1993).

\section{Models of Disorder}
\label{cha:model}

Before we can attempt to understand the experiments we have to define
our system in theoretical terms. We will make several simplifying
assumptions: (1) The electron system is strictly two-dimensional. (2)
The interactions between the electrons can be neglected. (3) The spin
degrees of freedom can be neglected. (4) Boundary effects, like edge
states, are not essential in understanding the critical behavior of
the system. The electron gas in real devices is only
quasi-two-dimensional, in the sense that the motion perpendicular to
the plane of the electron gas is quantized into electrical subbands.
If only the lowest electrical subband is occupied the electrons have
only two spatial degrees of freedom left and we may treat them as
two-dimensional, justifying our assumption (1).  Assumption (2)
lacks an a priori justification. The treatment of non-interacting
electrons will turn out to be quite successful in explaining the
experimental static-scaling results. While this might serve as an a
posteriori justification of our approach a theoretical assessment of
the relevance of electron-electron interactions is still missing.  For
an understanding of the observed dynamical scaling it seems to be
essential to go beyond the non-interacting electron approximation.
Assumption (3) seems justified since most experiments deal with
completely spin-split Landau level and we neglect interactions between
the electrons. For non-spin-split levels spin-orbit scattering might
be important and one would have to go beyond the spinless-electron
approximation. The last assumption (4) treats the phase transition in
the spirit of standard scaling theory as a bulk critical phenomenon.
We will make use of this assumption by choosing boundary conditions
most suitable to the calculation at hand.

We will thus describe our system by a single-particle Hamiltonian
\begin{equation}
  H = H_{0} + V({\bf r}),
  \label{H_general}
\end{equation}
where
\begin{equation}
  H_{0} = \frac{1}{2m}({\bf p} - e{\bf A})^2
  \label{H_kin}
\end{equation}
is the kinetic energy of the electron and $V({\bf r})$ is the disorder
potential. In the absence of disorder, $V({\bf r})=0$, the spectrum of
$H$ for an infinite system is a set equidistant levels,
\begin{equation}
  E_n = (n+1/2) \hbar\omega_c,
  \label{E_n}
\end{equation}
separated by the cyclotron energy $\hbar\omega_c=\hbar eB/m$. Each
Landau level $n$ is infinitely degenerate with a degeneracy of
\begin{equation}
  n_B=1/2\pi l_c^2
  \label{degeneracy}
\end{equation}
states per unit area. The length
$l_c=(\hbar/eB)^{1/2}$ is called the magnetic length and is
the classical cyclotron radius in the lowest Landau level $n=0$.

In order to fully define the model system we will now describe
different models of the disorder potential typically used in
calculations.

\subsection{Real-Space Models}
\label{sec:realmod}

The statistical properties of a disorder potential are completely
determined by the joint probability distribution $P[V]$ of the disorder
potential $V({\bf r})$ for all coordinates ${\bf r}$. Starting point
for most analytical calculations is the white-noise potential
distribution
\begin{equation}
  P[V] = {\cal N} \exp\left\{-\frac{1}{2V_0^2}\int d^2r [V({\bf
    r})]^2\right\},
  \label{white}
\end{equation}
where ${\cal N}$ normalizes the distribution. With this distribution
the potential at different coordinates is uncorrelated,
\begin{equation}
  \overline{V({\bf r})V({\bf r}')} = V_0^2 \delta({\bf r}-{\bf r}'),
  \label{deltacor}
\end{equation}
and has a Gaussian distribution at each point in space. The overbar
denotes the average with respect to the distribution $P[V]$ of the
disorder potential. This potential distribution is well suited for
analytic calculations since it allows one to perform the average over
disorder, due to the Gaussian dependence on the potential. For the
white-noise distribution Wegner (1983) showed that the
density of states in the lowest Landau level is given by
\begin{mathletters}%
  \label{Wegner}
\begin{equation}
  \rho(E) = \frac{\sqrt{2}}{\pi^2 l_c V_0}
  \frac{\exp({\nu^2})}
  {1+\left(2\pi^{1/2}{\displaystyle\int\nolimits_0^\nu}dx
      \exp({x^2})\right)^2},
\end{equation}
where
\begin{equation}
  \nu=\frac{\sqrt{2\pi}l_c}{V_0}\left(E-\frac{1}{2}\hbar\omega\right).
\end{equation}
\end{mathletters}%

If the integrand in the exponent of Eq.~(\ref{white}) contains a more
complicated function of $V({\bf r})$, but still depends only on the
potential at one point in space, the potential is still
$\delta$-correlated, but higher-order cumulants of the potential at
the same site have non-zero coefficients. Br\'ezin, Gross, and
Itzykson (1984) generalized Wegner's result (\ref{Wegner})
to this more general class of potentials.

For numerical calculations it is not practical to specify a random
value for the potential at every point in space. Instead one uses an
approximation to the white-noise distribution Eq.~(\ref{white}), the
$\delta$-scatterers potential,
\begin{equation}
  V({\bf r}) = \sum_{i=1}^N V_i \delta({\bf r}-{\bf r}_i),
  \label{sum_delta}
\end{equation}
where the sites ${\bf r}_i$ of the scatterers are randomly chosen and
there is an equal number of attractive $V_i=-V$ and repulsive
scatterers $V_i=+V$. This potential has the correlation function
(\ref{deltacor}) and approaches the white-noise potential
(\ref{white}) with $V_0^2=n_iV^2$ in the limit that the density of
scattering sites $n_i$ becomes infinite. Due to the symmetry of the
potential the density of states is symmetric with respect to the
center of the Landau level. The self-consistent Born approximation
(SCBA) for the density of states of this potential is a semi-circle,
\begin{equation}
  \rho(E) = \frac{1}{2\pi l_c^2} \frac{2}{\pi\Gamma} \left[ 1 -
  \left(\frac{E-E_n}{\Gamma}\right)^2\right]^{1/2},
  \label{scba_d_e}
\end{equation}
with the bandwidth
$\Gamma=(4n_iV^2/2\pi l_c^2)^{1/2}=2V_0/\sqrt{2\pi}l_c$ ({Ando and
  Uemura, }1974).
Since this is also the energy scale in Eq.~(\ref{Wegner}) we will
frequently use $\Gamma$ as a measure of the width of the Landau level.

In numerical calculations care has to be taken that the concentration
$c_i=2\pi l_c^2 n_i$ of scatterers is large enough in order to avoid
a singularity in the density of states due to wavefunctions avoiding
the scatterers ({Ando and Aoki, }1985). To see the origin of this
effect consider
a square of side $L$ containing an integral number of flux quanta
$N_S=L^2/2\pi l_c^2$. From Eq.~(\ref{degeneracy}) we see that $N_S$
is also the number of degenerate eigenstates of the system in the
absence of disorder. We can form a superposition of these $N_S$ states
that satisfies $N_S$ constraints. Thus if the number $N_i=n_il_c^2$ of
$\delta$-scatterers is less than $N_S$, the wavefunction can be made
to vanish at the site of each scatterer and hence the energy of this
state will not be influenced by the disorder potential. This leads to
a divergence of the density of states at $E=\hbar\omega_c/2$ even in
the presence of disorder. In terms of the concentration $c_i$, this
divergence will occur for $c_i<1$. Ando and Aoki state that for
$c_i=40$ a good approximation to the white-noise limit is achieved
({Ando and Aoki, }1985).

Disorder potentials with a finite correlation length can be generated
by replacing the $\delta$-functions in Eq.~(\ref{sum_delta}) by
functions with a finite range. For Gaussian scatterers,
\begin{eqnarray}
  V({\bf r}) &=& \sum_{i=1}^{N_S} v({\bf r}-{\bf r}_i),\\
  v({\bf r}) &\propto& \frac{1}{2\pi\sigma^2}e^{-|{\bf
      r}|^2/2\sigma^2},
  \label{sum_gauss}
\end{eqnarray}
the correlation function becomes Gaussian,
\begin{equation}
  \overline{V({\bf r})V({\bf r}')} \propto
  \frac{1}{2\pi\sigma^2}e^{-|{\bf r}-{\bf r}'|^2/2\sigma^2}.
  \label{gauss_cor}
\end{equation}
The concentration necessary to reach the high concentration limit is
much lower for Gaussian scatterers than for $\delta$-scatterers. The
number of impurities an electron can feel increases with the
range of the scatterers so that for Gaussian scatterers the effective
impurity concentration is enhanced by a factor $1+\sigma^2/l_c^2$
relative to $\delta$-scatterers ({Ando and Aoki, }1985). For scatterers of the
same strength $V$ the bandwidth is reduced by a factor of
$\beta=(1+\sigma^2/l_c^2)^{1/2}$ relative to the
$\delta$-scatterers, $\Gamma_\sigma = \Gamma/\beta$ ({Ando and Uemura, }1974).

Choosing different numbers of attractive and repulsive scatterers
allows for the study of potentials that are not particle-hole
symmetric on average. A potential with only repulsive scatterers
was used by Huo {\em et al.\/} (1993) to show that in this
case the critical energy is not simply related to features in the
density of states.

\subsection{Landau-Space Models}
\label{sec:lanmod}
This section gives an overview over models that specify the disorder
distribution in the space of the Landau states. While it provides the
reader with the basic ideas and results a more detailed description is
given in the Appendix~\ref{sec:rlmm}.

In a strong magnetic field it is helpful to choose a set of basis
states in which the kinetic energy part of the Hamiltonian
$H_{0}$ (\ref{H_kin}) is diagonal. In the Landau gauge
\begin{equation}
  {\bf A} = Bx\hat{{\rm e}}_y
  \label{landaug}
\end{equation}
and for a strip geometry of width $L_y$ with periodic boundary conditions
in the $y$-direction these states are the Landau states
\begin{equation}
  \psi_{nk}({\bf{r}}) \equiv \langle {\bf{r}}| nk\rangle =
  \frac{1}{\sqrt{L_y l_c}} e^{iky}
  \chi_n\left(\frac{x - k l_c^2}{l_c}\right),
  \label{Landaustates}
\end{equation}
where
\begin{equation}
  \chi_n(x) = \left(2^n n! \sqrt{\pi}\right)^{-1/2} H_n(x) e^{-x^2/2}
  \label{oscfn}
\end{equation}
are the harmonic-oscillator eigenfunctions, and $H_n(x)$ are
Hermite polynomials. In terms of these states the Hamiltonian including the
disorder potential $V({\bf{r}})$ can be written as
\begin{equation}
  H = \sum_{nk}\sum_{n'k'} |nk\rangle \langle nk|H|n'k'\rangle \langle n'k'|,
  \label{Ham_matrix}
\end{equation}
\begin{equation}
  \langle nk|H|n'k'\rangle = \left(n + \frac{1}{2}\right)\hbar\omega_{\rm c}
  \delta_{n,n'} \delta_{k,k'} + \langle nk|V|n'k' \rangle.
  \label{mat_element}
\end{equation}
For a particular realization of the real-space potential, the matrix
elements $\langle nk|V|n'k' \rangle$ can now be calculated.
Instead, one could try to calculate the distribution function of the
matrix elements given a particular distribution of the real space
potential. However, to study the universal critical behavior
of the Hamiltonian it is not necessary to know the whole distribution.
The universal properties presumably only depend on a few statistical
properties of the Hamiltonian, like the correlation function
\begin{equation}
  f({\bf{r}},{\bf{r}}')=\overline{V({\bf{r}})V({\bf{r}}')}.
  \label{frr}
\end{equation}
Under this assumption it is not necessary to construct the entire
distribution function of the matrix elements. By the same argument only
the second moment of the matrix elements is important for the
critical behavior. Thus it suffices to construct a random matrix
$\langle nk|V|n'k' \rangle$ [which we will call a random Landau matrix
(RLM)] with a second moment corresponding to (\ref{frr}). The
correlation function for the matrix elements can be calculated from
the real-space correlation function $f({\bf{r}},{\bf{r}}')$
\begin{eqnarray}
  \lefteqn{\overline{\langle n_1k_1|V|n_2k_2 \rangle\langle n_3k_3|V|n_4k_4
      \rangle}}\nonumber \\
  &=& \int d^2r d^2r' \langle n_1 k_1| {\bf{r}} \rangle \langle
  {\bf{r}} | n_2 k_2 \rangle \langle n_3 k_3| {\bf{r}}' \rangle \langle
  {\bf{r}}' | n_4 k_4 \rangle
  \overline{V({\bf{r}})V({\bf{r}}')},\nonumber\\
  &=& \int d^2r d^2r' \psi^*_{n_1k_1}({\bf{r}})
  \psi_{n_2k_2}({\bf{r}}) \psi^*_{n_3k_3}({\bf{r}}')
  \psi_{n_4k_4}({\bf{r}}') f({\bf{r}},{\bf{r}}').
  \label{matcor1}
\end{eqnarray}
If the real-space correlation function is translationally invariant,
\begin{equation}
  f({\bf{r}},{\bf{r}}') = g(|{\bf{r}}-{\bf{r}}'|),
  \label{inv}
\end{equation}
Eq.~(\ref{matcor1}) simplifies to
\begin{eqnarray}
  \overline{\langle n_1k_1|V|n_2k_2 \rangle\langle n_3k_3|V|n_4k_4
      \rangle} &=& \frac{1}{L_y^2 l_c^2}
  \delta_{k_1-k_2,k_4-k_3} \int d\rho_x \bar{g}(\rho_x,k_1-k_2) \int
  dX \nonumber \\
  & & \times \chi_{n_1}
  \left(\frac{X+\rho_x/2-k_1 l_c^2}{l_c}\right)
  \chi_{n_2}
  \left(\frac{X+\rho_x/2-k_2 l_c^2}{l_c}\right)
  \nonumber\\
  & & \times \chi_{n_3}
  \left(\frac{X-\rho_x/2-k_3 l_c^2}{l_c}\right)
  \chi_{n_4}
  \left(\frac{X-\rho_x/2-k_4 l_c^2}{l_c}\right),
  \label{gencor}
\end{eqnarray}
where $\bar{g}(x,k)$ is the Fourier transform of $g(x,y)$ with
respect to $y$,
\begin{equation}
  \bar{g}(x,k)=\int_{-L_y/2}^{L_y/2} dy\, g(x,y) e^{iky}.
  \label{fou}
\end{equation}
The $\delta$-function in Eq.~(\ref{gencor}) reflects the
translational invariance of the correlation function.
Equation (\ref{gencor}) describes the statistical properties of the RLM.
For magnetic fields much stronger than the disorder potential the
coupling between the states within one Landau level becomes much
stronger than the coupling between states in different Landau levels.
Neglecting the coupling between different Landau levels leads to the
single-band approximation in which the RLM becomes block diagonal in
the Landau level indices $n_i$. In this approximation the physics of
each Landau level can be discussed independently. For simplicity, let
us restrict our discussion of the statistical properties of the RLM in
the single-band approximation to the lowest Landau level
$n=0$ and the case of a random potential with Gaussian correlations
\begin{equation}
  g({\bf{r}}) = \frac{V_0^2}{2\pi\sigma^2} \exp
  \left(-\frac{|{\bf{r}}|^2}{2\sigma^2}\right).
  \label{g_gauss}
\end{equation}
The correlation function (\ref{gencor}) then becomes (up to terms
vanishing exponentially in the limit $L_y\to\infty$)
\begin{eqnarray}
  \lefteqn{\overline{\langle 0k_1|V|0k_2 \rangle\langle 0k_3|V|0k_4
      \rangle} = \frac{V_0^2}{2\sqrt{2\pi}\sigma L_y}
    \delta_{k_1-k_2,k_4-k_3}
  \exp\left(-\frac{(k_1-k_2)^2\sigma^2}{2}\right)\int dx \int dx' }
  \nonumber \\
  & & \times
  \exp\left(-\frac{(x-x')^2}{2\sigma^2}\right)
  \chi_{0}
  \left(\frac{x-k_1 l_c^2}{l_c}\right)
  \chi_{0}
  \left(\frac{x-k_2 l_c^2}{l_c}\right)
  \chi_{0}
  \left(\frac{x'-k_3 l_c^2}{l_c}\right)
  \chi_{0}
  \left(\frac{x'-k_4
    l_c^2}{l_c}\right),\nonumber\\
  & & = \frac{V_0^2}{\sqrt{2\pi}l_c L_y \beta}
  \exp\left(-\frac{1}{2} (k_1-k_2)^2l_c^2 \beta^2\right)
  \exp\left(-\frac{1}{2} (k_1-k_4)^2l_c^2
  \frac{1}{\beta^2}\right)\delta_{k_1-k_2,k_4-k_3},
  \label{delcor}
\end{eqnarray}
where $\beta^2=(\sigma^2+l_c^2)/l_c^2$
characterizes the correlation length of the potential projected onto
the lowest Landau level relative to the
magnetic length. Due to the Kronecker $\delta$ in Eq.~(\ref{delcor})
the matrix elements $\langle 0k_1|V|0k_2\rangle$ are correlated only if
they belong to the same diagonal $k_1-k_2 = \rm{const}$. More
precisely, if $\langle 0k_1|V|0k_2\rangle$ is split into its real
and imaginary parts, $\langle0k_1|V|0k_2\rangle_{\rm{R}}$ and
$\langle0k_1|V|0k_2\rangle_{\rm{I}}$, respectively,
\begin{equation}
  \langle 0k_1|V|0k_2\rangle = \langle 0k_1|V|0k_2\rangle_{\rm{R}}
  + i \langle 0k_1|V|0k_2\rangle_{\rm{I}},
  \label{realim}
\end{equation}
then, due to hermiticity,
\begin{eqnarray}
  \overline{\langle 0k_1|V|0k_2 \rangle_{\rm{R}}\langle 0k_3|V|0k_4
      \rangle_{\rm{R}}} &=& \frac{1}{2}
    (\delta_{k_1-k_2,k_4-k_3}+\delta_{k_1-k_2,k_3-k_4})
    p(0,0,0,0;k_1,k_2,k_3,k_4),\\
  \overline{\langle 0k_1|V|0k_2 \rangle_{\rm{I}}\langle 0k_3|V|0k_4
      \rangle_{\rm{I}}} &=& \frac{1}{2}
    (\delta_{k_1-k_2,k_4-k_3}-\delta_{k_1-k_2,k_3-k_4})
    p(0,0,0,0;k_1,k_2,k_3,k_4),
  \label{realimcor}
\end{eqnarray}
where $p(0,0,0,0;k_1,k_2,k_3,k_4)$ is the correlation function of
Eq.~(\ref{delcor}) without the Kronecker $\delta$, and real and
imaginary parts are uncorrelated ({Huckestein and Kramer, }1989). The
first exponential
factor in Eq.~(\ref{delcor}) leads to a band structure in the RLM, i.e.,
as $k_1-k_2$ becomes large the average magnitude of the matrix element
$\overline{|\langle 0k_1|V|0k_2\rangle|^2}$ becomes exponentially
small. The second exponential factor governs the strength of the
correlations along the diagonals. Note that even for $\sigma\to 0$,
i.e., for $\delta$-correlated potential, the range of the correlations
does not vanish due to the Gaussian spread of the Landau
wavefunctions.

We now want to express the correlated matrix elements $\langle n_1
k_1|V|n_2 k_2 \rangle$, which satisfy the correlation function
(\ref{gencor}), in terms of uncorrelated random variables, that can
relatively easily be generated on a computer. The
correlation function motivates the ansatz
\begin{eqnarray}
  \langle n_1 k_1|V|n_2 k_2 \rangle &=& \frac{1}{L_y
      l_c} \int dx dx'\, h(x-x',k_1-k_2)
    u_0(x',k_1-k_2)\nonumber\\
  & &\times
  \chi_{n_1}\left(\frac{x-k_1l_c^2}{l_c}\right)
  \chi_{n_2}\left(\frac{x-k_2l_c^2}{l_c}\right),
  \label{mat_el1}
\end{eqnarray}
where the $u_0(x,k)$ are uncorrelated, complex random variables
satisfying
\begin{equation}
  \overline{u_0(x,k)u_0(x',k')} = \delta(x-x') \delta_{k,-k'}.
  \label{u_cor}
\end{equation}
For the simple case of Gaussian correlations the intra-Landau-level
matrix elements in the lowest Landau level are given by
\begin{eqnarray}
  \langle 0 k_1|V|0 k_2 \rangle &=&
  \frac{V_0}{\sqrt{2\pi L_y}}
  \exp\left(\frac{(k_1-k_2)^2l_c^2\beta^2}{4}\right)
  \nonumber\\
  & &\times\int d\xi\, u_0\left(\beta\xi +
  \frac{(k_1+k_2)l_c}{2},(k_1-k_2)l_c\right)
  \exp(-\xi^2).
  \label{mat_el0}
\end{eqnarray}
Note that this is also the form of Eq.~(\ref{mat_elg}) for any Landau
level $n_1=n_2$ in the limit $\sigma/l_c\to\infty$ ({Huckestein, }1992).

It is worthwhile to compare the RLM model with the tight-binding model
of section \ref{sec:latmod}. The correlation function in
(\ref{gencor}) describes a random Hamiltonian on a one-dimensional
lattice where the lattice sites are labeled by the integers $i = k
L_y/2\pi$. The diagonal matrix elements $\langle n k|V|n k \rangle$
correspond to the random site energies $\epsilon_i$. The differences
to the Anderson Hamiltonian are in the off-diagonal matrix elements
that correspond to the hopping matrix elements $V_{ij}$. While in the
Anderson case the off-diagonal matrix elements are real
constants\footnote{In one dimension the hopping matrix elements can be
  made real by a global gauge transformation since closed paths can
  enclose no flux.} and
couple only nearest neighbors, they are random complex
variables in the RLM that couple all sites within the range $|i-j| \approx
L_y/2\pi l_c\beta$. As is shown in section \ref{cha:num} for finite
$L_y$, the RLM has only localized states. However, the largest
localization length is of the order of $L_y$ and in the
two-dimensional limit, $L_y\to\infty$, both the range of the hopping
matrix elements and the maximum localization length diverge.

\subsection{Lattice Models}
\label{sec:latmod}

The standard model of localization is Anderson's tight-binding
Hamiltonian ({Anderson, }1958)
\begin{equation}
  H = \sum_i \epsilon_i |i\rangle\langle i| + \sum_{\{i,j\}} V_{ij}
  |i\rangle\langle j|,
  \label{anderson}
\end{equation}
where $i$ labels the sites on a lattice and $\{i,j\}$ are nearest
neighbors. The site energies $\epsilon_i$ are independent random
variables and in the absence of a magnetic field the
hopping matrix elements are constant, $V_{ij}=V$, and define the energy
scale. With this Hamiltonian the Schr\"odinger equation takes the form
\begin{equation}
  \epsilon_i a_i + \sum_{j\atop \{i,j\}} V_{ij} a_j = E a_i,
  \label{ander_sch}
\end{equation}
where $a_i$ is the amplitudes of the wave function at site $i$. For
$\epsilon_i=0$ the Hamiltonian describes an electron moving in a
perfect crystal.

This model can be understood as the simplest form of a more general
tight binding model where the $V_{ij}$ are a function of $i$ and $j$.
Such a model can be derived from the Schr\"odinger equation for a free
particle in a periodic potential by expanding the  wavefunction in
terms of a complete set of orthonormal Wannier functions localized
at each atomic site ({Thouless, }1974)
\begin{equation}
  \psi({\bf r}) = \sum_i\sum_n a_i^{(n)} \phi^{(n)}({\bf r}- {\bf R}_i),
  \label{wannier}
\end{equation}
where ${\bf  R}_i$ is the position of atom $i$ and $n$ labels the
atomic orbitals. Taking into account only a single orbital per site
leads to an equation of the form (\ref{ander_sch}).

In two dimensions on a square lattice,
Eq.~(\ref{ander_sch}) has eigenvalues $E$ in the band
between $-4V$ and $+4V$. In the presence of disorder all states in the
system are localized ({Abrahams {\em et~al.\/}, }1979; {Wegner,
  }1979; {MacKinnon and Kramer, }1981).

When a magnetic field is applied the Hamiltonian (\ref{anderson})
has to be modified. To lowest order in the magnetic field the
influence of the magnetic field on the Wannier functions can be
neglected and the effect of the magnetic field is to change the phase
of the wavefunction between two sites. This
is achieved by replacing the constant hopping matrix elements by
\begin{equation}
  V_{ij} = V \exp\left( -\frac{ie}{\hbar} \int_{{\bf r}_i}^{{\bf
      r}_j} dr {\bf A}({\bf r}) \right),
  \label{vij_mag}
\end{equation}
a procedure known as Peierls substitution ({Peierls, }1933; {Luttinger, }1951).
In the absence of disorder this system shows a very rich band
structure, the self-similar ``Hofstadter butterfly''
({Azbel{'}, }1964; {Hofstadter, }1976). In particular, if the number of
flux quanta per
unit cell of the lattice, $a^2/2\pi l_c^2$, where $a$ is the lattice
constant, is a rational number $p/q$, the tight binding band
splits into $q$ magnetic subbands.

In the presence of disorder this model can quite accurately describe
the lowest Landau levels of a continuum model. This is the case in an
intermediate range of the disorder where the disorder broadening of
the magnetic subbands is large compared to their intrinsic width but
small compared to the separation of the magnetic subbands. Schweitzer
{\em et al.\/} (1984) have shown by a recursive Green
function method that in such a situation the density of states of the
lowest magnetic
subband agrees well with the exact result Eq.~(\ref{Wegner}) for the
continuum model. It should
be noted that this correspondence between the tight binding model and
free electrons holds only for the lowest magnetic subbands near the
edges of the spectrum.

\subsection{Semiclassical Approximation and Network Model}
\label{sec:network}

For potentials smooth on the scale of the magnetic length $l_c$ the
localization in the integer QHE can be discussed in terms of
semiclassical quantization and percolation.\footnote{Note that an
  expansion in $l_c^2$ is also an expansion in powers of $\hbar$.}
These ideas have been developed by Tsukada (1976),
Iordansky (1982), Kazarinov and Luryi (1982), Ono
(1982a), Prange and Joynt (1982), Trugman
(1983), Shapiro (1986), Wilkinson
(1987), Chalker and Coddington (1988), Fertig
(1988), Mil'nikov and Sokolov (1988), Jaeger
(1991), and Lee, Wang, and Kivelson (1993). In
the semiclassical limit, the motion of the electron separates into two
components with vastly different time and length scales. On the one
hand there is the slow motion of the guiding centers of the cyclotron
orbits along equipotential lines of the disorder potential that is
governed by classical drift equations. On the other hand there is the
rapid cyclotron motion of frequency $\omega_c$ around the guiding
center within the classical cyclotron radius $R_c=\sqrt{2n+1} l_c$
{}from the guiding center. The wavefunctions are non-zero on strips of
width $R_c$ around the path of the guiding center. They are thus
located at the edges of the regions where $V({\bf r}) < E$ (see
Fig.~\ref{fig:two_lakes}).
Near minima of the potential the wavefunctions encircle
these minima, while near maxima of the potential they encircle the maxima.
In both cases the orbits are closed and the states localized. Since
$V({\bf r})$ is a random potential the properties of the regions
$V({\bf r})<E$ are described by the continuum model of
percolation ({Trugman, }1983) with the regions bounded by the equipotential
lines corresponding to the clusters of the percolation model. At one
energy between the minima and maxima the largest percolation cluster
extends throughout the whole system and at this
energy the wave functions are extended.  For a disorder potential with
a distribution that is symmetric this energy corresponds to the band
center.

The rapid cyclotron motion around the guiding centers and hence the
extent of the wavefunctions perpendicular to the equipotential
line become important whenever two orbits approach each other on a
distance less than the cyclotron radius. This happens near saddle
points of the potential. There tunneling between the two
orbits becomes important and effects of quantum interference can
become observable. These effects are most important near the
percolation threshold because there the percolating equipotential line is
split up by saddle points into localized, closed contours. Thus while
tunneling and quantum interference have a negligible effect in the
tails of the density of states, they can in fact change the critical
behavior of the system.

Chalker and Coddington (1988) created a model that captures
the physics of percolation, tunneling and quantum interference near
the percolation threshold by introducing a network of saddle points
connected by equipotential lines. Due to its quantum nature this
network model shows critical behavior that is different from the
classical percolation problem but coincides with that
for short-ranged potentials (see section \ref{cha:num}).

In the following we will first introduce the semiclassical picture of
the QHE and then motivate the network model as a
simplification of this picture.

The semiclassical approximation is most conveniently derived by
replacing the coordinates $(x,y)$ by guiding center coordinates
$(X,Y)$ and relative coordinates $(\xi,\eta)$, given by
\begin{mathletters}%
\label{x_and_y}
\begin{eqnarray}
  x &=& X + \xi,\\
  y &=& Y + \eta,
\end{eqnarray}
\end{mathletters}%
and
\begin{mathletters}%
\label{xi_and_eta}
\begin{eqnarray}
  \xi &=& -\frac{1}{eB} (p_y - eA_y),\\
  \eta &=& \frac{1}{eB} (p_x -e A_x).
\end{eqnarray}
\end{mathletters}%
{}From the equations of motion for the pure system ($V({\bf r})=0$), we
get
\begin{mathletters}%
\label{eom_xi_eta}
\begin{eqnarray}
  \dot{\xi} &=& \frac{i}{\hbar} [H_0,\xi] =  \omega_c\eta,\\
  \dot{\eta} &=& \frac{i}{\hbar} [H_0,\eta] = - \omega_c\xi.
\end{eqnarray}
\end{mathletters}%
We see that $(\xi,\eta)$ indeed rotate with angular frequency
$\omega_c$ around the guiding center. Due to the commutation relations
of ${\bf p}$ and ${\bf r}$ both the guiding center coordinates and the
relative coordinates obey canonical commutation relations,
\begin{mathletters}%
\label{xi_X_comm}
\begin{eqnarray}
  [\xi,\eta] &=& il_c^2,\\
  {}[X,Y] &=& -il_c^2.
\end{eqnarray}
\end{mathletters}%

With Eq.~(\ref{xi_and_eta}), the Hamiltonian $H_0$ can be written in
terms of $\xi$ and $\eta$,
\begin{equation}
  H_0 = \frac{\hbar\omega_c}{2l_c^2} \left( \xi^2 + \eta^2 \right),
  \label{H_xi_eta}
\end{equation}
which expresses the degeneracy of the Landau levels as $H_0$ does not
depend on $X$ and $Y$.

In the presence of a disorder potential $V({\bf r})$ the degeneracy is
lifted. The equations of motion for the center coordinates are
\begin{mathletters}%
  \label{X_dot}
  \begin{eqnarray}
    \dot{X} &=& \frac{i}{\hbar} [H,X] = - \frac{l_c^2}{\hbar}
    \frac{\partial V}{\partial y},\\
    \dot{Y} &=& \frac{i}{\hbar} [H,Y] = \frac{l_c^2}{\hbar}
    \frac{\partial V}{\partial x}.
  \end{eqnarray}
\end{mathletters}%
If the potential $V({\bf r})$ is smooth on the scale $l_c$, we can
replace $V(x,y)$ by $V(X,Y)$ and obtain a drift of the guiding center
along equipotentials,
\begin{mathletters}%
    \label{center_drift}
  \begin{eqnarray}
    \dot{X} &=& - \frac{l_c^2}{\hbar} \frac{\partial V}{\partial Y},\\
    \dot{Y} &=& \frac{l_c^2}{\hbar} \frac{\partial V}{\partial X}.
  \end{eqnarray}
\end{mathletters}%
In this limit, the eigenenergies are
\begin{equation}
  E=(n+1/2)\hbar\omega_c + V(X,Y).
  \label{semi_class_ev}
\end{equation}
Due to the breaking of the time reversal
symmetry by the magnetic field the guiding centers can move only in
one direction determined by the magnetic field and
the slope of the potential (Fig.~\ref{fig:two_lakes}).

The electron eigenfunctions can be approximated by
\begin{equation}
  \psi(u,v) = C(u) \chi_n(v) e^{i\phi(u,v)},
  \label{eig_fun}
\end{equation}
where the variables $u$ and $v$ parametrize the distance along and
perpendicular to the equipotential line, respectively,
\begin{equation}
  C^2(u) \propto \frac{1}{|\nabla V(u,v)|_{v=0}},
  \label{c_u}
\end{equation}
and $\phi(u,v)$ is a gauge-dependent phase ({Trugman, }1983). The
semiclassical quantization condition is that the phase $\phi(u,v)$
must change by a multiple of $2\pi$ upon going around a closed
contour. This condition determines the allowed energies. From
Eq.~(\ref{eig_fun}) it follows that the overlap between different
states is exponentially small as long as the separation between the
corresponding contours is large compared to the magnetic length.

The problem of finding the wavefunction of largest extent is
equivalent to the problem of finding the largest connected cluster in
a continuum percolation problem ({Stauffer, }1979). The ``diameter''
$\xi_P(E)$ of this cluster grows on approaching the
percolation threshold $E_c$ as
\begin{equation}
  \xi_P(E) \propto |E-E_c|^{-\nu_{P}},
  \label{xi_cl}
\end{equation}
where the percolation critical exponent $\nu_P$ is 4/3 (den Nijs,
1979; Black and Emery, 1981).

In addition to the semiclassical drift of the guiding center
coordinates tunneling has to be considered near saddle points of the
potential, where two orbits approach each other. In order to study the
influence of tunneling on percolation Chalker and
Coddington(1988) introduced a
network model. To motivate this model, let us look at an electron
moving along an equipotential line far away, compared to $l_c$, from
other trajectories. Then its wavefunction is given by
Eq.~(\ref{eig_fun}). This wavefunction also contains information
about the degrees of freedom perpendicular to the trajectory that we
are not interested in. In its place, we consider a simpler quantity,
the complex function $Z(u)$ of the coordinate $u$ along the
equipotential line, defined by
\begin{equation}
  \arg(Z(u)) = \arg(\Psi(u,v=0))
  \label{z_arg}
\end{equation}
and
\begin{equation}
  |Z(u)|^2 = \int dv \,\psi^*(u,v)\hat{j}_u \psi(u,v),
  \label{z_mod}
\end{equation}
where $\hat{j}_u$ is the $u$-component of the current density
operator and $u$ increases in the direction of net current flow so that
the integral is positive. Since no current can escape perpendicular to
the equipotential line $Z(u)$ can only change its phase along the
trajectory. This phase change depends on the gauge and arc-length of
the trajectory in units of $l_c$.

Let us now consider the region near the saddle points
(Fig.~\ref{fig:network}). There two degenerate wavefunctions meet
and, due to tunneling, the description in terms of the $Z(u)$ breaks
down. Let us denote by $Z_1$, $Z_2$, $Z_3$, and $Z_4$ the values of
$Z(u)$ on the four arms some distance large compared to $l_c$ away
from the saddle point, where the $Z(u)$ are still fair descriptions of
the exact eigenfunctions. Due to the directedness of the trajectories
two of these describe ingoing currents (say, $Z_1$ and $Z_2$) and two
describe outgoing ($Z_3$ and $Z_4$). Solving the Schr\"odinger
equation near the saddle point gives a relation between the ingoing
and outgoing currents,
\begin{equation}
  {Z_1 \choose Z_3} = M {Z_4 \choose Z_2}.
  \label{in_out}
\end{equation}
Current conservation constrains the $2\times2$ matrix $M$ to unitarity
$|Z_1|^2+|Z_2|^2 = |Z_3|^2+|Z_4|^2$. This implies
\begin{equation}
  J = M^\dagger JM,
  \label{mjm}
\end{equation}
where
\begin{equation}
  J = \left( \begin{array}{cc}
              1 & 0\\ 0 & -1
             \end{array} \right),
  \label{j}
\end{equation}
which has the general solution
\begin{equation}
  M = \left( \begin{array}{cc}
              e^{i\phi_1} & 0 \\ 0 & e^{i\phi_2}
            \end{array} \right)
      \left( \begin{array}{cc}
              \cosh \theta & \sinh \theta \\
              \sinh \theta & \cosh \theta
            \end{array} \right)
      \left( \begin{array}{cc}
              e^{i\phi_3} & 0 \\ 0 & e^{i\phi_4}
            \end{array} \right)  ,
  \label{m_sol}
\end{equation}
with real $\theta$, $\phi$. The phases $\phi_i$ can be set to
zero by a suitable choice of gauge so that the saddle point is completely
characterized by a real parameter $\theta$. For $\theta\ll 1$ the
saddle point behaves classically and the electron almost always goes
from 1 to 4: $|Z_1|^2\approx |Z_4|^2$. For $\theta\gg 1$ it also
behaves classically, but now 1 goes to 3: $|Z_1|^2\approx |Z_3|^2$. In
between the saddle point behaves quantum mechanically. For a special
value $\theta_c$ the saddle point will be symmetric. At this value
the probability to scatter from 1 to 3 and to 4 are the same.
$\theta_c$ is given by
\begin{equation}
  \sinh\theta_c = 1,
  \label{theta_c}
\end{equation}
which has the solution $\theta_c = \ln(1+\sqrt{2}) \approx
0.8814$. Fertig (1988) showed how $\theta$ depends on the
energy,
\begin{equation}
  \sinh(\theta) = \exp(-\pi\gamma/2),
  \label{theta_eq}
\end{equation}
where
\begin{equation}
  \gamma = \frac{4(E-V_1)}{|V_{xx}V_{yy}|^{1/2}l_c^2},
  \label{gamma}
\end{equation}
$V_1$ is the saddle point potential, and $V_{ii}$ is the second
derivative of the potential in suitable coordinates.

The network model consists of a square lattice of saddle points, each
described by a $\theta$ parameter (see Fig.~\ref{fig:net_scheme}). Two
neighboring saddle points are connected by directed links representing
the equipotentials such that at every saddle point there are two
incoming and to outgoing links.  Each link is characterized by the
phase change of $Z(u)$ along the link. Since the lengths of the
links and hence the phase change depends on the random potential these
phases are taken to be uniformly distributed random variables. In the
original paper ({Chalker and Coddington, }1988) the $\theta$ parameter
was fixed for all
nodes in the network.  Generalizations of the model where the
$\theta$'s were random variables were studied by Chalker and
Eastmond (1993) and Lee {\em et al.\/} (1993).

In contrast to the other models of disorder the network model does not
deal with wavefunctions but with transmission probability amplitudes.
Thus it lends itself naturely to the study of properties of the
transmission matrix through the system but it is not obvious how to get
information about the multifractal properties of the system discussed
in section \ref{sec:multiscale}.

Mil'nikov and Sokolov (1988) employed the semiclassical
approximation to calculate the inverse localization length
$\xi(E)^{-1}$ in an intermediate energy range,
$\Gamma_\sigma(l_c/\sigma)^2\ll E \ll \Gamma_\sigma$. In this energy
range the equipotential lines differ significantly from the percolating
equipotential line only near the saddle points, where the percolating
equipotential line is split up. Due to the first strong inequality the
parameter $\gamma$ of Eq.~(\ref{gamma}) is large at these saddle
points and with Eqs.~(\ref{in_out}) and (\ref{theta_eq}) the
transmission probability through the saddle point is exponentially
small. Mil'nikov and Sokolov assume that in this limit a single path,
which consists of saddle points connected by parts of equipotential
lines, is important for the transmission probability between points
far apart in space. Then they treat this problem using the semiclassical
theory of tunneling through a one-dimensional barrier. The
solution to the eigenvalue equation (\ref{semi_class_ev}) defines a
dispersion relation $Y(E,X)$ similar to the dispersion relation of a
one-dimensional particle in a potential $k(E,x)$. The solutions
$Y(E,X)$ are real on the equipotential lines but are complex in the
classically forbidden regions near the saddle points. With
Eq.~(\ref{semi_class_ev}) the inverse localization length (\ref{xi_g})
can be written as
\begin{equation}
  \xi(E)^{-1} = \overline{\min_n |\text{Im}Y_n(E)|/l_c^2},
  \label{xi_ms}
\end{equation}
where $Y_n(E)$ are the roots of Eq.~(\ref{semi_class_ev}). The average
distance between saddle points is given by the diameter $\xi_P(E)$
(\ref{xi_cl}). Thus we can rewrite Eq.~(\ref{xi_ms}),
\begin{equation}
  \xi(E)^{-1} = \xi_P(E)^{-1} \overline{\int dX |\text{Im} Y(E,X)|/l_c^2},
  \label{xi_ms_2}
\end{equation}
where the integral runs over the classically forbidden range of $X$
near a saddle point. Since both the width of this region and the
maximum of $|\text{Im} Y(E,X)|$ are proportional to $E^{1/2}$, the
integral in Eq.~(\ref{xi_ms_2}) is proportional to $E$ and with
Eq.~(\ref{xi_cl}) we find
\begin{equation}
  \xi(E)^{-1} \propto E^{\nu_P+1} = E^{7/3}.
  \label{xi_ms_final}
\end{equation}

While the result $\nu=7/3$ agrees with the numerical value (\ref{nu})
it is not clear whether it describes the same physical situation. The
derivation of the result is based on the assumption that only a single
trajectory between two points is important. While this assumption seems
justified if $E\gg\Gamma_\sigma(l_c/\sigma)^2$ and each saddle point
transmits predominantly into a single link, closer to the percolation
threshold the probabilities for transmission into both possible links
become comparable. In this energy range the two-dimensional character
of the percolating equipotential line has to be taken into account. It
is this limit that is described by the network model, which has (see
Sec.~\ref{sec:lll}) the same scaling behavior as the random Landau
matrix model.

\section{Finite-Size Scaling}
\label{cha:scaling}

In this section we want to review the finite-size scaling theory as it
was developed for thermodynamic phase transitions. Since the
transitions between quantum Hall plateaus are believed to be
zero-temperature quantum phase transitions we want to analyze the
numerical data in section \ref{cha:num} in terms of this finite-size
scaling theory. We base our discussion on the review of finite-size
scaling by Barber (1983).

For infinite systems, thermodynamic quantities can diverge at phase
transitions. The ``usual'' scaling theory describes the
behavior of the thermodynamic quantities near the transition in terms
of scaling laws. In finite systems there are no phase transitions and
no singularities in thermodynamic quantities. Nevertheless in the
simultaneous limit that the temperature approaches the critical
temperature and the system size goes to infinity the scaling laws of
the infinite system are reflected in scaling laws for the finite
system involving the system size. It is this finite-size scaling
theory that we want to focus on in this section.

\subsection{Single-Parameter Scaling}
\label{sec:onepar}

Let us consider a $d$-dimensional thermodynamic system that is finite
in at least one direction. Since we are interested in the QHE in two
dimensions, let us assume $d=2$. Then the system could
have the geometry of a quasi-one-dimensional strip or cylinder,
depending on the boundary conditions in the transverse direction, or
it could be finite in both directions. Let us further assume that
the infinite $d$-dimensional system has a phase transition but that the
system with the finite length scale has no transition. An example of
such a system would be the Ising model.

Let $P_{\infty}(T)$ be a thermodynamic quantity of the infinite
system that diverges at the critical point $T_c$,
\begin{equation}
  P_\infty(T) \propto t^{-\rho},
  \label{p_infty}
\end{equation}
as $t=(T-T_c)/T_c \to 0$. For the $d=2$ Ising model this could be the
susceptibility and the parameter $T$ is the temperature. For a
finite system this divergence will be rounded and the function
$P_L(T)$ of the finite system will have a maximum at a position
$T_m(L)$ that will in general be different from $T_c$ but will
approach $T_c$ as $L\to\infty$. The approach is characterized by a
shift exponent $\lambda$ defined by
\begin{equation}
  \frac{T_m(L)-T_c}{T_c} \propto L^{-\lambda},\qquad L\to\infty,
  \label{lambda_def}
\end{equation}
where $L$ is measured in multiples of some microscopic length scale of
the system.

The central idea of finite-size scaling is that the magnitude of the
finite-size effects are determined by a single length scale $\xi(T)$
that diverges at the critical point,
\begin{equation}
  \xi(T) \propto t^{-\nu}, \qquad t\to 0.
  \label{xi_t}
\end{equation}
This length scale is identified with the correlation length of order
parameter fluctuations. Thermodynamic quantities will depend on the
ratio
\begin{equation}
  y = L/\xi(T).
  \label{l_xi}
\end{equation}
For $y\gg1$, the system does not feel the finite size and intensive
quantities will not depend on $L$. For $y\lesssim1$, the finite size
rounds off the thermodynamic singularities associated with the phase
transition. Thus $y=1$ defines another characteristic temperature
$T^*(L)$ of the system where finite-size effects become important.
{}From Eqs.~(\ref{xi_t},\ref{l_xi}) it follows that
\begin{equation}
  \frac{T^*(L)-T_c}{T_c} \propto L^{-1/\nu},\qquad L\to\infty.
  \label{t*}
\end{equation}
If $\xi(T)$ is the only relevant length scale the shift exponent
$\lambda$ must also be given by $\lambda=1/\nu$ ({Barber, }1983). In
the following we will not consider modifications to the finite-size
scaling analysis necessary to deal with the shift of the ``critical''
point in finite systems. The systems that we want to analyze exhibit a
symmetry such that the maxima of scaling functions are always located
at the critical point of the infinite system.

The finite-size scaling hypothesis states that for finite $L$ and
$T$ near $T_c$,
\begin{equation}
  P_L(T) = L^{\omega}Q(y).
  \label{single_par_scale}
\end{equation}
The exponent $\omega$ is determined by requiring that
Eq.~(\ref{single_par_scale}) reproduces Eq.~(\ref{p_infty}) in the
limit $L\to\infty$. This requires
\begin{equation}
  Q(y) \propto y^{-\omega},\qquad y\to\infty,
  \label{q_asym}
\end{equation}
and
\begin{equation}
  \omega=\rho/\nu.
  \label{omega_q}
\end{equation}
For $y\to0$ at fixed $L$, $P_L(T)$ has to go to a finite value and
hence
\begin{equation}
  Q(y)\to\text{const.},\qquad y\to0.
  \label{q_0}
\end{equation}

For the particular case of the correlation length $\xi_L$ of the
finite system the finite-size scaling relation
(\ref{single_par_scale}) becomes
\begin{equation}
  \xi_L(T) = L Q_\xi(y),
  \label{xi_single_scale}
\end{equation}
since comparing Eqs.~(\ref{p_infty}) and (\ref{xi_t}) yields
$\rho=\nu$ and hence $\omega=1$.

The phenomenological finite-size scaling theory for a particular
thermodynamic quantity outlined above can be related to and derived
from the scaling theory based on the renormalization group for {\em
  infinite\/} systems ({Suzuki, }1977; {Br{\'e}zin, }1982; {Barber,
  }1983). The renormalization
group establishes a
relation between the coupling constants ${\bf K}$ of a Hamiltonian $H$
and the coupling constants ${\bf K}'$ of a Hamiltonian $H'$, in which
all lengths have been rescaled by a factor $b$ relative to $H$,
\begin{equation}
  {\bf K}' = {\bf R}({\bf K}),
  \label{r_k}
\end{equation}
where ${\bf R}$ is an analytic function of ${\bf K}$. At a phase
transition there exists a non-trivial fixed point ${\bf K}^*$ of the
recursion Eq.~(\ref{r_k}), ${\bf K^*}={\bf R}({\bf K}^*)$. Near this
fixed point the transformations (\ref{r_k}) can be linearized and the
matrix $\partial R_\alpha/\partial K_\beta$ evaluated at ${\bf
  K}^*$ has the eigenvalues
\begin{equation}
  \lambda_\alpha=b^{y_\alpha}.
  \label{y_eigen}
\end{equation}
Using these eigenvalues Eq.~(\ref{r_k}) can be rewritten as
\begin{equation}
  \zeta_\alpha'= b^{y_\alpha}\zeta_\alpha,
  \label{r_zeta}
\end{equation}
where $\zeta_\alpha$ is a non-linear scaling field that is a
regular function of the coupling constants ${\bf K}$. In terms of the
scaling fields the fixed point ${\bf K}'={\bf K}$ is given by
$\zeta_\alpha=0$.

The singular part of the free energy per degree of freedom $f_s$,
responsible for the divergence of thermodynamic quantities at the
phase transition, can then be expressed in terms of the scaling fields
$\zeta_\alpha$ instead of the coupling constants ${\bf K}$.
$f_s(\zeta_1,\ldots;L)$ has the scaling form
\begin{equation}
  f_s(\zeta_1,\ldots;L) = L^{-d}F(L^{y_1}\zeta_1,\ldots),
  \label{f_scale}
\end{equation}
where $L$ is the length scale of the system that is reached after
iterating the renormalization group equation (\ref{r_k}) a number of
times starting from the microscopic Hamiltonian, and $F$ is an
analytic function of its arguments. From the free energy
(\ref{f_scale}) the scaling behavior of other thermodynamic functions
can be obtained by differentiating the free energy appropriately,
\begin{equation}
  P_L(\zeta_1,\ldots) =
  L^{\omega}\tilde{Q}(L^{y_1}\zeta_1,\ldots).
  \label{p_l_zeta}
\end{equation}
This expression is equivalent, in the thermodynamic limit, to the
single-parameter finite-size scaling hypothesis
(\ref{single_par_scale}) if
\begin{equation}
  \zeta_1=\zeta_t\propto t + O(t^2)
  \label{zeta_t}
\end{equation}
and
\begin{equation}
  y_1=1/\nu,
  \label{y_nu}
\end{equation}
where we have used Eq.~(\ref{xi_t}). Furthermore, all other scaling
fields $\zeta_i$, $i\geq 2$, have to be irrelevant, i.e., $y_i<0$,
$i\geq 2$. We can interpret this statement in the following way. The
Hamiltonian $H$ of a system depends in general on a large number of
coupling constants $K_i$. In the representation in terms of the
scaling field $\zeta_\alpha$ it depends on a large number of
scaling fields. However, the thermodynamic properties in the limit
$L\to\infty$ depend only on a small number of scaling fields
$\zeta_\alpha$ that have positive scaling indices $y_\alpha$. In
conventional phase transitions these are related to the temperature
$t$ (\ref{zeta_t}) and the symmetry breaking field $h$,
\begin{equation}
  \zeta_h\propto h + O(ht^2).
  \label{zeta_h}
\end{equation}

If a coupling constant $K$ shows single-parameter finite-size scaling
it means that the coupling constant $K(bL)$ of the rescaled system is
a function of the coupling constant $K(L)$ of the original system and
the scale factor $b$ only,
\begin{equation}
  K(bL) = f(b, K(L)).
  \label{g_b}
\end{equation}
In the continuum limit, $b\to1$, this property allows one to define a
$\beta$ function that is a function of $K$ only,
\begin{equation}
  \beta(K(L)) = \frac{d \ln K(L)}{d \ln L}.
  \label{beta}
\end{equation}
The fixed points of the renormalization group are given by the zeros
of the $\beta$ function (\ref{beta}). Near the fixed point the $\beta$
function can be linearized,
\begin{equation}
  \beta = y_K \ln(K/K^*),
  \label{beta_lin}
\end{equation}
where Eq.~(\ref{y_eigen}) was used. For $y_K>0$ ($y_K<0$) the fixed
point is repulsive (attractive). Comparison with Eq.~(\ref{y_nu})
shows that the localization length exponent $\nu$ is given by the
inverse slope of the $\beta$ function at the critical point.

\subsection{Corrections to Scaling}
\label{sec:correct}

The scaling fields $\zeta_\alpha$ with negative scaling indices
$y_\alpha<0$ are irrelevant in the thermodynamic limit
$L\to\infty$. However, for finite
systems they lead to corrections to the asymptotic scaling laws and it
can be necessary to take these into account when analyzing numerical
or experimental data for finite systems. For two scaling fields, a
relevant field $\zeta_1$ with $y_1>0$ and an irrelevant field
$\zeta_2$ with $y_2<0$, the renormalization flow of the ``coupling
constants'' $x_1=L^{y_1}\zeta_1$ and $x_2=L^{y_2}\zeta_2$ under
increase of the length scale $L$ is depicted in
Fig.~\ref{fig:saddle_flow}. The arrows indicate the direction of
change of $x_i$ when $L$ is increased. The point $x_1=x_2=0$ is the
fixed point of the renormalization group. The flow corresponds to the
down-hill flow near a saddle point in a potential.

Near the fixed point a thermodynamic quantity
$P_L(\zeta_1,\zeta_2,\ldots)$ can be expanded in the arguments
$x_\alpha$. From Eq.~(\ref{p_l_zeta}) we get
\begin{equation}
  P_L(\zeta_1,\zeta_2,\ldots) = L^{\omega} (\tilde{Q}_0 + a_1
  L^{y_1} \zeta_1 + a_2 L^{-|y_2|} \zeta_2 + \ldots).
  \label{p_expand}
\end{equation}
If the largest irrelevant scaling index $y_2$ is sufficiently large
compared to the next leading index it might be sufficient to take only
these two indices into account and one arrives at a two-parameter scaling
theory. If there are other scaling indices close to $y_2$ it might be
necessary to treat more scaling fields.

For the saddle point fixed point shown in Fig.~\ref{fig:saddle_flow}
the flow along the line $\zeta_2=0$ {\em away\/} from the fixed
point is described by the single-parameter scaling relation
\begin{equation}
  P_L(\zeta_1,0) = L^{\omega} \tilde{Q}(L^{y_1}\zeta_1,0),
  \label{sing_par_zeta}
\end{equation}
where $y_1=1/\nu$. The flow along the line $\zeta_1=0$ {\em
  towards\/} the fixed point is also described by a single-parameter
scaling relation
\begin{equation}
  P_L(0,\zeta_2) = L^{\omega} \tilde{Q}(0,L^{-|y_2|}\zeta_2).
  \label{sing_par_corr}
\end{equation}
Eq.~(\ref{sing_par_corr}) shows that the corrections to scaling due to
the irrelevant scaling field $\zeta_2$ exhibit finite-size scaling
at the fixed point $\zeta_1=0$.

\subsection{Dynamic Scaling}
\label{sec:dynamic}

So far we have focused our discussion on the static properties of
scaling quantities. We will now consider quantities varying in space
and time. Analogous to Eq.~(\ref{p_infty}), we expect in the infinite
system the wave vector $q$ and frequency $\omega$ dependence
\begin{equation}
  P_\infty (T,q,\omega) \propto t^{-\rho}
  Q\left(q\xi(T),\omega\tau_0(T)\right).
  \label{p_infty_dyn}
\end{equation}
The relaxation time $\tau_0(T)$ describes the rate at which the system
relaxes towards its equilibrium state. It diverges at the transition
and the dynamical exponent $z$ is defined by
\begin{equation}
  \tau_0(T) \propto \xi^z \propto t^{-\nu z}.
  \label{z_def}
\end{equation}
In a finite system fluctuation on long length-scales are cut off by
the system size $L$ and from Eq.~(\ref{p_infty_dyn}) we get the
dynamic finite-size scaling relation
\begin{mathletters}
  \label{dyn_sps}
\begin{eqnarray}
  P_L(T,\omega) &=& L^{\omega} Q\left(L/\xi(T),\omega\tau_0(T)\right),\\
  &=& L^{\omega} \tilde{Q}\left(L^{1/\nu}t,\omega^{-1/\nu z}t\right).
\end{eqnarray}
\end{mathletters}
In the limit $\omega\to0$ we recover the static scaling relation
Eq.~(\ref{single_par_scale}).

\section{Scaling Behavior and Field Theory}
\label{cha:ft}

In this section we discuss field-theoretical approaches to the
localization problem and their relation to scaling theory. While field
theory in the context of the integer QHE provides a
framework for a scaling analysis, it cannot at present provide
quantitative results suitable for a comparison with experiments. We
will therefore restrict ourselves to an outline of the motivation and
basic ideas. Thus this section lacks the technical details found in
the other sections.

\subsection{Anderson Localization}
\label{sec:phil}

We will start with a brief survey of the problem of
Anderson localization in the absence of strong magnetic fields. For a
more detailed review the reader is referred to Kramer and MacKinnon
(1993).

Using prior work of Thouless and co-workers ({Edwards and Thouless,
  }1972; {Thouless, }1974; {Licciardello and Thouless, }1975)
and Wegner (1976), Abrahams {\em et al.\/} (1979)
argued that the average dimensionless conductance $g=G \hbar/e^2$
should obey a single parameter scaling relation. They calculated the
$\beta$ function of the conductance,
\begin{equation}
  \frac{d \ln g}{d \ln L} = \beta(g),
  \label{beta_g}
\end{equation}
in the limits $g\ll 1$ and $g\gg1$. In the localized regime, $g\ll1$,
the conductance decreases exponentially with system size,
$g=g_0\exp(-\alpha L)$ and hence
\begin{equation}
  \lim_{g\to0} \beta(g)\propto \ln g.
  \label{g_ll1}
\end{equation}
For large conductance, macroscopic transport theory expresses for a
$d$-dimensional ``cube'' of side $L$ the conductance $G$ in terms of
the conductivity $\sigma$,
\begin{equation}
  G(L)=\sigma L^{d-2},
  \label{g_gg1}
\end{equation}
so that
\begin{equation}
  \lim_{g\to\infty} \beta(g) = d-2.
  \label{g_d_2}
\end{equation}
Assuming a continuous $\beta$ function Abrahams {\em et al.\/} arrived at
the flow diagram depicted in Fig.~\ref{fig:beta}. The $\beta$
function has a zero only in $d>2$ dimensions. They argued that the
$\beta$ function has no zero in $d=2$ dimensions since the first
correction to $\beta$ in $1/g$ is negative in perturbation theory
({Langer and Neal, }1966; {Abrahams {\em et~al.\/}, }1979).

The same behavior is obtained when the disordered electron problem is
mapped onto a non-linear $\sigma$ model ({Wegner, }1979; {Efetov {\em
    et~al.\/}, }1980). We will
briefly sketch how this model is derived. We start by representing
Green functions, like
\begin{equation}
  G^{+}({\bf r},{\bf r}',E) = \langle {\bf r}|(z-H)^{-1}|{\bf
      r}' \rangle,
  \label{g_r_r_z}
\end{equation}
$z=E+i\epsilon$, $\epsilon>0$, as path integrals
\begin{equation}
  G^+({\bf r},{\bf r}',E) =
  \frac{{\displaystyle\int} D[\bar{\Phi}^+] D[\Phi^+] \bar{\Phi}^+({\bf r})
    \Phi^+({\bf r}') \exp\left({\displaystyle\int} d^2r \bar{\Phi}^+({\bf
      r})(z-H)\Phi^+({\bf r})\right)}{\det(z-H)}.
  \label{g_s_s}
\end{equation}
The sign of $\epsilon$ ensures the convergence of the integral. To
represent advanced as well as retarded Green functions we need to
introduce two additional fields $\bar{\Phi}^-$ and $\Phi^-$ with the
opposite sign of $\epsilon$. Equation (\ref{g_s_s}) can be interpreted
as the statistical average $\langle \bar{\Phi}^+({\bf r})\Phi^+({\bf
  r}')\rangle$ with the partition function
\begin{equation}
  {\cal Z} = \int D[\bar{\Phi}^+] D[\Phi^+] \exp\left(\int d^2r
  \bar{\Phi}^+({\bf r})(z-H)\Phi^+({\bf r})\right).
  \label{z_sigma}
\end{equation}
Higher-order Green functions are given by higher-order correlation
functions of the fields.

In order to perform the average over disorder one needs to get rid of
the determinant in Eq.~(\ref{g_s_s}) that depends on the disorder
potential. One way of achieving this goal is the use of the replica
trick ({Wegner, }1979; {Sch{\"a}fer and Wegner, }1980; {Efetov {\em
    et~al.\/}, }1980). Here the fields are replicated $m$
times so that the denominator becomes $[\det(z-H)]^m$. In the formal
limit $m\to0$, performed at the end of the calculation, the
determinant becomes unity. Instead, the supersymmetry
method ({Efetov, }1983) can be employed where commuting as well as
anti-commuting (Grassmann) fields are used. Then the determinants for the
commuting and anti-commuting fields cancel exactly.

Using one of these schemes, the average over disorder in
Eq.~(\ref{z_sigma}) can readily be performed for the Gaussian
white-noise potential (\ref{white}). Since the disorder potential
couples two fields this leads to a term containing a product of
four fields in the exponential. These quartic terms can be
decoupled by a Hubbard-Stratonovic transformation to matrix fields
$Q({\bf r})$. The effective Lagrangian for this matrix is ({Pruisken
  and Sch{\"a}fer, }1981)
\begin{equation}
  L[Q]=\int d^2r \left[-\frac{1}{2V_0^2} {\rm Tr}\, Q^2({\bf r}) + {\rm Tr}
  \ln (E-H_0-Q)\right],
  \label{L_q}
\end{equation}
where we dropped convergence-ensuring terms.  This equation is the
starting point of the derivation of the non-linear $\sigma$ model.
It proceeds by evaluating the integral in Eq.~(\ref{L_q}) in
saddle-point approximation, suitably decomposing the $Q$-matrix, and
expanding the trace of the logarithm. The result is an effective
Lagrangian
\begin{equation}
  L(\tilde{Q}) = \int d^2r \left[-\frac{1}{t} {\rm Tr} (\partial
    \tilde{Q})^2\right],
   \label{L-eff}
\end{equation}
where the coupling constant $1/t$ is proportional to the
conductance.

The symmetry of the matrices $Q$ depends on the symmetry of the original
Hamiltonian and the choice of commuting or anti-commuting variables.

For the non-linear $\sigma$ model (\ref{L-eff}) the $\beta$
function for the coupling constant and hence for the conductivity can
be calculated ({Br{\'e}zin {\em et~al.\/}, }1980). The results depend
on the symmetry class
of the $Q$-fields. One distinguishes three universality classes: The
orthogonal class for random potential scattering, the symplectic class
in the presence of spin-orbit coupling and the unitary class in the
presence of a magnetic field. The results of Wegner (1989)
may be written in terms of the dimensionless conductance $g$ as
follows:
\begin{equation}
  \beta(g) = \left\{ \begin{array}{ll}
  (d-2) -g^{-1} - \frac{3}{4}\zeta(3)g^{-4} + {\cal O}(g^{-5}), &
  \text{orthogonal case}\\
  (d-2) -2g^{-2} -6g^{-4} + {\cal O}(g^{-6}), & \text{unitary case}\\
  (d-2) +g^{-1} - \frac{3}{4}\zeta(3)g^{-4} + {\cal O}(g^{-5}), &
  \text{symplectic case}
\end{array}\right.
  \label{beta_loop}
\end{equation}
Only in the symplectic case does the $\beta$ function have a zero in
$d=2$ dimensions while in the other cases all states are localized. In
particular, the presence of a magnetic field in the unitary case does
not lead to extended states necessary for explaining the QHE. We will
come back to this problem in the next section.

Further support for the scaling picture of Abrahams {\em et al.\/}
(1979) comes from numerical calculations
({MacKinnon and Kramer, }1981; {Pichard and Sarma, }1981; {MacKinnon and
  Kramer, }1983). For a recent review see Kramer and MacKinnon
(1993). Numerically it is easier to calculate the
localization length instead of the conductance. This can be done
recursively for strips or bars using either the Green-function method
of section \ref{sec:green} or the transfer-matrix method of section
\ref{sec:transfer}. The numerical results are compatible with the
finite-size scaling hypothesis (\ref{xi_single_scale}). In
one-dimensional systems no extended states are observed. In
two-dimensional systems a transition is only observed in the presence
of spin-orbit scattering. In three-dimensional system a
localization-delocalization transition is observed at a critical
disorder.

After the discovery of the absence of self-averaging in mesoscopic
systems ({Anderson {\em et~al.\/}, }1980; {Altshuler, }1985; {Lee and
  Stone, }1985) it became apparent that the description of
the metal-insulator transition in terms of the average conductance is
incomplete. Altshuler, Kravtsov, and Lerner (1986; 1989)
argued that the whole distribution of the conductance needs to be
considered. Single-parameter scaling in this context was discussed by
Shapiro (1987). His scaling parameter need not necessarily
be the average or typical conductance but any parameter that uniquely
determines the distribution function.

\subsection{Nonlinear Sigma Model and Topological Term}
\label{sec:pruisken}

In the previous section it was pointed out that the absence of
extended states in the non-linear $\sigma$ model for the unitary
two-dimensional case apparently contradicts the explanations for the
QHE ({Aoki and Ando, }1981; {Laughlin, }1981; {Halperin, }1982) which require
the
existence of extended states for sufficiently small disorder. Pruisken
(1984) and collaborators ({Levine {\em et~al.\/}, }1983; {Levine {\em
    et~al.\/}, }1984a; {Levine {\em et~al.\/}, }1984b; {Levine {\em
    et~al.\/}, }1984c)
argued that the failure of the non-linear $\sigma$ model is due to
the emergence of a topological term in the effective Lagrangian that
is not accessible to the perturbative treatment that leads to the
Lagrangian (\ref{L-eff}). A review of the discussion can be found in
Pruisken's chapter in the book by Prange and Girvin (1987).
Here we want to sketch the main features of this field theory.

The Lagrangian (\ref{L-eff}) has to be augmented by a term proportional
to the Hall conductivity,
\begin{equation}
  {\cal L} = \int d^2r \left(-\frac{1}{8}\sigma_{xx}^0 {\rm Tr}
  (\partial \tilde{Q})^2
  + \frac{1}{8}\sigma_{xy}^0{\rm Tr} \tilde{Q}\left[\partial_x
  \tilde{Q},\partial_y \tilde{Q}\right]\right).
  \label{L_pruisken}
\end{equation}
The values of the coupling constants $\sigma_{xx}^0$ and
$\sigma_{xy}^0$ in (\ref{L_pruisken}) are the mean-field values
corresponding to the conductivity tensor on short length scales.  They
are given by the SCBA ({Ando {\em et~al.\/}, }1975). Upon increase of
the length scale
the conductivities get renormalized to their physical values on long
length scales. We will use the correspondence between the scaling on
short length scales and the SCBA in our discussion of the corrections
to scaling in Sec.~(\ref{sec:hll}).

The occurrence of two coupling constants $\sigma_{xx}^0$ and
$\sigma_{xy}^0$ in the Lagrangian leads to a coupling of the
renormalizations of $\sigma_{xx}$ and $\sigma_{xy}$ and changes the
critical behavior compared to the non-linear $\sigma$ model for
Anderson localization. In a dilute-instanton-gas approximation for the
physically relevant field configurations in the weak-coupling limit
$\sigma_{xx}\gg1$ Pruisken derived $\beta$ functions for the
renormalization with system size $L$ of both $\sigma_{xx}$ and
$\sigma_{xy}$,
\begin{mathletters}%
  \label{beta_pruisken}
\begin{eqnarray}
  \beta_{xx} &=& \frac{\partial \sigma_{xx}}{\partial \ln L} =
  \frac{-1}{2\pi^2\sigma_{xx}} - \sigma_{xx}D
  e^{-2\pi\sigma_{xx}} \cos(2\pi\sigma_{xy}),\\
  \beta_{xy} &=& \frac{\partial \sigma_{xy}}{\partial \ln L} = -
  \sigma_{xx}D e^{-2\pi\sigma_{xx}} \sin(2\pi\sigma_{xy}) .
\end{eqnarray}
\end{mathletters}%
This leads to the renormalization-group flow diagram proposed by
Khmel'nitskii (1983) shown in Fig.~\ref{fig:sigma_flow}. It
has two kinds of fixed points: stable fixed points at
$\sigma_{xx}=0$, $\sigma_{xy}=n$ and fixed points at
half-integer values of $\sigma_{xy}$ with a finite conductivity
$\sigma_{xx}^*$. The former fixed points correspond to the localized
wave functions of the model in the absence of the topological term
while the latter correspond to the extended wave functions that carry
the current in the QHE. Furthermore the flow diagram
is periodic in $\sigma_{xy}$ so that the critical behavior is
independent of Landau level index, since ({Levine {\em et~al.\/}, }1983)
\begin{equation}
  \frac{\sigma_{xy}^0}{8}\int d^2r \text{Tr} \tilde{Q}\left[\partial_x
  \tilde{Q}, \partial_y \tilde{Q}\right] = 2\pi i q \sigma_{xy}^0.
  \label{topo_inv}
\end{equation}

The delocalization fixed points at half-integer $\sigma_{xy}$ are
partly attractive and partly repulsive like the fixed point discussed
in connection with Fig.~\ref{fig:saddle_flow}. They are characterized
by a relevant scaling index $y_1=1/\nu>0$ associated with the deviation
from $\sigma_{xy}^*=(2n+1)/2$ and an irrelevant scaling index
$y_2<0$ associated with the deviation from $\sigma_{xx}^*$.
The starting point for the flow in Fig.~\ref{fig:sigma_flow} at
microscopic length scales is given by the expressions for the
conductivity tensor in the self-consistent Born approximation (SCBA)
({Ando {\em et~al.\/}, }1975).

While the field theory with the Lagrangian (\ref{L_pruisken}) provides
a very appealing framework for the quantum Hall transition it has
major shortcomings. Up to date it has not led to quantitative
predictions, besides the existence of isolated critical fixed points.
It has not been possible to calculate the longitudinal conductivity
$\sigma_{xx}^*$ at the critical point, the localization length
exponent $\nu$, or the irrelevant exponent $y_2$. The validity of the
dilute instanton gas approximation that forms the basis of the
approximate flow equations (\ref{beta_pruisken}) has been questioned
by Weidenm\"uller and Zirnbauer (1988). Recently, Zirnbauer
(1994) has argued that the flow-diagram
Fig.~\ref{fig:sigma_flow} does not describe the flow of the coupling
constants in the nonlinear $\sigma$ model (\ref{L_pruisken}) and that
a different theory is called for. We conclude that at present the most
promising tool to obtain quantitative results about the critical
properties of the integer quantum Hall system seems to be numerical
simulation, on which we will now focus our attention.

\section{Localization Length}
\label{cha:loclen}

The localization length $\xi(E)$ is one of the simplest quantities to
calculate numerically that shows the occurrence of a metal-insulator
transition. It can be defined by the asymptotic behavior of the
modulus of the single-particle Green function
\begin{equation}
  G({\bf r},{\bf r}';E) = \langle {\bf r}|\frac{1}{E-H}|{\bf r}'\rangle.
  \label{g_e}
\end{equation}
The disorder average of $G({\bf r},{\bf r}';E)$ is a short-ranged
function due to the average over the phase fluctuations. It decays on
the scale of the elastic mean-free path and does not show critical
behavior. By averaging the modulus these phase cancellations are
avoided and $\overline{|G({\bf r},{\bf r}';E)|}$ becomes long-ranged
near the phase transition. $|G({\bf r},{\bf r}';E)|$ is not a
self-averaging quantity and has a broad distribution. The inverse
localization length
\begin{equation}
  \xi(E)^{-1} = - \lim_{|{\bf r}-{\bf r}'|\to\infty} \frac{1}{|{\bf
      r}-{\bf r}'|} \overline{\ln |G({\bf r},{\bf r}';E)|}
  \label{xi_g}
\end{equation}
on the other hand is a self-averaging quantity so that it takes on the
same value for almost all members of the disorder ensemble. The
self-averaging property makes it particularly useful for numerical
calculations. In Eq.~(\ref{xi_g}) the energy $E$ has to have an
imaginary part since after averaging $\overline{|G({\bf r},{\bf
    r}';E)|}$ is non-analytic on the real axis. Using the
self-averaging property and calculating the localization length for a
single disorder realization in a finite system this imaginary part can
be set to zero as long as $E$ is not an eigenenergy of the system.

In order to implement a finite-size scaling analysis of the
localization length one studies long strips (in $d=2$ dimensions) or
bars (in $d=3$ dimensions) and calculates the localization length
along the infinite direction of the system. This localization length
then depends on the perpendicular dimensions of the system
({MacKinnon and Kramer, }1981; {MacKinnon and Kramer, }1983). In
section \ref{sec:green} we review a method to
calculate recursively the localization length for systems of this
geometry and extend it to the case of the random Landau matrix.

Another way to define the localization length is to identify it
with the inverse of the smallest Lyapunov exponent of the transfer
matrix of the system ({MacKinnon and Kramer, }1981; {MacKinnon and
  Kramer, }1983; {Pichard and Sarma, }1981). We will return to this in
section \ref{sec:transfer}.

\subsection{Recursive Green Function Method}
\label{sec:green}

The recursive Green function method was developed by MacKinnon and
Kramer (1981; 1983) for the Anderson model (\ref{anderson}).
In one dimension the corresponding Schr\"odinger equation
(\ref{ander_sch}) reads
\begin{equation}
  a_{i+1}= (E-\epsilon_i) a_i - a_{i-1},
  \label{ander_sch1}
\end{equation}
where the hopping matrix element was set to unity.
Eq.~(\ref{ander_sch1}) allows one to calculate recursively the wave
function amplitudes $a_i$. In higher dimensions the corresponding
equation is
\begin{equation}
  A_{i+1}= (E-H_i) A_i - A_{i-1}.
  \label{ander_schd}
\end{equation}
$H_i$ is the Hamiltonian of the $i^{\text{th}}$ $(d-1)$-dimensional
slice of the $d$-dimensional ``bar'' and all quantities are
$(d-1)$-dimensional matrices. Eq.~(\ref{ander_schd}) can be rewritten in
terms of Green functions $G_{1,n}$ coupling sites in slice $1$ to
sites in slice $i$,
\begin{mathletters}%
  \label{ander_rec}
\begin{eqnarray}
  G_{1,i+1} &=& G_{1,i} G_{i+1,i+1},\\
  G_{i+1,i+1} &=& (E - H_{i+1} - G_{i,i})^{-1}.
\end{eqnarray}
\end{mathletters}%
Iterating Eq.~(\ref{ander_rec}) the matrix elements of the Green
function connecting sites at both ends of long ``bars'' can be
calculated without need for diagonalization of the Hamiltonian of the
whole system.

Eq.~(\ref{ander_rec}) is tailored to the Anderson model with constant
nearest-neighbor couplings. To apply this method to the
Hamiltonian (\ref{Ham_matrix}) it has to be modified. The random
Landau matrix $\langle nk_1|V|nk_2\rangle$ of Eq.~(\ref{ncor})
describes a random one-dimensional tight-binding Hamiltonian similar to
the Anderson Hamiltonian but with long-ranged hopping matrix
elements. To be able to apply recursion relations similar to
Eqs.~(\ref{ander_rec}) Aoki and Ando (1985; 1985)
reinterpreted the one-dimensional Hamiltonian $\langle
nk_1|V|nk_2\rangle$ as the Hamiltonian of a two-dimensional system
with couplings only between sites in neighboring slices. This is
possible since $|\langle nk_1|V|nk_2\rangle|^2$ is essentially zero
for $|k_1-k_2|>K$ and can be neglected, where $K$ is a constant of
order $(\beta l_c)^{-1}$ such that $N_K=K L_y/2\pi$ is an integer.
States with $k\in [(i-1)K+1,iK[$ belong to slice $i$ and are coupled
only to states in slices $i-1$, $i$, and $i+1$. $\langle
nk_1|V|nk_2\rangle$ can then be rewritten as a tri-diagonal matrix
with diagonal elements $H_{i,i}$ being the $N_K\times N_K$
submatrices of $\langle nk_1|V|nk_2\rangle$ connecting states within
slice $i$ and off-diagonal elements $H_{i,i+1}$ being the
$N_K\times N_K$ submatrices connecting states within slice $i$ with
those in slice $i+1$. The matrix elements in the off-diagonal matrices
$H_{i,i+1}$ are now random numbers in contrast to the case of the
two-dimensional Anderson model so that the recursion relation
(\ref{ander_rec}) has to be modified to
\begin{mathletters}%
  \label{aoki_rec}
\begin{eqnarray}
  G_{1,i+1} &=& G_{1,i} H_{i,i+1} G_{i+1,i+1},\\
  G_{i+1,i+1} &=& (E - H_{i+1} - H_{i+1,i} G_{i,i} H_{i,i+1})^{-1}.
\end{eqnarray}
\end{mathletters}%

A different approach was introduced by Huckestein (1990)
who generalized the one-dimensional Green function method to the
case of long-ranged hopping matrix elements. Consider the random
Landau matrix $H^{(K)}$ of dimension $K$ describing a system
containing states $k=2\pi i/L_y$, $i=1,\ldots,K$. In the following, we
will label the states by integers $k$ instead of quasi-momenta
$2\pi k/L_y$. Adding the state $K+1$ to the system leads to a
Hamiltonian $H^{({K}+1)}$ that can be decomposed into a
block-diagonal part $H_0$ and the couplings $H'$ of state
${K}+1$ to the rest of the system,
\begin{mathletters}%
  \label{H_decomp}
\begin{eqnarray}
  H^{({K}+1)} &=& \sum_{{k},{k}'=1}^{{K}+1}
  |k\rangle \langle k|V|k'\rangle \langle k'|,\\
  &=& H_0 + H',\\
  H_0 &=& H^{({K})} + |K\rangle \langle K|V|K\rangle \langle
  K|,\\
  H' &=& \sum_{{k}=1}^{{K}} (|k\rangle \langle
  k|V|K+1\rangle \langle K+1| + \text{h.c.}),
\end{eqnarray}
\end{mathletters}%
where we have suppressed the Landau level index.  The related Green
functions are $G^{({K})}=(E-H^{({K})})^{-1}$ and
$G_0=(E-H_0)^{-1}$ for which we obtain the recursion relation
\begin{mathletters}%
  \label{bodo_rec}
\begin{eqnarray}
  G_0 &=& G^{({K})} + \frac{|{K}+1\rangle\langle K+1|}{E-\langle
    K+1|V|K+1\rangle}, \\
  G^{K+1} &=& G_0 + G_0 H' G^{K+1}.
\end{eqnarray}
\end{mathletters}%
The last relation can be used to calculate $G^{K+1}$ owing to the fact
that the matrix $H'$ only couples to site $K+1$. Using the property
that the random Landau matrix is a banded matrix the dimensions of the
matrices in Eqs.~(\ref{bodo_rec}) can be kept finite even in the limit
$K\to\infty$. Choosing matrices of dimension $2M/\beta$, where
$M=L_y/\sqrt{2\pi}l_c$, results in neglecting matrix elements of
order $\exp(-4\pi)$ compared to the diagonal matrix elements. Let
$N$ be the dimension of these matrices, then the recursion relation
(\ref{bodo_rec}) for the matrix elements are
\begin{eqnarray}
  \lefteqn{\langle K+1|G^{(K+1)}|K+1 \rangle =}\nonumber\\
  & & \left( E - \langle K+1|V|K+1
  \rangle - \sum_{i,j=K-N+2}^K \langle K+1 |V| i\rangle \langle i|
  G^{(K)} | j\rangle \langle j|V|K+1 \rangle \right)^{-1}
  \label{bodo_rec_kk}
\end{eqnarray}
for the diagonal matrix element,
\begin{equation}
  \langle i|G^{(K+1)}|K+1 \rangle = \langle K+1 | G^{(K+1)}| K+1
  \rangle \sum_{j=K-N+2}^{K} \langle i| G^{(K)} | j\rangle \langle
  j|V|K+1 \rangle
  \label{bodo_rec_kj}
\end{equation}
for $i\leq K$, and
\begin{equation}
  \langle i| G^{(K+1)} | j \rangle = \langle i| G^{(K)} | j \rangle +
  \sum_{k=K-N+2}^{K} \langle i| G^{(K)}| k \rangle \langle k|V|K+1
  \rangle \langle K+1 | G^{(K+1)} | j \rangle
  \label{bodo_rec_ij}
\end{equation}
for $i\leq K$ and $j\leq K$.

The localization length $\lambda_M$ of a system with width
$M=L_y/\sqrt{2\pi}l_c$ and length $K$ is given by
\begin{equation}
  \lambda_M^{-1}(E;K) = - \frac{L_y}{K 2\pi l_c^2} \ln |\langle 1
  | G^{(K)}(E) | K \rangle|.
  \label{lambda_M}
\end{equation}
In calculating $\langle 1 | G^{(K)} | K \rangle$, we are faced with the
difficulty that this is an exponentially decreasing quantity. To
circumvent this problem we introduce variables $q_K = \langle 1 |
G^{(K)} | K \rangle / \langle 1 | G^{(K-1)} | K-1 \rangle$ for
the change in $\langle 1 | G^{(K)} | K \rangle$ when the system size
is increased by one state. In terms of the $q_k$, Eq.~(\ref{lambda_M})
is given by
\begin{mathletters}%
  \label{lambda_M_q}
\begin{eqnarray}
  \lambda_M^{-1}(E;K) &=& - \frac{M}{K \sqrt{2\pi}l_c} \ln
  \left| \frac{\langle 1 | G^{(K)} | K \rangle}{\langle 1 | G^{(K-1)} | K-1
    \rangle}
  \frac{\langle 1 | G^{(K-1)} | K-1 \rangle}{\langle 1 | G^{(K-2)} |
    K-2 \rangle}
  \cdots
  \langle 1 | G^{(1)} | 1 \rangle\right| ,\\
  &=& - \frac{M}{K \sqrt{2\pi}l_c}
  \sum_{i=1}^K \ln|q_i|,
\end{eqnarray}
\end{mathletters}%
where we set $q_1=\langle 1 | G^{(1)} | 1 \rangle = (E-\langle 1|V|1
\rangle)^{-1}$. From Eq.~(\ref{bodo_rec_kj}) we get a recursion
relation for $q_{K+1}$,
\begin{equation}
  q_{K+1} = \langle K+1 | G^{(K+1)}| K+1 \rangle \sum_{j=K-N+2}^{K}
  \langle 1| g^{(K)} | j\rangle \langle j|V|K+1 \rangle
  \label{rec_q},
\end{equation}
where $\langle 1| g^{(K)} | j\rangle = \langle 1| G^{(K)} | j\rangle /
\langle 1| G^{(K)} | K\rangle$ is normalized to compensate for the
exponential decay. With Eq.~(\ref{bodo_rec_ij}), we obtain the recursion
relation for $\langle 1| g^{(K+1)} | j\rangle$,
\begin{eqnarray}
  \lefteqn{\langle 1| g^{(K+1)} | j\rangle = \frac{1}{q_{K+1}} \Biggl(
    \langle 1 | g^{(K)} | j\rangle}\nonumber\\
  & & + \sum_{k=K-N+2}^{K} \langle 1| g^{(K)} | k
  \rangle \langle k|V|K+1 \rangle \langle K+1 | G^{(K+1)} | j \rangle
  \Biggr).
  \label{rec_g}
\end{eqnarray}

\subsection{Transfer-Matrix Method}
\label{sec:transfer}

The transfer-matrix method deals with a physical situation where a long
narrow disordered system is connected at both ends to semi-infinite
ideal conductors. The transfer matrix $T$ relates the amplitudes on
the right-hand side of the system to those on the left-hand side. For the
Anderson model (\ref{ander_schd}) the transfer matrix $T_i$ of slice
$i$ is given by
\begin{equation}
  T_i = \left ( \begin{array}{cc}
  E - H_i & -I \\
  I & 0
  \end{array} \right ),
  \label{transfer_ander}
\end{equation}
where $I$ is the unit matrix, since Eq.~(\ref{ander_schd}) can be
rewritten as
\begin{equation}
  {{A_{i+1}} \choose {A_i}} = T_i {A_{i} \choose A_{i-1}}.
  \label{ander_trans}
\end{equation}
For Chalker's and Coddington's network model of section \ref{sec:network}
the transfer matrix $T$ consists of a product of four matrices, $T = ABCD$,
where $A$ and $C$ are diagonal matrices with random phases,
corresponding to the links connecting the nodes of the network, and
$B$ and $D$ are block-diagonal, consisting of the $2\times 2$
matrices $M$ of Eq.~(\ref{m_sol}). Note that the network of
Fig.~\ref{fig:net_scheme} is periodic when using two columns of nodes
as a unit.

{}From the definition of the transfer matrix it follows immediately that
the transfer matrix of a long system is just the product of the
transfer matrices of the slices composing the system,
\begin{equation}
  M_n = \prod_{i=1}^n T_i,
  \label{transfer_prod}
\end{equation}
where $M_n$ is the transfer matrix of the system consisting of $n$
slices and $T_i$ is the transfer matrix of slice $i$. Oseledec's
theorem ({Oseledec, }1968) states that a limiting matrix $\Gamma$ exists,
\begin{equation}
  \Gamma = \lim_{n\to\infty} (M_n^* M_n)^{1/2n},
  \label{oseledec}
\end{equation}
with eigenvalues $\exp\gamma_1$, \ldots, $\exp\gamma_{2m}$, where
$2m$ is the dimension of the transfer matrix. The $\gamma_i$ are
the characteristic Lyapunov exponents of $M_n$. The inverse
localization length $\xi^{-1}$ is given by the Lyapunov exponent of
smallest absolute magnitude, $\xi^{-1} = \text{min}|\gamma_i|$.

\section{Other Scaling Quantities}
\label{cha:others}

The localization length is by no means the only quantity that can be
used in finite-size scaling studies. In fact, early numerical
calculation on the QHE used the size-dependence of the
Thouless number to obtain information about the critical behavior
({Ando, }1983; {Ando, }1984a; {Ando, }1984b). Information about the
localization
properties of eigenstates can also be obtained from their topological
character. Huo and Bhatt (1992) used the density of states
with non-zero Chern number as a finite-size scaling quantity.

\subsection{Thouless Number}
\label{sec:thouless}

Edwards and Thouless (1972) have argued that the sensitivity
of eigenenergies of finite systems to changes in the boundary
conditions can be used to distinguish between extended and localized
states. For localized states the eigenenergies should be insensitive
against the change in boundary conditions if the system size is large
compared to the localization length. Extended states should feel the
change in the boundary conditions and the shift in energy should be of
the order of $\hbar/\tau$, where $\tau$ is the time it takes
the electron to diffuse to the boundary ({Thouless, }1974). In second-order
perturbation theory the change in energy $\delta E$ when periodic
boundary conditions are replaced by anti-periodic boundary conditions
is approximately related to the conductivity $\sigma$ by
({Licciardello and Thouless, }1975)
\begin{equation}
  \sigma L^{d-2} = \frac{e^2}{h} f \frac{\delta E}{\Delta E} =
  \frac{e^2}{h} f g(L),
  \label{thouless_number}
\end{equation}
where $\Delta E$ is the mean level spacing at the Fermi energy and
$f$ is a constant of order unity depending on details of the
microscopic model. $g(L)$ is called the Thouless number. In the
presence of a strong magnetic field Ando (1983) argued that
the constant $f$ is $\pi/2$. The Thouless number $g(L)$ for a particular
level is a strongly fluctuating quantity as a function of level index.
To get a less fluctuating quantity typically the geometric mean
$\delta E = \exp(\langle \ln \delta E_i \rangle_{\text{ave}})$
is used, where the average is taken over a given energy interval. On
the other hand, the very fact that $g(L)$ is strongly fluctuating was
used by Fastenrath, Jan\ss{}en and Pook (1992) in a
multifractal analysis of the integer quantum Hall transition
(Sec.~\ref{sec:multiscale}).

\subsection{Chern Number}
\label{sec:chern}

While the Thouless number relates to the longitudinal conductivity
$\sigma_{xx}$, the ability of an eigenstate to contribute to the Hall
conductivity $\sigma_{xy}$ serves as another means of distinguishing
extended from localized states. For a system with generalized periodic
boundary conditions,
\begin{equation}
  t({\bf L}_j)|\psi\rangle =
\exp(i\theta_j)|\psi\rangle \qquad (j=1,2),
  \label{gen_per_bound}
\end{equation}
where $t({\bf L}_j)$ is the magnetic translation operator
({Zak, }1964; {Haldane and Rezayi, }1985; {Arovas {\em et~al.\/},
  }1988), the Hall conductivity may be
written as a sum over all occupied states below the Fermi energy $E$
({Thouless {\em et~al.\/}, }1982; {Pook and Hajdu, }1987)
\begin{eqnarray}
  \sigma_{xy}(E) &=& \frac{e^2}{h} \sum_m\frac{i}{2\pi} \int_{T^*}
  d^2\theta\, \Theta(E-E_m)\nonumber\\
  & & \times \left(\left\langle \frac{\partial \tilde{\psi}_m({\bf
      \theta})}{\partial \theta_1}\right|\left. \frac{\partial
    \tilde{\psi}_m({\bf \theta})}{\partial \theta_2}\right\rangle -
  \left\langle \frac{\partial \tilde{\psi}_m({\bf
      \theta})}{\partial \theta_2}\right|\left. \frac{\partial
    \tilde{\psi}_m({\bf \theta})}{\partial
    \theta_1}\right\rangle\right),\\
  |\tilde{\psi}_m\rangle &=& \exp[-i(\theta_1 x/L_x + \theta_2
  y/L_y)] |\psi_m\rangle,
  \label{sigma_thouless}
\end{eqnarray}
where $T^*$ is the torus $\theta_i\in[0,2\pi[$, $i=1,2$.
It has been shown by Thouless and co-workers ({Thouless {\em
    et~al.\/}, }1982; {Niu {\em et~al.\/}, }1985) and
others ({Avron {\em et~al.\/}, }1983; {Kohmoto, }1985; {Aoki and Ando,
  }1986; {Pook and Hajdu, }1987) that the integral in
Eq.~(\ref{sigma_thouless}) is $2\pi$ times an integer $C_1(m)$, the
first Chern index, which is a topological invariant characterizing the
topological properties of the wave functions. Arovas {\em et al.\/}
(1988) have shown that for states with finite $C_1(m)$ the zeros
of the wave function can be moved to any position in real space by a
suitable choice of the boundary angles $\theta_1,\theta_2$, while
states with zero $C_1(m)$ have zeros that are confined in space. This
leads to a natural distinction between extended and localized wave
function. Again, as in the discussion of the Thouless number, the
sensitivity to boundary conditions is used to distinguish extended
from localized states. Since the first Chern index is an integer it
allows a clear definition of extended versus localized states even
for finite systems.

\section{Numerical Calculations}
\label{cha:num}

First numerical calculations of the critical properties for
two-dimensional systems in a strong magnetic field were performed by
Aoki (1977; 1978; 1985), Ando
(1983; 1984a; 1984b), and Schweitzer, Kramer, and MacKinnon
(1984). Mostly due to the insufficient computing power
available at the time these studies were not able to obtain accurate
information about the behavior of the localization length or the
Thouless number. Ando (1983; 1984a; 1984b) interpreted his
results obtained from the size dependence of the Thouless number as
being compatible with an extended state at a single energy in each
Landau level, a result previously obtained by Ono (1982b) by
summing a certain class of diagrams analytically.  However, the
divergence of the localization length obtained numerically was much
weaker than Ono's result $\xi(E)\propto \exp(1/E^2)$, where the
energy is measured relative to the critical energy. While Ando was not
able to extract a result for the localization length exponent $\nu$
from his data he found first signs of a single-parameter scaling law
since his numerically constructed $\beta$ function for the
conductance depended only the conductance itself. Schweitzer, Kramer,
and MacKinnon (1984) on the other hand interpreted their
results for the localization length as indicating a band of extended
states separated from localized states in the band tails by two
mobility edges.

In the following we will concentrate on more recent results that shed
light on the critical behavior near the band centers of the Landau
levels. In our discussion we have to keep in mind that numerical
calculations always deal with finite systems while strictly speaking
phase transitions exist only in infinite systems. Thus, we need a
theoretical framework in which to analyze data for finite systems in
order to extrapolate results to infinite system size. The finite-size
scaling theory outlined in section \ref{cha:scaling} provides such
a framework.

We will first review calculations for Hamiltonians projected onto a
single Landau level. In the strong magnetic field limit,
$\hbar\omega_c\gg\Gamma$, this seems to be a reasonable
approximation. The effects of Landau level mixing will be discussed in
Sec.\ \ref{sec:coupl} while the influence of a periodic potential on a
single Landau level is considered in Sec.\ \ref{sec:period}.

\subsection{Lowest Landau Level}
\label{sec:lll}

The first calculation that produced reasonable estimates of the
critical exponents for the quantum Hall system was performed by Aoki
and Ando (1985)({Ando and Aoki, }1985). They used a real-space model
for the disorder with the random potential given by a sum over either
short-range scatterers (\ref{sum_delta}) or Gaussian scatterers
(\ref{sum_gauss}). Using the recursive Green function method of
section \ref{sec:green} they calculated the localization length for
strips of width 1.25 to $40.1\,l_c$. In the direction perpendicular to
the strip periodic boundary conditions were employed to avoid the
effects of edge states extended along the edges of the strip. Aoki and
Ando tried to interpret their data in terms of Pruisken's two-parameter
scaling theory (Sec.~\ref{sec:pruisken}). However, in view of later
calculations ({Huckestein and Kramer, }1989; {Huckestein and Kramer,
  }1990) it has to be realized that their systems
were too small to be in the critical region. Nevertheless, they found that
the localization length diverges at a single energy at the center of
the Landau band and estimated the localization-length exponent to be
$\nu\approx 2$ in the lowest Landau level.

Huckestein and Kramer (1990) extended these calculations by
introducing the random Landau-matrix model (Sec.\ \ref{sec:lanmod}).
Comparing their results for the localization length with the results
of Aoki and Ando shows ({Huckestein and Kramer, }1989) that the
different random
potentials give the same results for a width of $40\,l_c$, the largest
calculated by Aoki and Ando, while for smaller widths differences are
observed. Microscopic details of the disorder potential therefore seem
to become irrelevant for the behavior of the localization length on
this length scale for $\delta$-correlated potentials.
In Fig.~\ref{fig:lll_0} the normalized localization length
$\lambda_M(E)/M$ is plotted as a function of the system width
$M=L_y/(2\pi)^{1/2}l_c$ for energies between 0.01 and
1.0$\,\Gamma$ relative to the band center. The
length of the systems was $2.5\times 10^5l_c$ ($2.5\times 10^4l_c$
for the widest systems) and the width ranged up to $321\,l_c$. The
data within the dotted region are compatible with a single-parameter
scaling law
\begin{mathletters}
  \label{lambda_xi}
\begin{eqnarray}
  \lambda_M(E) &=& M \Lambda(M/\xi(E)),\label{lambda_xi_a}\\
  \xi(E) &=& \xi_0 \left|\frac{E-E_c}{\Gamma_\sigma}\right|^{-\nu}.
\end{eqnarray}
\end{mathletters}
At this point we want to elaborate on the significance of this
statement. In Sec.\ \ref{sec:green} it was argued that the
inverse localization length calculated from the Green function is
self averaging for system length $L\to\infty$. Numerically, it can
only be calculated for finite $L$. The value obtained fluctuates
within the disorder ensemble. The statistical error can be estimated
from sampling different realizations of the disorder. In addition,
systematic errors of the order of $\xi(E)/L$ have to be expected. In
the present calculation the statistical errors are typically less than
1\%, with possible systematic errors of the same order. Using a
statistical test ({Huckestein, }1990), it was determined that the data in the
dotted region are compatible with both scaling relations in
Eq.~(\ref{lambda_xi}) within the statistical errors. Corrections to
Eq.~(\ref{lambda_xi}) larger than the statistical errors are ruled
out by this procedure. The deviations from scaling behavior for
smaller width in Fig.~\ref{fig:lll_0} are not only due to irrelevant
scaling fields, as discussed in Sec.\ \ref{sec:correct}, but are also
associated with the particular one-dimensional limit of the system. As
was pointed out by Ando and Aoki (1985) for $M\to0$ the
localization length can be calculated analytically. In this limit the
distance $\Delta X=\Delta k l_c^2= 2\pi l_c^2/L_y$ between the
center coordinates of neighboring Landau states becomes large compared
to $l_c$ so that the matrix elements $\langle nk|V|nk'\rangle$ fall
off rapidly as a function of $|k-k'|$. The random Landau matrix
becomes a tri-diagonal Hermitean matrix corresponding to a
tight-binding model with nearest-neighbor hopping. Applying the exact
result for the localization length of that model ({Thouless, }1974) we get
\begin{equation}
  \lim_{M\to0}\frac{\lambda_M}{M} = \frac{2}{\pi}\frac{1}{\beta^2},
  \label{1d_lim}
\end{equation}
independent of energy.

The critical region starts at system width $M=16$, the same size at
which the differences between the real-space model of Aoki and Ando
and the random Landau-matrix model vanish. For energies larger than
$0.5\Gamma$ no scaling is observed since the localization length
does not decrease arbitrarily in the band tails but levels off at
about one magnetic length. Fig.~\ref{fig:scale_curve} shows the data
from the dotted region of Fig.~\ref{fig:lll_0} scaled according to
Eq.~(\ref{lambda_xi_a}). Note that this figure contains data not only
from Fig.~\ref{fig:lll_0} but also from Figs.~\ref{fig:lll_1},
\ref{fig:nll_1} and from the network model (Sec.\
\ref{sec:netcalc}). The scaling relations Eq.~(\ref{lambda_xi})
determine $\xi(E)$ only up to a constant prefactor. This prefactor can
be fixed by observing that in the localized regime, $M/\xi\gg 1$,
$\lambda_M(E)$ converges to $\xi(E)$ for large $M$ so that
\begin{equation}
  \lim_{x\to\infty} \Lambda(x) = x^{-1}.
  \label{lim_lam}
\end{equation}

The divergence of the localization length in Fig.~\ref{fig:xi_E} is
given by the exponent
\begin{equation}
  \nu=2.35\pm0.03.
  \label{nu}
\end{equation}
This value for the exponent is in remarkable agreement with the
experimental values $2.3\pm0.1$ ({Koch {\em et~al.\/}, }1991b) and $2.4\pm0.1$
({Wei {\em et~al.\/}, }1994). By performing similar calculations for
smaller systems of
up to $150l_c$  width Mieck (1990) obtained an exponent of
$\nu=2.3\pm0.08$.

In the previous analysis the critical energy $E_c$ was taken to be the
band center so that the localization length diverges only at a single
energy. The absence of extended states within a finite range of energies
is consistent with the field-theoretic picture of Sec.\
\ref{sec:pruisken} and was proven for single Landau levels by Chalker
(1987).

The influence of a correlation length $\sigma=l_c$ on the behavior
of the localization length can be seen in Fig.~\ref{fig:lll_1}
({Huckestein {\em et~al.\/}, }1992). Single-parameter scaling behavior
is again observed for
the data in the dotted region. The necessary width increases by a
factor of 4 relative to the $\delta$-correlated potential. One
reason for this behavior is the increase in the effective length scale
by a factor $\beta^2=(l_c^2+\sigma^2)/l_c^2$. Furthermore, due to
the one-dimensional limit Eq.~(\ref{1d_lim}) the flow in
Fig.~\ref{fig:lll_1} starts at values of $\lambda_M/M$ smaller by
a factor of $\beta^2$ compared to $\sigma=0$. For large systems,
$M\geq 64$, the scaling behavior is the same as for $\sigma=0$.
Consequently, the data can be fitted to the same scaling curve in
Fig.~\ref{fig:scale_curve} and the localization length exponent is
again given by Eq.~(\ref{nu}).  The prefactor of the localization
length is roughly independent of the correlation length if the
correlation-length dependence of the bandwidth $\Gamma_\sigma=
\Gamma_0/\beta$ ({Ando and Uemura, }1974) is taken into account. This
result is
in contrast to calculations by Ando (1989b) that showed a
strong dependence of the prefactor on the correlation length.
Presumably, the difference can be attributed to the limited system
size $M\leq20$ of that calculation.

For significantly larger correlation length it seems difficult at
present to reach the system width necessary to observe scaling
behavior. For $\sigma=4l_c$ Mieck (1990) estimated a
value for $\nu$ of about $2.3$.

In another approach to the localization problem Huo and Bhatt
(1992) calculated the Chern numbers of the eigenstates in
finite systems using Eq.~(\ref{sigma_thouless}). By studying the size
dependence of the disorder-averaged density of states with non-zero
Chern numbers they found that a finite portion of the states in the
band center is extended in the thermodynamic limit. Away from the
band center they found that the width $\Delta E$ of the band of
states with non-zero Chern number decreased with the area $A$ of the
system as
\begin{equation}
  \Delta E \propto A^{-1/2\nu},
  \label{huo_bhatt}
\end{equation}
with $\nu=2.4\pm0.1$. Eq.~(\ref{huo_bhatt}) is obtained by assuming
that the states are extended in an energy range where the localization
length $\xi(E)\propto |E-E_c|^{-\nu}$ exceeds the linear system size
$L$, and $A=L^2$. They obtain the same exponent from the scaling of average
number $N_c$ of states with non-zero Chern number, $N_c\propto A^y$,
with $y=1-1/2\nu$.

Ando (1992) studied the transmission through a disordered
system with edges connected to two ideal leads. Using a lattice model
for the system he calculated the transmission probability through the
system as a function of the width of the system. He found that the
energy range $\Delta E$ over which the transmission probability rises
from zero to unity scales with the width $M$ of the system as
$\Delta E\propto M^{-1/\nu}$ with $\nu=2.2\pm0.1$. In the
edge-state picture the finite conductivity near the center of the
Landau levels is due to backscattering between edge states located at
opposite sides of the sample. The backscattering is facilitated by
extended bulk states and hence scales with the bulk critical exponent
$\nu$.

Chalker and Daniell (1988) calculated numerically the
two-particle spectral function $S(q,\omega;E)$ near the critical
energy $E_c$. $S(q,\omega;E_c)$ is the Fourier transform of
\begin{equation}
  S({\bf r},\omega;E_c) =
  \overline{\sum_{\alpha,\beta}
    \delta(E_c-\hbar\omega/2-E_\alpha)
    \delta(E_c+\hbar\omega/2-E_\beta)
    \psi_\alpha({\bf 0})\psi_\alpha^*({\bf r})
    \psi_\beta({\bf r})\psi_\beta^*({\bf 0})}.
  \label{S_r}
\end{equation}
Employing current conservation $S(q,\omega;E)$ can be expressed in
terms of a wave-vector and frequency dependent diffusion coefficient
$D(q,\omega)$,
\begin{equation}
  S(q,\omega;E) =
  \frac{\rho(E)\hbar}{\pi}\frac{q^2D(q,\omega)}{\omega^2 +
    (q^2D(q,\omega))^2}.
  \label{velicky}
\end{equation}
For a scale-invariant system, i.e., at the critical energy $E_c$, the
diffusion coefficient can only depend on the dimensionless quantity
$qL_\omega$, where $L_\omega=(\hbar\rho(E_c)\omega)^{-1/2}$ is
the size of a system with mean level spacing $\omega$. Comparing this
result, which is supported by numerical calculations ({Chalker and
  Daniell, }1988; {Huckestein and Schweitzer, }1994),
with the dynamic scaling relation (\ref{p_infty_dyn}) applied to the
spectral function we can conclude that the dynamic critical exponent
$z=2$.

In the limit $q,\omega\to0$ the diffusion coefficient $D(q,\omega)$
becomes the diffusion constant $D_0$ that is related to the conductivity
by the Einstein relation
\begin{equation}
  \sigma_{xx}^c = e^2 \rho(E_c) D_0.
  \label{Einstein}
\end{equation}
Chalker (1987) has shown that due to a sum rule the
diffusion coefficient $D(qL_\omega)$ at large values of
$qL_\omega$ is reduced,
\begin{equation}
  D(qL_\omega) \propto (qL_\omega)^{-\eta}.
  \label{D_eta}
\end{equation}
A non-zero value for $\eta$ is related to the multifractal properties
of eigenfunctions at the mobility edge discussed in Sec.\
\ref{cha:multi}. Chalker and Daniell obtain values of $\eta=0.38\pm0.04$
and $\sigma_{xx}^c=0.45 e^2/h$.

Huo, Hetzel and Bhatt (1993) improved the calculation of
the conductivity at the critical point using the same method as
Chalker and Coddington. They found a universal value of
$\sigma_{xx}^c=0.5e^2/h$ within uncertainties of about 10\% for
seven different kinds of disorder potentials. Using the Kubo formula
({Aoki and Ando, }1981) they calculated the Hall conductivity to obtain
$\sigma_{xy}^*=0.5e^2/h$. While this result is trivial for random
potentials, which are particle-hole symmetric on the average, they
showed that it also holds for potentials with an asymmetric density of
states, again with uncertainties of about 10\%.

\subsection{Higher Landau Levels}
\label{sec:hll}

While the behavior of the localization length in the lowest Landau
level seems to be well described by a single-parameter scaling
relation the situation in higher Landau levels seems to be less clear.
Published values for the localization-length exponent $\nu$ in the
second Landau level $n=1$ range from 2.3 ({Mieck, }1990) to 6.2
({Mieck, }1993). Indeed, it has been argued that the critical exponent
$\nu$ should be nonuniversal in higher Landau levels and should
depend on details of the disorder potential, in particular the
correlation length $\sigma$ of the disorder potential
({Mieck, }1990; {Mieck, }1993; {Liu and {Das Sarma}, }1993; {Liu and {Das
    Sarma}, }1994). Only recently have these discrepancies
been reconciled with a universal value of $\nu$ by considering
corrections to single-parameter scaling behavior ({Huckestein, }1994).

The first estimate for the exponent $\nu$ in the second Landau level
was again given by Aoki and Ando (1985) who found
$\nu\approx 4$. The disorder potential consisted of short-range
scatterers in this case.  The first systematic study of the effect of
the correlation length of the disorder was performed by Mieck
(1990; 1993). He noticed that the correlations of the
matrix elements Eq.~(\ref{ncor}) in the random Landau model become
independent of the Landau level index for large correlation length
$\sigma$ ({Huckestein, }1992). Mieck obtained values of $\nu\approx 6.2$
for $\sigma=0$, $\nu\approx 3.7$ for $\sigma=0.5l_c$, and
$\nu\approx 2.3$ for $\sigma=4l_c$. In performing a fit to the
single-parameter scaling relation (\ref{lambda_xi}) for the short
correlation lengths he used data from the energy interval between
$0.5\Gamma_\sigma$ and $0.8\Gamma_\sigma$.

Huckestein (1992) showed that for a correlation length
$\sigma=l_c$ the scaling behavior of the system is indeed the same as
in the lowest Landau level. Fig.~\ref{fig:nll_1} shows the normalized
localization length. The data from the dotted region can be fitted to
the scaling curve of Fig.~\ref{fig:scale_curve}. Note that
Fig.~\ref{fig:nll_1} differs from Fig.~\ref{fig:lll_1} for the
lowest Landau level only for small width. The behavior for short-ranged
correlations with $\sigma=0$ as shown in Fig.~\ref{fig:nll_0} is
drastically different from both the behavior in the lowest Landau level
(Fig.~\ref{fig:lll_0}) and the correlated potential
(Fig.~\ref{fig:nll_1}). The important features of
Fig.~\ref{fig:nll_0} are the slow but noticeable decrease of
$\lambda_M/M$ as a function of $M$ in the band center, the very weak
energy dependence near the band center ($E<0.3\Gamma$), and the strong
energy dependence towards the band tails ($E>0.3\Gamma$). The first
observation shows that the system width is too small to reach the
critical region of single-parameter scaling as
$\lambda_M(E_c)/M=\Lambda_c$ is constant at the critical point.
This prohibits an analysis of the data in terms of a single-parameter
scaling relation. If the strong energy dependence in the tails of the
band is still analyzed using a single-parameter scaling relation large,
nonuniversal values for $\nu$ are obtained
({Aoki and Ando, }1985; {Mieck, }1990; {Huckestein, }1992; {Mieck,
  }1993; {Liu and {Das Sarma}, }1993; {Liu and {Das Sarma}, }1994). In
view of the absence of scale
invariance in the band center and the fact that these energies are
mostly outside the critical energy range in the lowest Landau level it
seems highly doubtful that these exponents describe the critical
behavior of the system.

The weak system-size dependence in the band center is due to
corrections to the single-parameter scaling laws as discussed in Sec.\
\ref{sec:correct}. To analyze these corrections we have to
connect the behavior in the band center for short correlation length
to the universal single-parameter scaling at larger correlation length
where the corrections are too small to be observed. From the
dependence of $\lambda_M(E_c)/M$ on $\beta^2$ and $M$ in
Fig.~\ref{fig:corr_beta} we see that the asymptotic single-parameter
scaling regime is reached at about $\beta^2=1.8$. The corrections
for smaller values of $\beta^2$ show single-parameter scaling as a
function of $\beta^2$ and $M$ as expected from two-parameter
finite-size scaling with one relevant and one irrelevant scaling index
(see Sec.\ \ref{sec:correct}) ({Huckestein, }1994). Eq.~(\ref{sing_par_corr})
for the localization length $\lambda_M$ reads
\begin{equation}
  \lambda_M(E_c,\beta^2) = M \Lambda\left(0,M^{-|y_{\text{irr}}|}
  \zeta_{\text{irr}}\right),
  \label{lambda_corr}
\end{equation}
where $\Lambda((M/\xi(E))^{1/\nu},0)$ is the single-parameter
scaling function of Eq.~(\ref{lambda_xi}). The physical nature of the
irrelevant scaling field $\zeta_{\text{irr}}$ is unclear, except
that it depends in some way on the disorder potential, particularly on
the correlation length parameter $\beta^2$. In the present context
Eq.~(\ref{lambda_corr}) is an ansatz that describes the numerical data
well.

Relating the unknown irrelevant scaling field $\zeta_{\text{irr}}$
to an irrelevant length scale $\xi_{\text{irr}}\propto
\zeta_{\text{irr}}^{1/|y_{\text{irr}}|}$ we can rewrite
Eq.~(\ref{lambda_corr}) as
\begin{equation}
  \lambda_M(E_c,\beta^2) = M
  \Lambda\left(0,(M/\xi_{\text{irr}})^{-|y_{\text{irr}}|}\right).
  \label{lambda_corr_xi}
\end{equation}
This is a single-parameter scaling law similar to
Eqs.~(\ref{lambda_xi}) but this time describing the flow towards the
critical point as indicated by the negative scaling index. By applying
the same fitting procedure as in Sec.\ \ref{sec:lll} we can obtain the
scaling function $\Lambda(0,x)$ and the dependence of
$\xi_{\text{irr}}$ on the correlation length of the disorder potential.
Fig.~\ref{fig:scale_corr} shows $\lambda_M/M$ as a function of
$M/\xi_{\text{irr}}$. The irrelevant length scale $\xi_{\text{irr}}$
increases by four orders of magnitude when the correlation length of
the disorder potential is reduced from $0.8l_c$ to 0
(Fig.~\ref{fig:xi_irr}). The overall scale of $\xi_{\text{irr}}$ is
not given by the fitting procedure. It is approximately known by
observing that is becomes of the order of the magnetic length $l_c$ as
the correlation length approaches $l_c$.
The asymptotic single-parameter scaling governed by the localization
length $\xi$ can
only be observed for length scales larger than the irrelevant length
scale $\xi_{\text{irr}}$. Systems numerically accessible at present are
3 to 4 orders of magnitude too narrow to reach the asymptotic
single-parameter scaling regime. The effects seen in all previous
numerical calculations for short-ranged potentials in the second
Landau level do not reflect the asymptotic single-parameter scaling
regime but are dominated by corrections due to irrelevant scaling
fields. The fact that these corrections can be analyzed using a
two-parameter scaling relation shows that the system sizes accessible
in numerical simulations are in fact sufficient to reach the critical
regime. It is just the description in terms of single-parameter
scaling laws that becomes valid only after the irrelevant length scale
$\xi_{\text{irr}}$ is exceeded.

For small values of $M^{-|y_{\text{irr}}|}\zeta_{\text{irr}}$ we can
expand the function $\Lambda$ in Eq.~(\ref{lambda_corr}) to get
\begin{equation}
  \Lambda = \Lambda_c + b M^{-|y_{\text{irr}}|}
  \zeta_{\text{irr}} + \cdots.
  \label{corr_dev_scale}
\end{equation}
{}From the width dependence of $\Lambda-\Lambda_c$ in
Fig.~\ref{fig:scale_dev_1} we get the irrelevant scaling index
$y_{\text{irr}}=-0.38\pm0.04$ ({Huckestein, }1994). For
$M/\xi_{\text{irr}}\gtrsim 0.1$ this exponent describes the data well.
This suggests that, at least in this regime, corrections due to
further irrelevant scaling fields with scaling indices less than
$y_{\text{irr}}$ are negligible.

The structure of the fixed point described by the two-parameter
scaling relation
\begin{equation}
  \lambda_M(\Delta E,\zeta_{\text{irr}}) = M
  \Lambda\left(M^{1/\nu}\Delta E,
  M^{-|y_{\text{irr}}|}\zeta_{\text{irr}}\right)
  \label{xi_two_par}
\end{equation}
is identical to the one expected by Pruisken (1987) for the
delocalization fixed points at half-integer $\sigma_{xy}$ (see
Fig.~\ref{fig:sigma_flow}),
\begin{equation}
  \lambda_M(\Delta\sigma_{xy},\Delta\sigma_{xx}) = M
  \Lambda\left(M^{1/\nu}\Delta\sigma_{xy},
  M^{-|\tilde{y}|}\Delta\sigma_{xx}\right).
  \label{xi_pruisken}
\end{equation}
Here $\Delta\sigma_{ij}$ are the linear deviations from the
fixed-point values of $\sigma_{ij}$. Near the critical point
$\Delta\sigma_{xy}$ is proportional to $\Delta E$ so that the
scaling with respect to the first variables in Eqs.~(\ref{xi_two_par})
and (\ref{xi_pruisken}) is the same. For the second variables,
$\zeta_{\text{irr}}$ and $\Delta\sigma_{xx}$ a direct relation
is not obvious. Though both quantities depend on the correlation
length of the potential it is not clear that
$\zeta_{\text{irr}}$ is proportional to $\Delta\sigma_{xx}$.

We can understand the dependences of the corrections to
single-parameter scaling on the correlation length of the disorder
potential and the Landau level index in the field-theoretical framework.
As discussed in Sec.~\ref{sec:pruisken} the mean-field approximation
to the field theory is given by the non-critical SCBA ({Pruisken, }1987). To
observe the localization-delocalization fixed point brought about by
fluctuations around the mean-field saddle point the system size has to
exceed the localization length in SCBA. The scaling in this
approximation is described by the corresponding unitary non-linear
$\sigma$-model without topological term and is given by the $\beta$
function in Eq.~(\ref{beta_loop}). The localization length is obtained
by integrating the leading term in Eq.~(\ref{beta_loop}) for the
unitary case. In terms of the conductivity $\sigma_{xx}^0$ on small
length scales it is given by (Wei {\em et al.}, 1985)
\begin{equation}
  \xi_0 = l_c \exp\left(\pi^2\sigma^0_{xx}{}^2\right).
  \label{xi_scba}
\end{equation}
Using the result for the conductivity in SCBA for short-range
scatterers ({Ando and Uemura, }1974) the dependence on Landau level
index $n$ is
\begin{equation}
  \xi_0(n) = l_c \exp\left[(2n+1)^2\right].
  \label{xi_scba_0}
\end{equation}
While this length scale is of the order of the magnetic length in the
lowest Landau level it is more than 3 orders of magnitude larger for
$n=1$. This explains why the corrections are not observed in the
lowest Landau level even for $\delta$-correlated potentials. The
strong variation of the irrelevant length scale in
Fig.~\ref{fig:xi_irr} can also be understood from the range
dependence of the conductivity in SCBA and Eq.~(\ref{xi_scba}). For
$n=1$ the conductivity decreases by about a factor of 6 between
$\sigma=0$ and $\sigma=l_c$. The corresponding length scale
$\xi_0$ decreases from about $10^4l_c$ to about one magnetic length.
In the lowest Landau level the conductivity changes by a factor of 2
and $\xi_0$ decreases only by about the same factor.

Direct calculations of the critical value of the longitudinal
conductivity $\sigma_{xx}$ are missing in higher Landau levels.
Ando's results for the Thouless numbers for a lattice system with
several Landau levels do not show a strong Landau level dependence as
expected from SCBA ({Ando, }1989a). It has been suggested that
zero-temperature quantum phase transitions in two dimensions should
have a universal value for the conductivity at the transition
({Fisher {\em et~al.\/}, }1990; {Fisher, }1990; {Lee {\em et~al.\/},
  }1992). Lee, Wang, and Kivelson (1993)
argued that the critical value $\sigma_{xx}^*$ is related to the
fixed-point scaling amplitude $\Lambda_c$. They obtain
$\sigma_{xx}^*=1/2$ for $\Lambda_c=1/\ln(1+\sqrt{2})$ which is
very close to the best-fit result $\Lambda_c=1.14\pm0.02$
({Huckestein, }1994). Since the scaling curve Fig.~\ref{fig:scale_curve},
and in particular $\Lambda_c$, is independent of Landau level index,
this would support the notion of a universal value of
$\sigma_{xx}^*=1/2$ in accordance with the numerical calculations in
the lowest Landau level ({Huo {\em et~al.\/}, }1993). Landau level
independence of the critical conductivity is also a feature of
Khmel'nitskii's flow diagram Fig.~\ref{fig:sigma_flow}.

\subsection{Network model}
\label{sec:netcalc}

The first single-parameter scaling function for the QHE was obtained
by Chalker and Coddington (1988). They studied the scaling
behavior of the network model described in Sec.\ \ref{sec:network}
using the transfer-matrix method of Sec.\ \ref{sec:transfer}. Their
scaling function is not reproduced here as it is identical to
Fig.~\ref{fig:scale_curve}. In fact, Fig.~\ref{fig:scale_curve}
was obtained by jointly fitting data from the network model and the
random Landau-matrix model for $n=0$, $\sigma=0$, $n=1$,
$\sigma=l_c$, and $n=1$, $\sigma=l_c$, to a single scaling
function. The exponent $\nu$ was estimated by Chalker and Coddington
to be about $2.5\pm0.5$. The large uncertainties are due to the
non-linear dependence of the parameter $\theta$ on energy,
Eq.~(\ref{theta_eq}). From the fact that the scaling functions of the
models are identical it follows that the exponents are identical, too.

The network model has the remarkable feature that it shows very small
finite-size corrections to single-parameter scaling. In contrast to
the systems discussed above the data are close to the scaling function
Fig.~\ref{fig:scale_curve} already for small system width. This
behavior can be understood from the different one-dimensional limits
of the models. In this limit the normalized localization length of the
network model is given by ({Jaeger, }1991)
\begin{equation}
  \frac{\lambda_M(E)}{M} = \left(\ln 2 + \ln\cosh
  \left(\frac{\pi\gamma}{2}\right)\right)^{-1},
  \label{1d_lim_net}
\end{equation}
where $\gamma$ is related to the energy $E$ by Eq.~(\ref{gamma}). In
contrast to Eq.~(\ref{1d_lim}) the network model retains an energy
dependence in the one-dimensional limit and the value at the critical
energy $\lambda_M(E_c)/M=1/\ln 2$ is only about 20\% larger than the
fixed point value $\Lambda_c$.

A generalization of the network model in which the $\theta$'s are
taken to be random variables was discussed by Chalker and Eastmond
(1993) and Lee, Wang, and Kivelson (1993). From
finite-size scaling studies they find that this generalization is an
irrelevant perturbation of the original model. If the distribution of
the $\theta$'s is narrow the scaling behavior is unchanged. Lee,
Wang, and Kivelson obtain $\nu=2.43\pm0.18$ in this case. In the
limit that the width of the distribution of $\theta$'s goes to
infinity the system behaves classically since each saddle point
transmits only into one of the two possible links. Lee, Wang, and
Kivelson find single-parameter scaling with $\nu=1.29\pm0.05$
compatible with the classical percolation exponent $4/3$. The
classical percolation fixed point is unstable against quantum effects
and the system flows towards the quantum fixed point described by
$\nu=2.35\pm0.03$. Close to this fixed point the corrections to
single-parameter scaling due to the finite width of the distribution
of $\theta$'s can be analyzed in the same way as discussed above. It
was in this context that Chalker and Eastmond (1993)
pioneered the analysis in terms of irrelevant scaling fields in the
context of the QHE. They obtained the irrelevant scaling index
$y_{\text{irr}}=-0.38\pm0.02$ in agreement with the results in the
previous section. However, for the network model the corrections to
the fixed-point value $\Lambda_c$ are negative as opposed to
Fig.~\ref{fig:scale_corr} where they are positive.

\subsection{Effects of Landau Level Coupling and Spin-Orbit Interaction}
\label{sec:coupl}

When the strength of the disorder is not negligible compared to the cyclotron
energy $\hbar\omega_c$ one has to go beyond the single-Landau-level
approximation and include the effect of Landau level coupling by
the disorder potential. In this context two different
questions are important. For weak Landau level coupling such that each
Landau level still contains a critical energy, does the Landau level
coupling change the critical behavior at these critical energies? On
the other hand, for strong disorder the system is expected to
become insulating eventually (cf.\ Sec.\ \ref{sec:phil}). In this limit
the question is: How do the divergences in the localization length
disappear with increasing disorder? Numerically these question are hard
to answer since the inclusion of several Landau levels necessitates
much smaller systems so that finite-size effects become very
prominent. It is also not obvious exactly how many Landau levels have
to be taken into account to capture the essential physics.

The first question has been approached by Liu and Das Sarma
(1994). They considered the two lowest Landau levels and
studied the scaling behavior of the localization length as a function
of the strength of the Landau level coupling for short correlation
length of the potential. They find that while in the lowest Landau
level the critical exponent $\nu$ does not change when the Landau
level coupling is included and is close to the value of Eq.~(\ref{nu}),
in the $n=1$ Landau level there is a change in the exponent. While Liu
and Das Sarma erroneously concluded from these data that Landau level
coupling is essential for universal scaling for short correlation
length, we can understand this behavior by considering the influence
of Landau level coupling on the magnitude of corrections to scaling.
With increasing Landau level coupling the inter-Landau level
scattering increases and the conductivity on short length scales
decreases. From Eq.~(\ref{xi_scba}) we see that this drastically
decreases the characteristic length scale for single-parameter
scaling. The system thus moves closer to the quantum
critical point and finite-size corrections become smaller.

The dependence of the critical energies on Landau level mixing has
been studied by Ando, both for continuous systems (1984b)
and for lattice systems (1989a), using the Thouless number.
For $\delta$-function scatterers and a Hamiltonian projected onto
the three lowest Landau levels he finds that the localization length
diverges at critical energies in each Landau level and that these
energies move upward in energy when the Landau level coupling is
increased. It also has to be pointed out that when the level
broadening becomes comparable to the Landau level separation the
localization length becomes very large, even for energies between the
original Landau levels. For lattice systems the critical energies also
move upward and eventually vanish. The square-lattice system studied
by Ando is symmetric with respect to the center of the tight-binding
band and for every Landau level below this center there exists another
Landau level above with the opposite Hall conductivity. When two
Landau levels merge in the center of the tight-binding band their
contribution to the Hall conductivity vanishes and so does the
associated critical energy. Due to this symmetry for the lattice
system it is not clear whether the upward motion of the critical
energy with increasing Landau level coupling is a generic feature.

The influence of spin degeneracy on the critical behavior has been
studied by Lee and Chalker (1994) and Wang, Lee, and Wen
(1994b) in an extension of the network model. To account for
the spin degree of freedom two quantum-mechanical fluxes are carried
by each link. The random phases on the links are then replaced by
random $U(2)$ matrices. The numerical results of both calculations
support the existence of two separate plateau transitions even in the
spin-degenerate case. The critical exponent of each transition is in
agreement with Eq.~(\ref{nu}) for single Landau levels. If, on the
other hand, the data are analyzed assuming a single critical energy
over a limited range in energy a fit with an exponent $\nu\approx
5.8$ is possible.

\subsection{Presence of a Periodic Potential}
\label{sec:period}

In this section we want to consider the influence of an additional
periodic potential on the scaling behavior of the integer QHE. For
definiteness, let us consider a periodic potential of the
form
\begin{equation}
  V({\bf r}) = 4 E_0
  \cos\left(\sqrt{2}\pi x/a\right)\cos\left(\sqrt{2}\pi y/a\right).
  \label{per_pot}
\end{equation}
In the absence of disorder each Landau level splits into $p$ subbands
if $\alpha=2\pi l_c^2/a^2=q/p$ ({Azbel{'}, }1964; {Hofstadter, }1976)
and the Hall
conductivity in the gaps between these subbands is quantized in
integer multiples of $e^2/h$ ({Thouless {\em et~al.\/}, }1982). While
the contribution of
each subband to the Hall conductivity takes on large positive and
negative values if $p$ is large, the sum of these contributions over a
whole Landau level equals unity. We see that at least in the absence
of disorder the additional periodic potential is quite relevant in
that it completely changes the phase diagram and produces phases with
new quantized values of the Hall conductivity. Under the influence of
disorder the subband gaps will eventually be filled in and the
intricate structure of the phase diagram will evolve into the simple
phase diagram of the integer QHE Fig.~\ref{fig:sigma_flow}.

The phase diagram in Fig.~\ref{fig:period_phase} was obtained by
performing finite-size scaling studies for a system with
$\alpha=3/5$ as a function of the ratio $\Gamma/E_0$ of the
strength of disorder to periodic potential ({Huckestein, }1993). The Hall
conductivity was calculated from the Diophantine equation
\begin{equation}
  p\sigma + q s = t,
  \label{diophantine}
\end{equation}
where $\sigma$ is the Hall conductivity in units of $e^2/h$, $t$
labels the subband gap, and $s$ is an integer ({Thouless {\em
    et~al.\/}, }1982). In the
presence of disorder the Hall conductivity can only change when a
critical energy moves through the Fermi energy. For strong disorder
there is only one critical energy at the center of the Landau level
where the Hall conductivity changes from zero to one, while for weak
disorder all phases of the non-disordered system are recovered.

The preceding discussion raises the question whether or not a periodic
potential is a relevant perturbation of the fixed point at the center
of the Landau level. For fixed system width the ratio
$\lambda_M(E_c)/M$ increases with increasing strength of the
periodic potential and decreases with increasing system width for
fixed $E_0$ ({Huckestein, }1994). The analysis of this behavior shows
that it is
compatible with the scaling due to an irrelevant scaling field
according to Eq.~(\ref{lambda_corr_xi}). The scaling index obtained in
this way is within error bars the same as in the second Landau
level and in the network model, $y_{\text{irr}}=-0.38\pm0.04$. Thus the
periodic potential is irrelevant at the fixed point at the center
of the Landau level. The universality of the irrelevant scaling index
suggests that further scaling indices are sufficiently small compared
to $y_{\text{irr}}$ so that they are not observed in the numerical
simulations.

The critical behavior at the other fixed points introduced by the
periodic potential is still an open question. Numerically it is a hard
problem since the finite-size corrections to single-parameter scaling
become very strong and the position of the critical energy is not
known a priori as it is for the critical state at the center of the
Landau level. Since the new fixed points are connected to the
band-center fixed point\footnote{At least for certain values of
  $\alpha$.} (see Fig.~\ref{fig:period_phase}) one could conjecture
that the critical behavior is the same for all fixed points.

\section{Multifractal Analysis}
\label{cha:multi}

So far our discussion focused on quantities, like the
localization length, that are self-averaging in the thermodynamic
limit, even at the localization-delocalization transition. These
quantities are a rather special class. In general, phase transitions
are accompanied by large fluctuations in physical observables. At the
transition where the natural length scale, the localization length,
diverges these fluctuations appear on all length scales. In this
section we want to analyze critical fluctuations of local observables
in terms of multifractal measures. We will first outline the language
of multifractal analysis, then apply this apparatus to computer
simulations, and finally put this discussion into the context of the
scaling theory.

\subsection{Generalized Dimensions and Singularity Strengths}
\label{sec:lingo}

The multifractal analysis of measures was pioneered by Mandelbrot
(1983) and further developed by, among others, Hentschel
and Procaccia (1983), Halsey {\em et al.\/}
(1986), and Chhabra and Jensen (1989). A
recent review in the context of the localization-delocalization
transition was written by Jan\ss{}en (1994).

Consider a normalized measure on a two-dimensional square of linear
dimension $L$,
\begin{equation}
  \int\limits_{L^2}d\mu = 1.
  \label{norm_meas}
\end{equation}
An example of such a measure is the local density, i.e., the squared
modulus $|\psi({\bf r})|^2$ of an eigenfunction of the Hamiltonian.
The large fluctuations of this quantity can be seen in
Fig.~\ref{fig:wave} where the local density for a state at the
center of the lowest Landau level is shown.

With this measure we can define box-probabilities $p(l)$ by
integrating the measure over squares $\Omega(l)$ of linear dimension
$l$,
\begin{equation}
  p(l) = \int\limits_{\Omega(l)} d\mu = \int\limits_{\Omega(l)} d^2r
  |\psi({\bf r})|^2.
  \label{box_prob}
\end{equation}
In our example $p(l)$ is just the probability to find an electron in
the box $\Omega(l)$. The dimension of the support of the
wavefunction is $d=2$, since the wavefunction has only isolated
zeros. By covering the system with squares
$\Omega(l)$ we get $N(\lambda)=\lambda^{-d}$ different
box-probabilities $p_i(l)$, where $\lambda=l/L$. Due to the
normalization $\sum_{i=1}^{N(\lambda)} p_i(l) = 1$ the average
box-probability scales with the ratio $\lambda$ as
\begin{equation}
  \langle p(l) \rangle_L \propto \lambda^{d},
  \label{p_ave}
\end{equation}
where the average is defined as
\begin{equation}
  \langle A \rangle_L = \frac{1}{N(\lambda)}
  \sum_{i=1}^{N(\lambda)} A_i.
  \label{ave_def}
\end{equation}

The scaling law (\ref{p_ave}) is not useful in distinguishing between
localized, extended, and critical wavefunctions as it holds for all
normalized wavefunctions. This situation changes if we look at the
scaling of moments of the box-probabilities. In the absence of length
scales these will also show power law scaling but with nontrivial
exponents $\tau(q)$,
\begin{equation}
  \langle p^q(l) \rangle_L \propto \lambda^{d+\tau(q)}.
  \label{p_moment}
\end{equation}
Here $q$ is real, but not necessarily integer. For a homogeneous
measure, $p_i(l) = p(l) = 1/N(\lambda)$, one finds $\tau(q) =
(q-1)d$. We therefore define generalized dimensions $D(q)$, such that
\begin{equation}
  \tau(q) = (q-1) D(q).
  \label{tau_d}
\end{equation}

We need to clarify under which conditions we expect power law scaling
as in Eq.~(\ref{p_moment}) to hold. As already pointed out, the
localization length $\xi$ must not set a relevant length scale. For finite
system size $L$ this gives the condition $L\ll \xi$. For short length
scales microscopic lengths become important. In the quantum Hall
system wavefunctions are smooth on scales of the magnetic length or
the correlation length of the disorder. In order to see power law
scaling we therefore need $l_m \ll l$, where $l_m$ is a microscopic
length. In summary, the condition for power law scaling is
\begin{equation}
  l_m \ll l < L \ll \xi.
  \label{scale_cond}
\end{equation}
In the limit $L\to\infty$ this condition is fulfilled only at the
critical energy where the localization length diverges. In this limit
the generalized dimensions are given by
\begin{equation}
  (q-1) D(q) = \lim_{\lambda\to0} \frac{\ln \langle p^q(\lambda)
    \rangle_L}{\ln \lambda} - d.
  \label{gen_dim}
\end{equation}
For finite system size the condition (\ref{scale_cond}) is met over a
finite range of energies. In this case the generalized dimensions can
be estimated from the slope of $\ln\langle p^q(\lambda)\rangle_L$ vs.\
$\ln \lambda$ over a range of parameters that satisfies
Eq.~(\ref{scale_cond}).

For $q=0$ we find from Eq.~(\ref{p_moment}) that $D(0)=d$ is the
dimension of the support of the measure. The exponent $D(2)$ is related
to the scaling exponent $d^*$ of the inverse participation ratio,
defined by Wegner (1980)
\begin{equation}
  P^{(2)}=\int d^2r |\psi({\bf r})|^4 \propto L^{-d^*}.
  \label{ipr_def}
\end{equation}
Formally, this equation is obtained from Eq.~(\ref{p_moment}) for
$q=2$ in the limit $l\to0$. In this limit $d^*=D(2)$. In the limit
$q\to 1$, $p^{(q-1)}$ can be expanded in Eq.~(\ref{gen_dim}) and $D_1$
is given by
\begin{equation}
  D(1) = \lim_{\lambda\to0} \frac{\langle p(\lambda) \ln
    p(\lambda)\rangle_L}{\langle p(\lambda)\rangle_L \ln \lambda}.
  \label{D_1}
\end{equation}
$D(1)$ is sometimes called the information dimension.

The function $D(q)$ has the following analytic properties that we will
only sketch here. For a complete discussion see ({Hentschel and
  Procaccia, }1983; {Halsey {\em et~al.\/}, }1986).
$D(q)$ is a monotonously decreasing, positive function of $q$ bounded
from above and below by $D_{-\infty}=D(-\infty)$ and
$D_{\infty}=D(\infty)$, respectively, the generalized dimensions
that describe the scaling of the minima and maxima of the measure.

Another way of expressing the multifractal properties of a measure is
the spectrum of singularity strengths $f(\alpha)$. It is related to
the function $\tau(q)$ by a Legendre transform, i.e., we change from
the variable $q$ to the variable
\begin{equation}
  \alpha(q) = \frac{d \tau(q)}{dq}.
  \label{alpha_q}
\end{equation}
$f(\alpha(q))$ is then implicitly given by
\begin{equation}
  f(\alpha(q)) = \alpha(q) q -\tau(q).
  \label{Legendre}
\end{equation}

{}From the analytical properties of the Legendre transform and the
function $\tau(q)$ one finds ({Chhabra and Jensen, }1989) that
$f(\alpha)$ is a
positive function of negative curvature with a single maximum.
$\alpha$ ranges from $D_{\infty}$ to $D_{-\infty}$. At these
points $f(\alpha)$ vanishes with infinite slope. For $q=1$,
$f(\alpha)$ has slope 1 and $f(\alpha(1)) = \alpha(1) = D(1)$.
It takes on its maximum value $f(\alpha_0)=d$ for $q=0$, where
$\alpha_0=\alpha(0) > d$.

Chhabra and Jensen (1989) introduced a method to calculate
$f(\alpha)$ directly. It has the advantage to be numerically more
stable than calculating $\tau(q)$ and performing the Legendre
transform (\ref{Legendre}). In addition to the box-probabilities we
need to define a one-parameter family of normalized probabilities
\begin{equation}
  \mu_i(q,l) = [p_i(l)]^q / \sum_i [p_i(l)]^q,
  \label{q_mic}
\end{equation}
that act as $q$-microscopes on the original measure. For each value of
$q$ the corresponding values of $\alpha(q)$ and
$\tilde{f}(q)=f(\alpha(q))$ can be calculated from
\begin{equation}
  \alpha(q) = \frac{\sum_i \mu_i \ln p_i}{\ln \lambda}
  \label{alpha_cj}
\end{equation}
and
\begin{equation}
  \tilde{f}(q) = \frac{\sum_i \mu_i \ln\mu_i}{\ln \lambda}.
  \label{f_q_cj}
\end{equation}

Geometrically, the $f(\alpha)$ has a quite intuitive interpretation. If a
certain box-probability $p_i(l)$ scales with a singularity strength
$\alpha$, i.e., $p_i(l) \propto l^\alpha$, then $f(\alpha)$ is
the fractal dimension of the set of boxes with singularity strength
between $\alpha$ and $\alpha+d\alpha$.

Another aspect of the $f(\alpha)$ spectrum becomes clear when the
average (\ref{ave_def}) is interpreted as an average over an
ensemble instead of an average over the whole measure
({Pook and Jan\ss{}en, }1991; {Fastenrath {\em et~al.\/},
  }1992; {Jan\ss{}en, }1994). In terms of the distribution function
$P(p,\lambda)$ the average of a function $A(p)$ can be written as
\begin{equation}
  \langle A(p,\lambda) \rangle = \int_0^1 dp P(p,\lambda) A(p).
  \label{ave_dist}
\end{equation}
Changing the integration variable to $\alpha=\ln p/ \ln
\lambda$ and defining $\tilde{P}(\alpha,\lambda) d\alpha =
P(p,\lambda)dp$, Eq.~(\ref{ave_dist}) becomes
\begin{equation}
  \langle A(p,\lambda) \rangle = \int_0^{\infty} d\alpha
  \tilde{P}(\alpha,\lambda) P(\lambda^\alpha).
  \label{ave_dist_2}
\end{equation}
{}From this we recover Eq.~(\ref{p_moment}) for the scaling of the
moments of $p$ if ({Pook and Jan\ss{}en, }1991)
\begin{equation}
  \tilde{P}(\alpha,\lambda) \propto \lambda^{-f(\alpha)+d},
  \label{p_scale}
\end{equation}
where $f(\alpha)$ is given by Eq.~(\ref{Legendre}). We see that
$f(\alpha)$ describes the scaling of the whole distribution of
box-probabilities, while the $\tau(q)$ describe the scaling of
certain moments of that distribution.

The connection between the shape of the $f(\alpha)$ curve and the
distribution function $\tilde{P}(\alpha,\lambda)$ becomes more
transparent if we approximate the maximum of the $f(\alpha)$
spectrum by a parabola,
\begin{equation}
  f(\alpha) = d - \frac{(\alpha-\alpha_0)^2}{4(\alpha_0-d)}.
  \label{para_approx}
\end{equation}
Note that this parabolic approximation depends only on one parameter
$\alpha_0$, besides the dimension $d$ of the support of the measure,
which follows from the property $f(\alpha(1))=\alpha(1)$. The
corresponding probability distribution of the box-probabilities is
log-normal,
\begin{equation}
  \tilde{P}(\ln P/\ln \lambda, \lambda) \propto
  \exp\left(\frac{(\ln P -\alpha_0\ln \lambda)^2}{4(\alpha_0-d)\ln
    \lambda}\right).
  \label{log_dist}
\end{equation}
We see that the absence of length scales that is reflected in the
$f(\alpha)$ spectrum corresponds to broad distributions of local
quantities like the box-probabilities. The parameter $\alpha_0$ that
defines the parabolic approximation is the value of $\alpha$ at the
maximum of $f(\alpha)$ and describes the scaling of the typical
value $p_{\text{typ}}(\lambda)$ of the box-probability defined by
\begin{equation}
  p_{\text{typ}}(\lambda) = \exp(\langle \ln p(\lambda) \rangle)
  \propto \lambda^{\alpha_0}.
  \label{p_typ}
\end{equation}

The multifractal density fluctuations lead to anomalous diffusive
behavior ({Chalker and Daniell, }1988). This is reflected in power-law
decay of density
correlations and slow decay of temporal wave-packet autocorrelations
({Huckestein and Schweitzer, }1994). The exponent $\eta$ of density
correlations,
\begin{equation}
  \overline{|\psi({\bf r})\psi({\bf r}+{\bf R})|^2} \propto
  |{\bf R}|^{-\eta},
  \label{dens_corr}
\end{equation}
is related to the generalized dimension $D(2)$. If we replace in
Eq.~(\ref{ave_def}) the average by an ensemble average and express
$\overline{p^2(l)}$ as an integral over (\ref{dens_corr}), we get
\begin{equation}
  \overline{p^2(l)} \propto l^{2d-\eta}.
  \label{p_2_res}
\end{equation}
Comparing with Eqs.~(\ref{p_moment}) and (\ref{tau_d}) yields the result
\begin{equation}
  \eta = d - D(2).
  \label{D_2_eta}
\end{equation}
At finite wave vector and frequency the exponent $\eta$ describes the
decay of the wave-vector and frequency-dependent diffusion coefficient
(Eq.~(\ref{D_eta})). The value $\eta=0.38\pm0.04$ obtained by Chalker
and Daniell (1988) from this quantity agrees well with
Eq.~(\ref{D_2_eta}) and $D(2)=1.62\pm0.02$ obtained by Huckestein and
Schweitzer (1994).

The anomalous diffusive behavior can directly be seen in the
spreading of wave packets ({Chalker and Daniell, }1988; {Huckestein and
  Schweitzer, }1994). While the variance
$R(2,t)=\int d^2r |{\bf r}|^2 |\phi({\bf r},t)|^2$ of a wave packet
$\phi({\bf r},t)$ at the critical point grows linearly in time, which
merely reflects the scale invariance of the system, the return
probability $p(t)=|\phi({\bf 0},t)|^2$ to the origin is enhanced and
scales like
\begin{equation}
  p(t) \propto \frac{1}{t^{1-\eta/2}}=\frac{1}{t^{D(2)/2}}.
  \label{p_t_scale}
\end{equation}

A nonzero value of $\eta$ should be observable in measurements of the
Coulomb drag in double layer systems ({Shimshoni and Sondhi, }1994)
and in hot-electron
relaxation rates and electron-electron and electron-phonon scattering
rates ({Brandes {\em et~al.\/}, }1994).

The scaling of the two-particle spectral function $S(q,\omega)$ with
$q^2/\omega$ allows to relate the correlation dimension $D(2)$ of the
local density to the correlation dimension $\tilde{D}(2)$ of the
spectral measure associated with the local density ({Ketzmerick {\em
    et~al.\/}, }1992; {Huckestein and Schweitzer, }1994).
In two dimensions, $\tilde{D}(2)=D(2)/2$.

\subsection{Universality of Multifractal Spectra}
\label{sec:spectra}

In this section we review some numerical results about $f(\alpha)$
spectra for different models of the quantum Hall system and different
physical quantities. We start with our prototype measure, the local
density $|\psi({\bf r})|^2$. The $f(\alpha)$ spectrum in
Fig.~\ref{fig:f_alpha_density} is calculated for the density shown
in Fig.~\ref{fig:wave}. In addition to the numerical data the
parabolic approximation (\ref{para_approx}) is shown with
$\alpha_0=2.29\pm0.02$ ({Huckestein {\em et~al.\/}, }1992). The
variation of $\alpha_0$
for different eigenstates in the energy range where $\xi(E)\gg L$ is
of the same order. This shows that the $f(\alpha)$ spectrum not only
describes a single eigenstate but is a universal feature of the system
studied. While these calculations were performed for a tight-binding
Hamiltonian, the results agree with calculations by Pook and
Jan{\ss{}}en (1991) for a real space potential on a square of
linear size $70l_c$ who obtained $\alpha_0=2.3\pm0.07$,
$D_{\infty}=0.95\pm0.1$, and $D_{-\infty}=3.7\pm0.1$ ({Jan\ss{}en, }1994).

These calculations give a strong indication that the $f(\alpha)$
spectrum of the local density is universal at the critical point and
that exponents, such as $\alpha_0$, are suited to describe the
phase transition. It is not clear to what extend this universality
holds for other physical observables.  Calculations by Huckestein and
Schweitzer (1992a) for the equilibrium current density $|{\bf
  j}({\bf r})|$ as the local observable (see
Fig.~\ref{fig:f_alpha_current}), by Schweitzer (1992) for
the local magnetization $|{\bf m}({\bf r})|=|{\bf r}\times{\bf j}|$,
and by Fastenrath, Jan{\ss{}}en, and Pook (1992) for the
Thouless numbers give the same spectrum within the error bars and
support the notion of universality. On the other hand, it is easy to
construct multifractals with different $f(\alpha)$ spectra. Given a
normalized multifractal measure $p$ the normalized moments $p^m$
generate new measures with
$\alpha^{[p^m]}(q)=m\alpha(mq)-\tau(m)$ and
$f^{[p^m]}(\alpha^{[p^m]})=f(\alpha(qm))$ ({Jan\ss{}en, }1994). While this
$f(\alpha)$ spectrum is simply related to the spectrum of $p$ it is
not identical.

\subsection{Multifractality and Scaling}
\label{sec:multiscale}

We now want to relate the results about the multifractal properties at
the quantum Hall transition to the discussion about scaling in the
previous sections.

Bauer, Chang, and Skinner (1990) considered the scaling of
generalized inverse participation ratios [cf.\ Eq.~(\ref{ipr_def})]
\begin{equation}
  P^{(q)}=\int d^2r |\psi({\bf r})|^{2q}
  \label{gen_ipr}
\end{equation}
as a function of system size $L$ and distance from the critical energy
$\Delta E$. Wegner (1980) introduced critical exponents
$\pi(q)$ defined by
\begin{equation}
  P^{(q)} \propto \Delta E^{\pi(q)}.
  \label{pi_def}
\end{equation}
Using a standard finite-size scaling ansatz for $P^{(q)}$,
\begin{equation}
  P^{(q)}(L,E) \propto L^{-\tau(q)} F_q(L^{1/\nu}\Delta E),
  \label{p_q_ansatz}
\end{equation}
the condition that $P^{(q)}$ is finite for $E\ne E_c$ leads to the
scaling relation
\begin{equation}
  \pi(q) = \nu \tau(q).
  \label{tau_pi}
\end{equation}
In particular, for the inverse participation ratio one finds
$\pi(2)=\nu D(2)$. Hikami (1986) calculated the exponent
$\pi(2)=3.8\pm0.4$ for the lowest Landau level. Neglecting the
multifractal structure of the wavefunctions and assuming
$P^{(2)}\propto \xi^{-2}$ he concluded $\nu=\pi(2)/2=1.9\pm0.2$.
Using $D(2)=1.62\pm0.02$ obtained from numerical calculations
({Huckestein and Schweitzer, }1994) and the correct relation
(\ref{tau_pi}), we get
\begin{equation}
  \nu = 2.4 \pm 0.3,
  \label{nu_hikami}
\end{equation}
in agreement with the result (\ref{nu}) obtained from finite-size
scaling of the localization length.

In the spirit of Shapiro (1987) one can try to study the
scaling of the whole distribution of box-probabilities $P(p,\lambda)$.
At the critical point this distribution becomes universal and is
determined by the $f(\alpha)$ spectrum. The bulk of the distribution
is log-normal with deviations in the tails of the distribution that
account for the differences between the parabolic approximation and
the true $f(\alpha)$ curve in Fig.~\ref{fig:f_alpha_density}. In
this limit the parameter $\alpha_0$ characterizes the whole
distribution. Away from the critical energy, as long as $\xi(E)\gg L$,
the distribution will still show power-law scaling (\ref{p_scale}).
Since the system moves towards the localized regime the typical
box-probability $p_{\text{typ}}$ will decrease and by
Eq.~(\ref{p_typ}) $\alpha_0$ will increase. If single-parameter
scaling of the distribution holds, $\alpha_0$ has to obey a
finite-size scaling law,
\begin{equation}
  \alpha_0(E) = A(L^{1/\nu} \Delta E) = \alpha_0(E_c) + a
  L^{1/\nu} |\Delta E| + \cdots.
  \label{alpha_scale}
\end{equation}
The linear energy dependence is compatible with the numerical
simulations presented in Fig.~\ref{fig:alpha_scale} ({Huckestein and
  Schweitzer, }1992b), but these calculations do not rule out a
quadratic energy dependence.
For larger distances from the critical energy, when $\xi(E)$ becomes
smaller than the system size $L$, power-law scaling breaks down on
length scales larger than $\xi(E)$. Eq.~(\ref{p_scale}) no longer
describes the whole distribution $P(p,\lambda)$ and $\alpha_0$
loses its meaning as the single parameter describing the
distribution.

Employing conformal invariance in two-dimensional systems Cardy
(1984) was able to show that the scaling amplitude
{}$\Lambda_c^{-1}=\lim_{M\to\infty}M/\lambda_M$ of the inverse
correlation length $\lambda_M^{-1}$ of the correlation function of an
operator on a cylinder is related to the scaling dimension $x$ of the
operator in the plane by
\begin{equation}
  \Lambda_c^{-1} = 2\pi x.
  \label{cardy}
\end{equation}
For the local density one is tempted to identify $\Lambda_c$ with the
amplitude derived from the localization length on the cylinder and $x$
with the exponent $\eta/2$. This argument neglects the broad
distribution of the local density and the corresponding broad
distribution of the correlation functions. Instead of mapping a single
correlation function from the plane onto the cylinder, in the presence
of broad distributions one needs to map the whole distribution
function. Since the scaling amplitude $\Lambda_c$ was extracted from
averaging the logarithm of the density correlation function the
corresponding scaling dimension describes the scaling of the typical
correlations $x=\alpha_0-d$ ({Jan\ss{}en, }1994). Noting that, by
convention, the localization length is defined by the absolute value
of the Green function, we get
\begin{equation}
  \Lambda_c = \frac{1}{\pi(\alpha_0-d)},
  \label{Lambda_alpha}
\end{equation}
which yields $\Lambda_c=1.1\pm0.1$ in accordance with the
finite-size scaling result. A similar relation has been derived by
Ludwig (1990) for a random 2D ferromagnet.

The discussion of multifractals can be extended to the description of
non-normalized observables $Q$ ({Ludwig, }1990; {Duplantier and Ludwig,
  }1991; {Pook and Jan\ss{}en, }1991; {Jan\ss{}en, }1994). If the
normalized observables $p^{[Q]}=Q/\sum_i^N Q$ form a multifractal
measure the distribution of $Q$ can be described by the
$f^{[Q]}(\alpha^{[Q]})$ spectrum of $p^{[Q]}$ and the normalization
exponent $X^{[Q]}$, defined by
\begin{equation}
  \sum_i^N Q(l) \propto \lambda^{X^{[Q]}}.
  \label{norm_exp}
\end{equation}
Analogous to Eq.~(\ref{p_typ}) the scaling of the typical value
$Q_{\text{typ}}$ is then given by
({Jan\ss{}en, }1994)
\begin{equation}
  Q_{\text{typ}}(l) = \exp(\ln\langle Q(l)\rangle) \propto
  l^{\alpha^{[Q]}_0 + X^{[Q]}}.
  \label{q_typ}
\end{equation}

Jan{\ss{}}en (1994) interpreted $Q_{\text{typ}}(l)$ as
the value of $Q$ for a system of size $l$ and the average in
Eq.~(\ref{q_typ}) as an average over an ensemble of systems. If
single-parameter finite-size scaling holds for $Q_{\text{typ}}$ and
$Q_{\text{typ}}$ is scale invariant at the critical point then
close to the critical point
\begin{equation}
  Q_{\text{typ}}(l) = Q^*_{\text{typ}} + a l^{1/\nu} t.
  \label{q_typ_sps}
\end{equation}
The box-observable
$\tilde{Q}_{\text{typ}}=Q_{\text{typ}}-Q^*_{\text{typ}}$ thus scales
like
\begin{equation}
  \tilde{Q}_{\text{typ}} \propto l^{1/\nu}
  \label{q_tilde_scale}
\end{equation}
close to the critical point. Identifying Eqs.~(\ref{q_typ}) and
(\ref{q_tilde_scale}) relates the critical exponent $\nu$ of the
localization length to the multifractal exponents $\alpha^{[Q]}_0$ and
$X^{[Q]}$,
\begin{equation}
  1/\nu = \alpha^{[Q]}_0 + X^{[Q]}.
  \label{nu_alpha}
\end{equation}
Fastenrath, Jan{\ss{}}en, and Pook (1992) calculated Thouless
numbers near the center of the lowest Landau level for systems of
varying size. They found that the typical
Thouless number $g_{\text{typ}}(l)=\exp[\langle \ln g(l) \rangle]$
scales according to Eq.~(\ref{q_typ}) with
$\alpha_0^{[g]}=2.25\pm0.05$ and $X^{[g]}=-1.75\pm0.05$. Using
Eq.~(\ref{nu_alpha}), $\nu=2.2\pm0.3$ in accordance with
Eq.~(\ref{nu}).

\section{Discussion and Conclusions}
\label{cha:conc}

Having reviewed the experimental and theoretical results we are now in
a position to compare both. The most important result of the
comparison is that both experiments and numerical simulations show
similar scaling behavior.  The plateau transition in the integer QHE
can thus be understood as a continuous phase transition of
noninteracting electrons with a single diverging length $\xi$. If we
focus on spin-split Landau levels there is a remarkable agreement in
the value of the exponent $\nu$, describing the divergence of the
localization length $\xi$, between experiment [$2.3\pm0.1$
({Koch {\em et~al.\/}, }1991b), $2.4\pm0.1$ ({Wang {\em et~al.\/},
  }1994a)] and numerical simulation
[$2.35\pm0.03$ ({Huckestein, }1992)]. This suggests that the temperatures
in these experiments are in fact low enough to observe the critical
behavior and influences of the finite width of the Fermi-Dirac
distribution are negligible.

{}From the value of $\nu$ and the value of $1/\nu z=0.41\pm0.04$
measured in the dynamic scaling experiment of Engel {\em et al.\/}
(1993) we get the dynamic critical exponent $z=1.0\pm0.1$. Dynamical
scaling can also be used to explain the temperature dependence of the
transition. In the approach outlined in Sec.~\ref{sec:temp} the
temperature sets the effective system size $L_{\text{eff}} \propto
T^{-p/2}$ that has to be compared to the localization length
$\xi\propto|\Delta E|^{-\nu}$. In this interpretation the measured
temperature exponent $\kappa=p/2\nu$. In the dynamic scaling
interpretation the temperature enters as the frequency range
$\omega=k_B T/\hbar$ over which excitations are present in the
system. Scaling as a function of $\omega\tau$ with $\tau\propto\xi^z$
leads to the identification $\kappa=1/\nu z$, in agreement with the
results of
Wei {\em et al.\/} (1988; 1990) from experiments on
InGaAs/InP heterostructures and on AlGaAs/GaAs heterostructures
(1992). Koch {\em et al.\/} (1991a; 1991b),
using AlGaAs/GaAs heterostructures, however, find values of
$\kappa$, which vary between samples. Wei {\em et al.\/}
(1992) argue that there is a characteristic temperature
below which universal scaling behavior is observed and that this
temperature is much lower in AlGaAs/GaAs than in InGaAs/InP. The
agreement between the dynamic scaling experiment and the temperature
scaling below the characteristic temperature suggests that these
experiments are indeed probing the asymptotic low-temperature
regime described by the dynamic exponent $z$. Which interpretation
describes the experiments correctly presumably depends on the relative
size of $L_{\text{eff}}/\xi$ and $k_BT\tau/\hbar$.

We then need to understand why other experiments see power-law scaling
with temperature but with nonuniversal exponents. This difference
could be due to the different ranges of the disorder potentials in the
different materials. In InGaAs/InP the electron gas is formed in the
InGaAs, which is an alloy with a potential presumably varying on an
atomic length scale. By contrast, in AlGaAs/GaAs heterostructures the
electron gas is situated in the GaAs, which forms a perfect crystal.
In conventional heterostructures, doped in the AlGaAs to increase the
mobility, the major scattering mechanism is remote ionized donor
scattering. The samples used in the experiments by Koch {\em et al.\/}
were additionally doped in the plane of the electron gas. Since this
introduces charged impurities into the quasi-two-dimensional gas it is
not obvious that the range of the scatterers is actually sufficiently
different in both materials. An increase in the correlation length of
the disorder potential leads to an increase in the irrelevant length
scale $\xi_{\text{irr}}$ and hence to an increase in finite-size
effects.

The temperature dependence of different phase-breaking mechanism could
also account for different characteristic temperatures. At
sufficiently low temperatures phase-breaking occurs presumably due to
a single process, Coulomb interaction between the electrons,
associated with the dynamical exponent $z$. At higher temperatures
other processes, like the electron-phonon interaction, become
important. Since these processes will, in general, have different
temperature dependences the observed nonuniversality of $\kappa$
might be a crossover phenomenon. The characteristic temperature of the
crossover to the asymptotic, universal low-temperature regime is
nonuniversal and might show a strong sample dependence.

In addition to the relevant length scale $\xi$ the numerical
simulations show the influence of an irrelevant length scale
$\xi_{\text{irr}}$. For higher Landau levels and short correlation
length of the disorder the irrelevant length can become orders of
magnitude larger than the magnetic length and hence comparable to the
phase coherence length in experiments. On length scales smaller than
$\xi_{\text{irr}}$ the scaling is described by a two-parameter flow
according to Eq.~(\ref{xi_two_par}) while the asymptotic scaling
behavior for large length scales is given by a single-parameter
scaling relation. In a $\sigma_{xx}$ vs.\ $\sigma_{xy}$ flow
diagram we expect the renormalization group flow to produce a set of
different curves in the former regime and a single scaling curve in
the latter. The differences in the flow diagrams of Wei {\em et al.\/}
(1987), of Yamane {\em et al.\/} (1989), of
Kravchenko {\em et al.\/} (1990), and of McEuen {\em et
  al.\/} (1990) might have their origin in the different
size of the crossover length scale.

The value $z=1$ extracted from the experiments does not agree with the
result $z=2$ for noninteracting electrons. This means that another
mechanism is necessary to describe the dynamics of the system. It has
been argued by Polyakov and Shklovskii (1993) and by Lee,
Wang, and Kivelson (1993) that the Coulomb interaction
between the electrons gives rise to $z=1$.  This can be motivated by the
naive scaling argument that the relevant energy scale is the Coulomb
energy on the length scale $\xi$, $E\propto e^2/\xi$, which
immediately provides $z=1$. If Coulomb interactions have to be taken
into account to understand the dynamic scaling behavior, this raises
the question whether these interactions are relevant for the static
scaling exponent $\nu$, too. The agreement between experiments and
simulations for noninteracting electrons suggests that $\nu$ is not
changed by Coulomb interactions. However, no calculation of $\nu$
for interacting electrons under quantum Hall conditions has been
performed so far, although Lee, Wang, and Kivelson (1993)
claim that the edge states of interacting quantum Hall droplets behave
like noninteracting particles and that the results of the network
model apply in presence of interactions as well.

For spin-degenerate Landau levels the measured values of $\nu$ do
not agree with the numerical value $\nu=2.35$ ({Koch {\em et~al.\/},
  }1991b). Also
the measured value of $\kappa$ is much smaller in the spin-degenerate
case ({Wei {\em et~al.\/}, }1988; {Koch {\em et~al.\/}, }1991a; {Engel
  {\em et~al.\/}, }1993). The data analysis in all
experimental investigations assumed a single critical energy in the
center of the spin-degenerate Landau level. Polyakov and Shklovskii
(1993), Lee and Chalker (1994), and Wang, Lee, and Wen (1994b) argued
recently for the existence of two separate critical energies even in
the completely spin-degenerate case. Lee and Chalker (1994), and Wang,
Lee, and Wen (1994b) supported their arguments
with numerical results that were compatible with two critical energies
and $\nu=2.35$. When fitting their data with a single critical
energy, they observed a divergence of $\xi$ with an exponent $\approx
5.8$ in some intermediate range of energies. It has to be seen whether
experimental data can also be fitted with two critical energies and
$\nu=2.35$.

As far as the value of the conductivity tensor at the transition is
concerned there is not much agreement between experiment and theory.
While theoretical arguments and calculations seem to favor $\sigma_{xx}=0.5
e^2/h$ and $\sigma_{xy}=(n+1/2)e^2/h$ (at least in the lowest
Landau level), the experimental situation is completely unclear.

So far, there exists no direct experimental evidence for the
multifractal structure of eigenfunctions at the transition. Since the
multifractality leads to an anomalous behavior of the density-response
function signatures of multifractality might be observable in the
temperature dependence of transport processes ({Chalker and Daniell,
  }1988; {Brandes {\em et~al.\/}, }1994).

In summary, we have shown that numerical simulations are a suitable
instrument to calculate the critical behavior of noninteracting
electrons in the quantum Hall system. The observed behavior agrees
with the description of the plateau transitions as continuous phase
transitions with a single diverging length scale. The numerically
determined critical exponent $\nu=2.35\pm0.03$ agrees well with
experimental results. In addition to the relevant scaling field
driving the transition an irrelevant scaling field is observed in the
numerical simulations. The associated crossover length can become
macroscopic, in which case the irrelevant scaling field has to be
included in the scaling flow.

\section*{Acknowledgments}
\label{sec:ack}

I would like to thank everybody who had part in the completion of this
work: The colleagues and friends from whom I learned so much. My
collaborators who helped me perform the calculations presented here.
Hans A. Weidenm\"uller for giving me the opportunity and encouraging
me to write this review. Gregor Hackenbroich, Felix von Oppen, Ludwig
Schweitzer, and Petra Wild for their help in the final stages of this
work. And last, but not least, my family. Without their patience,
support, and love I could never have done it.

\appendix

\section*{Random-Landau-Matrix-Model}
\label{sec:rlmm}
The generalization of the result for the correlation function
Eq.~(\ref{delcor}) to arbitrary Landau level index $n$ was given by
Mieck (1993),
\begin{eqnarray}
  \lefteqn{\overline{\langle nk_1|V|nk_2 \rangle\langle nk_3|V|nk_4
      \rangle} = \frac{V_0^2}{\sqrt{2\pi}l_c L_y \beta}
  \exp\left(-\frac{1}{2} (k_1-k_2)^2l_c^2 \beta^2\right)
  \exp\left(-\frac{1}{2} (k_1-k_4)^2l_c^2
  \frac{1}{\beta^2}\right)}\nonumber\\
  &&\delta_{k_1-k_2,k_4-k_3}
  \times\sum_{k=0}^n\sum_{m=0}^n\sum_{l=0}^{\min(2k,2m)} 2^{-l}
 {2k\choose l} {2m\choose l} \frac{l!(k+m-l)!}{k!m!}
 \beta^{-2(m+k-l)}\nonumber\\
 &&\times
 L_{n-k}^{-1/2}\left(\frac{(k_1-k_2)^2l_c^2}{2}\right)
 L_{n-m}^{-1/2}\left(\frac{(k_1-k_2)^2l_c^2}{2}\right)
 L_{k+m-l}^{-1/2}\left(-\frac{1}{2}
 \frac{(k_1-k_4)^2l_c^2}{\beta^2(\beta^2-1)}\right).
 \label{ncor}
\end{eqnarray}

It is instructive to study the dependence of the correlation function
on the correlation length of the disorder potential and the Landau
level index. As the correlation length and hence $\beta$ increases,
the off-diagonal matrix elements become smaller but the range of the
correlations along the diagonals increases. In higher Landau levels
additional polynomials enter the correlation function (\ref{ncor}) due
to the presence of the Hermite polynomials in the harmonic oscillator
functions. While in the lowest Landau level the correlations
(\ref{delcor}) are always positive, in higher Landau levels the
correlations start to oscillate for short correlation lengths and
small differences $k_1-k_4$. In the limit of large $\sigma/l_c$, the
correlations become independent of the Landau level and take on the
universal form given by the expression for the lowest Landau level,
Eq.~(\ref{delcor}). This limit corresponds to the
semiclassical limit discussed in Section~\ref{sec:network} that is
independent of the Landau level index.

In order to use the Ansatz~(\ref{mat_el1}) for the matrix elements
$\langle n_1 k_1|V|n_2 k_2 \rangle$ the weight function $h(x,k)$
appearing in it has to be expressed through the correlation function
$\bar{g}(x,k)$ of the disorder potential. Using the translational
invariance of the disorder-potential correlation function and the fact
that the disorder potential is real, the weight function $h(x,k)$ is
seen to be a real symmetric function,
\begin{equation}
  h(x,k) = h(-x,k) = h(x,-k) = h(-x,-k).
  \label{h_prop}
\end{equation}
With Eq.~(\ref{gencor}), $h(x,k)$ is related to the
correlation function $\bar{g}(x,k)$ by
\begin{equation}
  \bar{g}(x_1-x_2,k) = \int dx'\, h(x_1-x',k) h(x_2-x',k),
  \label{h_g}
\end{equation}
which can easily be solved by Fourier transformation to give
\begin{equation}
  \tilde{g}(q,k) = \left[\tilde{h}(q,k)\right]^2,
  \label{h_of_g}
\end{equation}
where $\tilde{g}(q,k)$ and $\tilde{h}(q,k)$ are the Fourier transforms
with respect to $x$ of $\bar{g}(x,k)$ and $h(x,k)$, respectively. For
the simple case of Gaussian correlations we find that $h(x,k)$ is
given by
\begin{equation}
  h(x,k) = \frac{V_0}{\sqrt{2\pi
      L_y}\sigma}e^{-k^2\sigma^2/4}e^{-x^2/\sigma^2},
  \label{h_gauss}
\end{equation}
and Eq.~(\ref{mat_el1}) becomes
\begin{eqnarray}
  \lefteqn{\langle n_1 k_1|V|n_2 k_2 \rangle =
  \frac{V_0\beta l_c}{\sqrt{2\pi L_y}\sigma}
  \exp\left(\frac{(k_1-k_2)^2l_c^2\beta^2}{4}\right)
  \left(2^{(n_1+n_2)}n_1!n_2!\pi\right)^{-1/2}}\nonumber\\
  & & \times\int d\xi\, u_0\left(\beta\xi +
  \frac{(k_1+k_2)l_c}{2},(k_1-k_2)l_c\right)
  \exp(-\xi^2)\nonumber\\
  & &\times \int d\eta
  \exp\left(-\frac{l_c^2+\sigma^2}{\sigma^2}\eta\right)
  H_{n_1}\left(\eta +
  \frac{\xi}{\beta}-\frac{(k_1-k_2)l_c}{2}\right)
  H_{n_2}\left(\eta +
  \frac{\xi}{\beta}+\frac{(k_1-k_2)l_c}{2}\right),
  \label{mat_elg}
\end{eqnarray}
which gives Eq.~(\ref{mat_el0}) for $n_1=n_2=0$.

For numerical calculations, the integration in Eq.~(\ref{mat_el0})
needs to be discretized. Huckestein and Kramer (1989) chose a
step width of $\Delta X/2l_c$, where $\Delta X=\Delta k
l_c^2= 2\pi l_c^2/L_y$ is the difference in the
center coordinate of neighboring Landau wavefunctions. The discrete
version of Eq.~(\ref{mat_el0}) then reads
\begin{eqnarray}
  \langle 0 k_1|V|0 k_2 \rangle &=&
  \frac{V_0}{\left(\sqrt{2\pi}l_cL_y\Sigma\right)^{1/2}}
  \exp\left(\frac{(k_1-k_2)^2l_c^2\beta^2}{4}\right)
  \nonumber\\
  & &\times\sum_i u_{2k_1+i,k_2-k_1}
  \exp\left(-\frac{\pi^2 l_c^2}{L_y^2\beta^2}i^2\right),
  \label{mat_dis}
\end{eqnarray}
with
\begin{equation}
  \Sigma=\sum_i
  \exp\left(-2\frac{\pi^2 l_c^2}{L_y^2\beta^2}i^2\right).
  \label{sigma}
\end{equation}
The limits on the summation over $i$ can be chosen such that the
influence of the neglected terms is less than the statistical
fluctuation due to disorder.

It remains to specify the distribution function of the random
variables $u_{x,k}$. To correspond to a Gaussian distribution in real
space they should also be taken from a Gaussian distribution.
Numerically, it is faster to generate random numbers with a uniform
distribution. While the distribution function of the matrix elements
seems to strongly influence finite-size effects for the
three-dimensional Anderson model ({Kramer {\em et~al.\/},
  }1990; {MacKinnon, }1994), in the present
case no noticeable differences could be observed between Gaussian and
uniform distributions of the same variance.

A similar approach to that presented above was developed by
Weidenm\"uller and Mieck ({Weidenm{\"u}ller, }1987; {Mieck,
  }1990; {Mieck and Weidenm{\"u}ller, }1991; {Mieck, }1993). Central to
their idea is the observation that operators projected onto a single
Landau level can be expanded in matrix polynomials that are related to
the magnetic translation operators ({Weidenm{\"u}ller, }1987). For the
correlation function of the disorder potential this expansion proceeds
as follows. Eq.~(\ref{matcor1}) can be rewritten in terms of the
Fourier transform $\tilde{g}({\bf{p}})$ of
$g({\bf{r}}-{\bf{r}}')$,\footnote{For simplicity, we discuss the limit
  $L_y\to\infty$ in the following.}
\begin{eqnarray}
  \lefteqn{\overline{\langle n_1k_1|V|n_2k_2 \rangle\langle n_3k_3|V|n_4k_4
      \rangle} = \int \frac{d^2p}{2\pi}
    \tilde{g}(|{\bf{p}}|)}\nonumber\\
  & & \times
  \langle n_1k_1|\exp[i{\bf{p}}\cdot{\bf{r}}_1]|n_2k_2 \rangle
  \langle n_3k_3|\exp[-i{\bf{p}}\cdot{\bf{r}}_2]|n_4k_4 \rangle.
  \label{gencorf}
\end{eqnarray}
The matrix elements of the plane waves can be expressed using the
magnetic translation operators. These are generated by
the operator $\cal{P}$ that complements the canonical momentum
$\bf{\Pi}={\bf p}-e{\bf A}$. In the gauge
${\bf{A}}=B(-(1-\alpha)y\hat{\rm{e}}_x +
\alpha x\hat{\rm{e}}_y)$ these two operators are given by
\begin{equation}
  {\bf{\Pi}}={p_x + eB(1-\alpha)y \choose p_y -eB\alpha x},\quad
  {{\cal P}} = {p_x - eB\alpha y \choose p_y + eB(1-\alpha)x}.
  \label{twoops}
\end{equation}
The commutation relations for these operators are those of the
two-dimensional harmonic oscillator where the canonical momentum
$\bf{\Pi}$ is diagonal in the degenerate, internal states $k$ of
each Landau level while ${\cal P}$ is diagonal between the Landau
bands $n$,
\begin{equation}
  [\Pi_x,\Pi_y] = ieB\hbar,\; [{\cal P}_x,{\cal P}_y]=-ieB\hbar,\;
  [{\bf{\Pi}},{\cal P}]=0.
  \label{commrels}
\end{equation}
With these operators $i{\bf p}\cdot{\bf r}$ can be expressed
as
\begin{equation}
  i{\bf p}\cdot{\bf r} = \frac{i}{eB} (\hat{\rm{e}}_z
  \times{\bf{p}}) ({\cal P}-{\bf{\Pi}}).
  \label{ipr}
\end{equation}
Using the commutation relations Eq.~(\ref{commrels}) this can be
exponentiated to give
\begin{eqnarray}
  \exp[i{\bf p}\cdot{\bf r}] &=& \exp[(i/eB)(\hat{\rm{e}}_z
  \times{\bf{p}}) ({\cal P}-{\bf{\Pi}})]\nonumber\\
  &=& \exp[(i/eB) (\hat{\rm{e}}_z\times{\bf{p}}){\cal P}]
  \exp[-(i/eB) (\hat{\rm{e}}_z\times{\bf{p}}){\bf{\Pi}}].
  \label{expipr}
\end{eqnarray}
The matrix elements of $\exp[i{\bf p}\cdot{\bf r}]$ are then
given by the matrix elements $U_{k_1,k_2}({\bf{p}}) = \langle n_1
k_1| \exp[i/eB (\hat{\rm{e}}_z\times{\bf{p}}){\cal P}]| n_2
k_2\rangle$ and $W_{n_1,n_2}({\bf{p}}) = \langle n_1
k_1|\exp[-i/eB (\hat{\rm{e}}_z\times{\bf{p}}){\bf{\Pi}}]| n_2
k_2 \rangle$,
\begin{equation}
  \langle n_1 k_1 | \exp[i{\bf p}\cdot{\bf r}] | n_2 k_2
  \rangle = U_{k_1,k_2}({\bf{p}}) W_{n_1,n_2}({\bf{p}}).
  \label{uw}
\end{equation}
Note that due to the commutation relations,
$U_{k_1,k_2}({\bf{p}})$ does not depend on the Landau-level index,
and $W_{n_1,n_2}({\bf{p}})$ does not depend on the internal
quantum numbers $k$. Using Eq.~(\ref{expipr}), the correlation function
(\ref{gencorf}) can now be expressed in terms of the matrix polynomials
\begin{eqnarray}
  \overline{\langle n_1k_1|V|n_2k_2 \rangle\langle n_3k_3|V|n_4k_4
      \rangle} &=& \int \frac{d^2p}{2\pi}
    \tilde{g}(|{\bf{p}}|) \nonumber\\
  & & \times U_{k_1,k_2}({\bf{p}}) W_{n_1,n_2}({\bf{p}})
  U_{k_3,k_4}(-{\bf{p}}) W_{n_3,n_4}(-{\bf{p}}).
  \label{gencorfmat}
\end{eqnarray}
The particular form of the matrix polynomials depends on the choice of
the gauge. For the Landau gauge (\ref{landaug}) they are given by
({Mieck, }1993)
\begin{eqnarray}
  U_{k,k'}({\bf{p}})&=&\exp[ip_x(k+k')/2] \delta(p_x-k+k')\\
  W_{n,n'}({\bf{p}})&=&i^{|\Delta n|}
  \left(\frac{\nu!}{(\nu+|\Delta n|)!}\right)^{1/2}
  e^{-|{\bf{p}}|^2/4}\left(|{\bf{p}}|/\sqrt{2}\right)^{|\Delta n|}
  L_{\nu}^{|\Delta n|}(|{\bf{p}}|^2/2)
  e^{-i(n'-n)\phi},
  \label{uandw}
\end{eqnarray}
\begin{equation}
  \nu=\min(n,n'),\;\tan\phi=p_y/p_x.
  \label{uandwstuff}
\end{equation}
The delta function in the matrix polynomial
$U_{k,k'}({\bf{p}})$ reproduces the delta function that we already
observed in the correlation function (\ref{gencor}).

The matrix-polynomial representation (\ref{uw}) of the plane-wave
operator can also be used to generate the matrix elements $\langle n_1
k_1 | V | n_2 k_2 \rangle$. For this, we express the real space
potential $V({\bf r})$ by its Fourier transform $\tilde{V}({\bf k})$,
\begin{equation}
  V({\bf r}) = \int \frac{d^2k}{2\pi} e^{-i{\bf k}\cdot{\bf r}}
  \tilde{V}({\bf k}).
  \label{vr_fourier}
\end{equation}
Using Eq.~(\ref{frr}) the second moment of the Fourier transform is
given by
\begin{equation}
  \overline{\tilde{V}({\bf k_1})\tilde{V}({\bf k_2})} =
  2\pi \tilde{g}(|{\bf k_1}|) \delta({\bf k}_1 +
  {\bf k}_2).
  \label{fkk}
\end{equation}
Thus the Fourier transform of the random potential is
$\delta$-correlated and can be expressed by Gaussian random
variables $\tilde{v}({\bf k})$ of unit variance,
\begin{equation}
  \tilde{V}({\bf k}) = (2\pi)^{1/2} \left[\tilde{g}(|{\bf
    k}|)\right]^{1/2} \tilde{v}({\bf k}).
  \label{v_k}
\end{equation}
The potential $V({\bf r})$ is then determined by $\tilde{v}({\bf k})$,
\begin{equation}
  V({\bf r}) =  \int \frac{d^2k}{(2\pi)^{1/2}} \left[\tilde{g}(|{\bf
    k}|)\right]^{1/2} e^{-i{\bf k}\cdot{\bf r}} \tilde{v}({\bf k}),
  \label{vr_vk}
\end{equation}
where the second moment of $\tilde{v}({\bf k})$ is given
by\footnote{Note that since $V({\bf r})$ is real $\tilde{v}^*({\bf
    k})=\tilde{v}(-{\bf k})$.}
\begin{equation}
  \overline{\tilde{v}^*({\bf k}_1)\tilde{v}({\bf k}_2)} =
  \delta({\bf k}_1 - {\bf k}_2)
  \label{vkvk}
\end{equation}

Using Eq.~(\ref{uw}) the matrix elements of the disorder potential can
now be expressed in terms of the random variables $\tilde{v}({\bf k})$
and the matrix polynomials,
\begin{equation}
  \langle n_1 k_1 | V | n_2 k_2 \rangle =
  \int \frac{d^2p}{(2\pi)^{1/2}}  \left[\tilde{g}(|{\bf
    p}|)\right]^{1/2} U_{n_1n_2}(-{\bf p}) W_{k_1k_2}(-{\bf p})
  \tilde{v}({\bf p}).
  \label{v_uw}
\end{equation}

\begin{figure}[tbp]
  \caption[Kawaji paper]{The temperature dependence of the
    longitudinal and  Hall conductivity [from ({Kawaji and
      Wakabayashi, }1987)].}
  \label{fig:temp_dep}
\end{figure}

\begin{figure}[tbp]
  \caption[Scaling in Japan]{Longitudinal conductivity $\sigma_{xx}$
    vs.\ Hall conductivity $\sigma_{xy}$ in a Si-MOSFET for
    temperatures between 0.35 K and 1.5 K [from ({Yamane {\em
        et~al.\/}, }1989)].}
  \label{fig:two_scale_jpn}
\end{figure}

\begin{figure}[tbp]
  \caption[Scaling in Princeton]{Longitudinal conductivity $\sigma_{xx}$
    vs.\ Hall conductivity $\sigma_{xy}$ in an InGaAs-InP
    heterostructure for temperatures between 50 mK and 770 mK [from
    ({Wei {\em et~al.\/}, }1987)].}
  \label{fig:two_scale_pu}
\end{figure}

\begin{figure}[tbp]
  \caption[Wei, the first]{Hall resistivity $\rho_{xy}$, longitudinal
    resistivity $\rho_{xx}$, and the derivative $d\rho_{xy}/dB$ of
    an InGaAs-InP heterostructure [from ({Wei {\em et~al.\/}, }1988)].}
  \label{fig:wei_fig1}
\end{figure}

\begin{figure}[tbp]
  \caption[Wei, the second]{The temperature dependence of the maxima
    of $d\rho_{xy}/dB$ and the width $\Delta{}B$ of the
    $\rho_{xx}$ peaks from Fig.~\ref{fig:wei_fig1}. The slope of
    the straight lines gives $(d\rho_{xy}/dB)^{\text{max}}\propto
    T^{-\kappa}$ and $\Delta{}B\propto T^{\kappa}$ with
    $\kappa=0.42\pm0.04$ [from ({Wei {\em et~al.\/}, }1988)].}
  \label{fig:wei_fig2}
\end{figure}

\begin{figure}[tbp]
  \caption[Koch's experiment]{Temperature dependence of the peak width
    $\Delta{}B$ for samples of different size [from ({Koch {\em
        et~al.\/}, }1991a)].}
  \label{fig:koch_exp}
\end{figure}

\begin{figure}[tbp]
  \caption[Lloyds figure 3]{Real part of the dynamical conductivity
    Re$(\sigma_{xx})$ at different frequency and temperatures [from
    ({Engel {\em et~al.\/}, }1993)].}
  \label{fig:engels_fig}
\end{figure}

\begin{figure}[tbp]
  \caption[Two lakes]{Regions where $V({\bf r})<E$ are shaded. The
    eigenfunctions
    are localized within a distance $l_c$ of the edge of the regions.
    The arrow indicates the direction of guiding center
    motion [after ({Trugman, }1983)].}
  \label{fig:two_lakes}
\end{figure}

\begin{figure}[tbp]
  \caption[Network]{Equipotential lines of a smooth random potential. The thick
    lines correspond to the links in the network model. The circles
    enclose the saddle points [after ({Chalker and Coddington, }1988)].}
  \label{fig:network}
\end{figure}

\begin{figure}[tbp]
  \caption[Network model]{The network model of Chalker and
    Coddington. Each node represents a saddle point and each link an
    equipotential line of the random potential [after ({Chalker and
      Coddington, }1988)].}
  \label{fig:net_scheme}
\end{figure}

\begin{figure}[tbp]
  \caption{The renormalization-group flow near a saddle-point fixed
    point $x_1=x_2=0$.}
  \label{fig:saddle_flow}
\end{figure}

\begin{figure}[tbp]
  \caption[Gang of four]{$\beta$ function according to Abrahams et al.\
    (1979). The slope of the dashed line is $1/\nu$ [cf.\
    Eq.~(\ref{beta_lin})]. }
  \label{fig:beta}
\end{figure}

\begin{figure}[tbp]
  \caption[sigam flow diagram]{Renormalization flow diagram of
    $\sigma_{xx}$ vs. $\sigma_{xy}$ [after ({Khmel'nitskii, }1983)].}
  \label{fig:sigma_flow}
\end{figure}

\begin{figure}[tbp]
  \caption[Localization length]{Normalized localization length
    $\lambda_M/M$ in the lowest Landau level for $\delta$-correlated
    potential as a function of system width $M$ (in units of
    $\sqrt{2\pi}l_c$) for energies 0.01 ($\ast$), 0.05
    (${\scriptstyle\bigtriangledown}$), 0.07
    (${\scriptstyle\bigtriangleup}$), 0.1 (${\scriptstyle\Box}$), 0.18
    ($\diamond$), 0.30 ($\star$), 0.5 ($\circ$), and 1.0 ($\bullet$) (in
    units of $\Gamma$) [after ({Huckestein and Kramer, }1990)].}
  \label{fig:lll_0}
\end{figure}

\begin{figure}[tbp]
  \caption[Scaling function]{The scaling function $\lambda_M(E)/M =
    \Lambda(M/\xi(E))$ [from ({Huckestein, }1992)].}
  \label{fig:scale_curve}
\end{figure}

\begin{figure}[tbp]
  \caption[xi_E]{The localization length $\xi(E)$ as a function of energy
    normalized by the band width $\Gamma_\sigma$ for $n=0$,
    $\sigma=0$ (${\scriptstyle\bigtriangledown}$), $n=0$,
    $\sigma=l_c$ (${\scriptstyle\bigtriangleup}$), and $n=1$,
    $\sigma=l_c$ (${\scriptstyle\Box}$) [after ({Huckestein, }1992)].}
  \label{fig:xi_E}
\end{figure}

\begin{figure}[tbp]
  \caption[Localization length 2]{Normalized localization length
    $\lambda_M/M$ in the lowest Landau level for correlated
    potential ($\sigma=l_c$) as a function of system width $M$ (in
    units of $\sqrt{2\pi}l_c$) for energies 0 ($\ast$), 0.03
    (${\scriptstyle\times}$), 0.05
    (${\scriptstyle\bigtriangledown}$), 0.07
    (${\scriptstyle\bigtriangleup}$), 0.1
    (${\scriptstyle\Box}$), 0.14 ($\star$), 0.18 ($\diamond$), 0.3
    ($\circ$), and 0.5 ($\bullet$) (in units of $\Gamma$) [after
    ({Huckestein {\em et~al.\/}, }1992)].}
  \label{fig:lll_1}
\end{figure}

\begin{figure}[tbp]
  \caption[Localization length 3]{Normalized localization length
    $\lambda_M/M$ in the second ($n=1$) Landau level for correlated
    potential ($\sigma=l_c$) as a function of system width $M$ (in
    units of $\sqrt{2\pi}l_c$) for energies 0.01 ($\ast$), 0.03
    (${\scriptstyle\bigtriangledown}$), 0.05
    (${\scriptstyle\bigtriangleup}$), 0.07 (${\scriptstyle\Box}$), 0.1
    ($\diamond$), 0.18 ($\star$), 0.3 ($\circ$), and 0.5 ($\bullet$) (in
    units of $\Gamma$) [from ({Huckestein, }1992)].}
  \label{fig:nll_1}
\end{figure}

\begin{figure}[tbp]
  \caption[Localization length 4]{Normalized localization length
    $\lambda_M/M$ in the second ($n=1$) Landau level for uncorrelated
    potential ($\sigma=0$) as a function of system width $M$ (in
    units of $\sqrt{2\pi}l_c$) for energies 0.01 ($\ast$), 0.1
    (${\scriptstyle\bigtriangledown}$), 0.18
    (${\scriptstyle\bigtriangleup}$), 0.3 (${\scriptstyle\Box}$), 0.5
    ($\diamond$), 0.65 ($\star$), 0.8 ($\circ$), and 1.0 ($\bullet$) (in
    units of $\Gamma$) [from ({Huckestein, }1992)].}
  \label{fig:nll_0}
\end{figure}

\begin{figure}[tbp]
  \caption[Localization length vs. correlation length]{The normalized
    localization length $\lambda_M/M$ at the band center of the
    second Landau level as function of
    $\beta^2=(\sigma^2+l_c^2)/l_c^2$ for system width $M=16$
    ($\circ$), $M=32$ ($\bullet$), $M=64$ ($\diamond$), and $M=128$
    ($\ast$) [after ({Huckestein, }1992)].}
  \label{fig:corr_beta}
\end{figure}

\begin{figure}[tbp]
  \caption[Scaling of corrections to scaling]{Scaling towards the
    fixed point value $\Lambda_c=1.14$ as a function of
    $M/\xi_{\text{irr}}$ for the data shown in
    Fig.~\ref{fig:corr_beta} [from ({Huckestein, }1994)].}
  \label{fig:scale_corr}
\end{figure}

\begin{figure}[tbp]
  \caption[Irrelevant length scale]{The irrelevant length scale
    $\xi_{\text{irr}}$ (in units of $\sqrt{2\pi}l_c$) as a function
    of $\beta^2$ [from ({Huckestein, }1994)].}
  \label{fig:xi_irr}
\end{figure}

\begin{figure}[tbp]
  \caption[Deviations from fixed point]{Deviations of the scaling
    function $\Lambda$ from its asymptotic value $\Lambda_c$ [from
    ({Huckestein, }1994)].}
  \label{fig:scale_dev_1}
\end{figure}

\begin{figure}[tbp]
  \caption[Phase diagram]{Phase diagram of the lowest Landau level in
    the presence of a periodic potential with $\alpha=3/5$ unit
    cells per flux quantum. Indicated are the positions of the
    critical energies and the Hall conductivity in the different
    phases. The bars at zero $\Gamma/E_0$ show the subbands in the
    absence of disorder.}
  \label{fig:period_phase}
\end{figure}

\begin{figure}[tbp]
  \caption[Wavefunction]{The local density $|\psi({\bf r})|^2$ of an
    eigenstate at the center of the lowest Landau level of a lattice
    system with $150\times150$ sites [from ({Huckestein and
      Schweitzer, }1992a)].}
  \label{fig:wave}
\end{figure}

\begin{figure}[tbp]
  \caption[f(alpha) of density]{The $f(\alpha)$ spectrum of the local
    density of the eigenstate shown in Fig.~\ref{fig:wave}. Crosses
    show the errors in $f(\alpha(q))$ and $\alpha(q)$ for $q=-2$,
    -1, 1, and 2 (from right to left) [from ({Huckestein {\em
        et~al.\/}, }1992)].}
  \label{fig:f_alpha_density}
\end{figure}

\begin{figure}[tbp]
  \caption[f(alpha) of current]{The $f(\alpha)$ spectrum of the local
    equilibrium current density of the eigenstate shown in
    Fig.~\ref{fig:wave}. Crosses show the errors in $f(\alpha(q))$
    and $\alpha(q)$ for $q=-2$, -1, 1, and 2 (from right to left)
    [from ({Huckestein and Schweitzer, }1992a)].}
  \label{fig:f_alpha_current}
\end{figure}

\begin{figure}[tbp]
  \caption[Energy dependence of alpha0]{The energy dependence of
    $\alpha_0(E)$ near the center of the lowest Landau band. The
    straight line is a guide to the eye [from ({Huckestein and
      Schweitzer, }1992b)].}
  \label{fig:alpha_scale}
\end{figure}

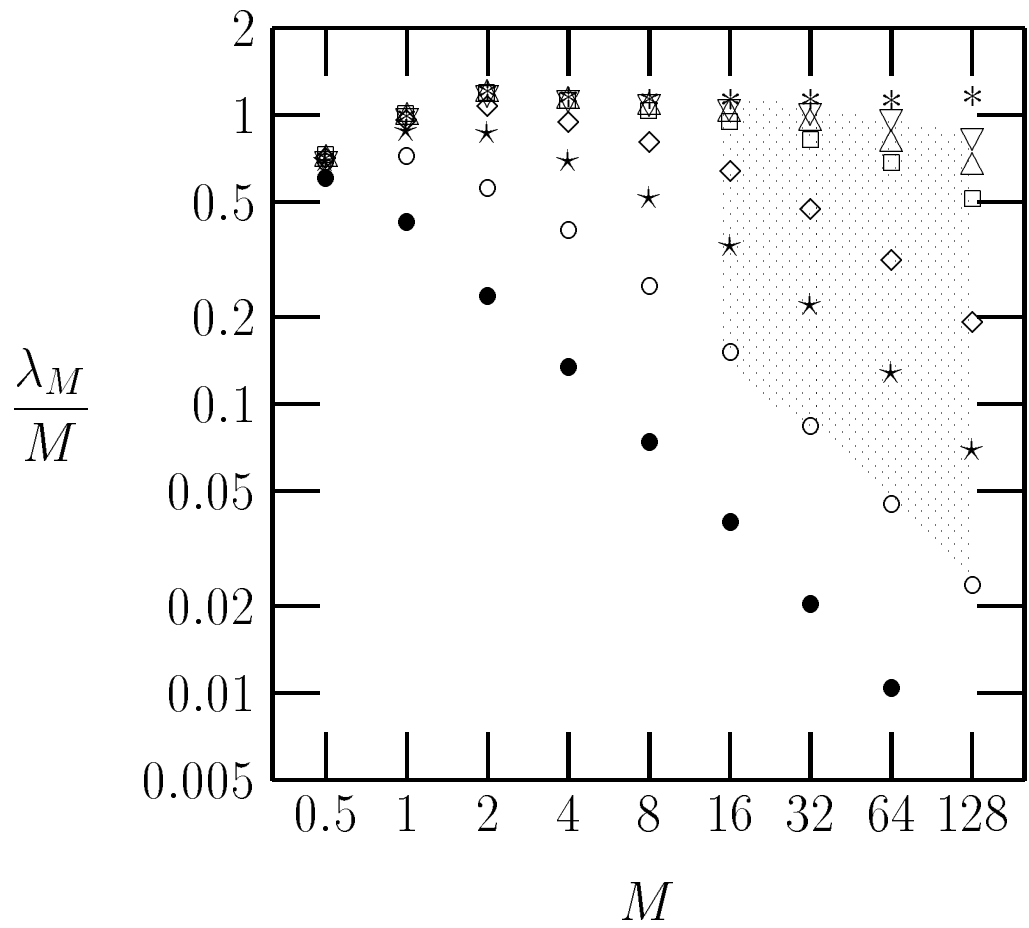
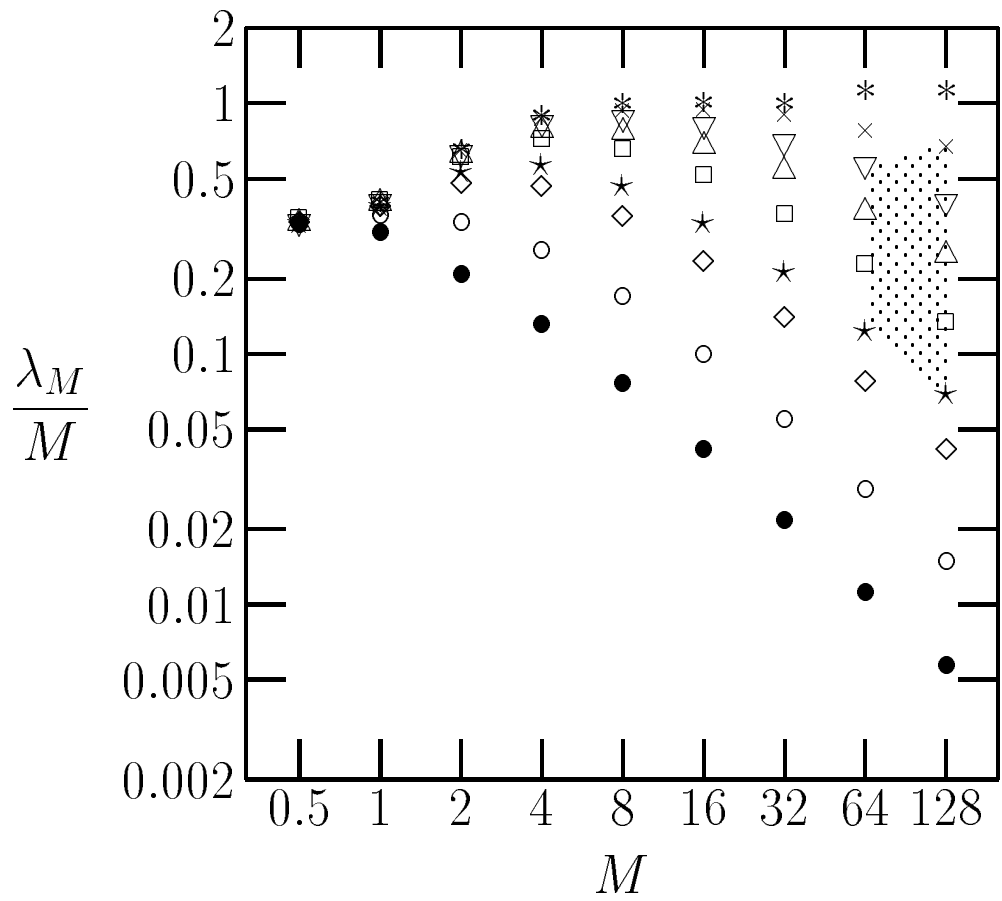
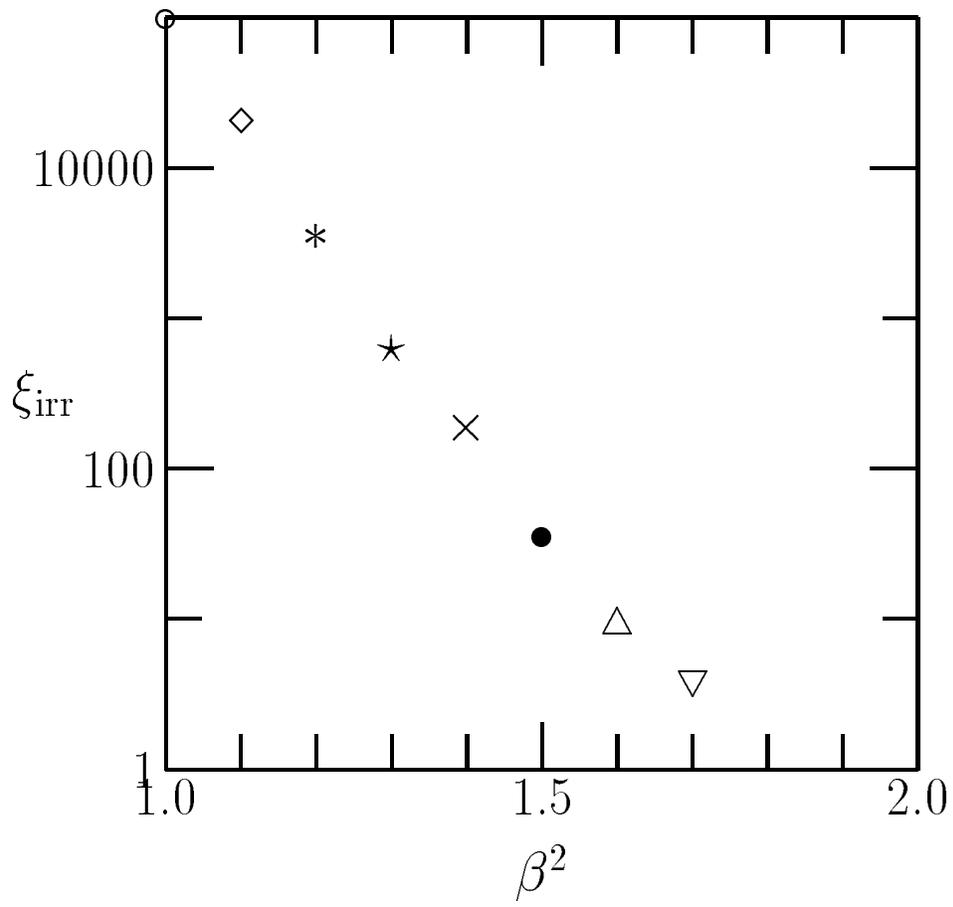
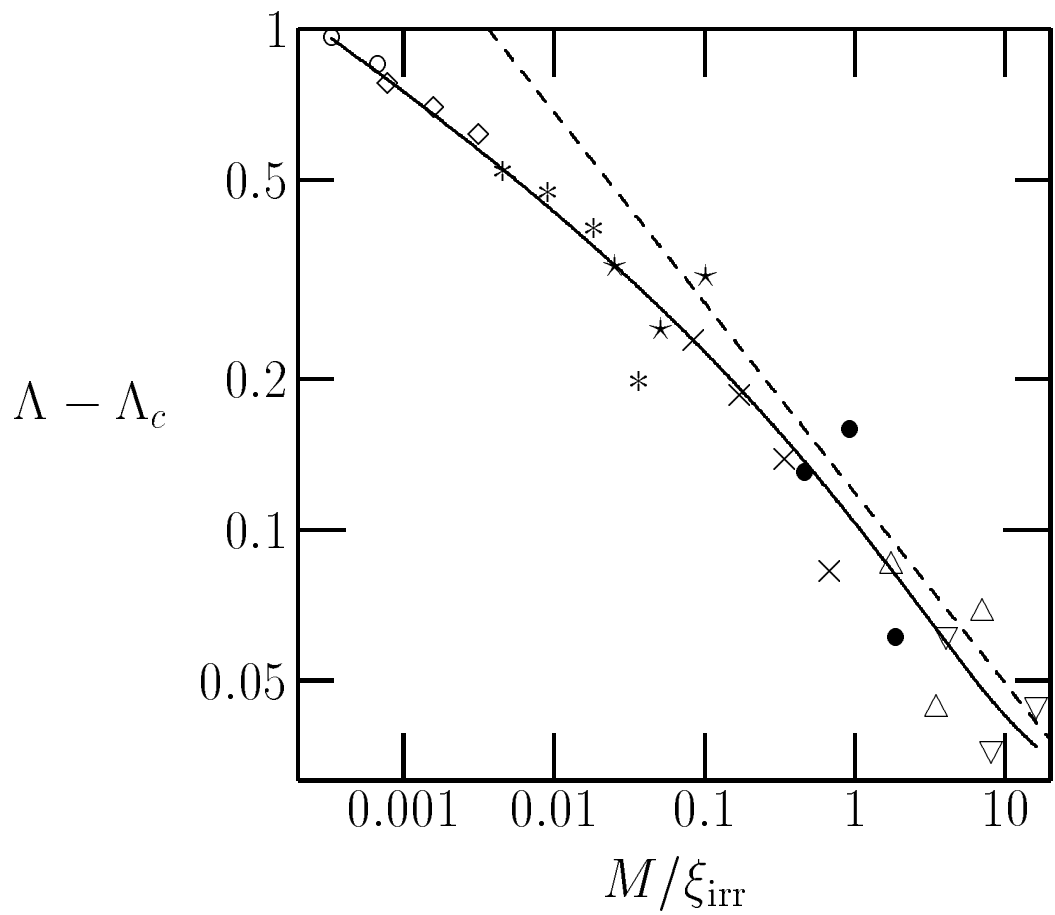
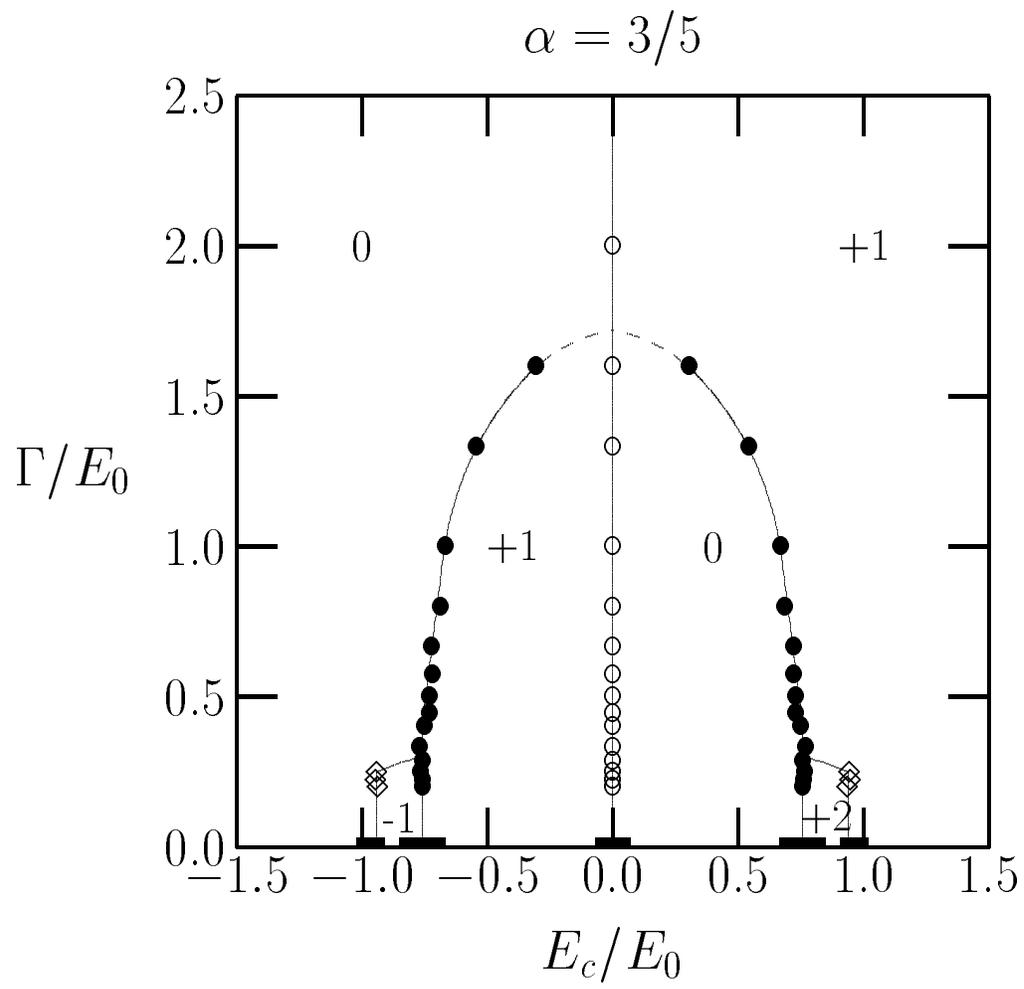
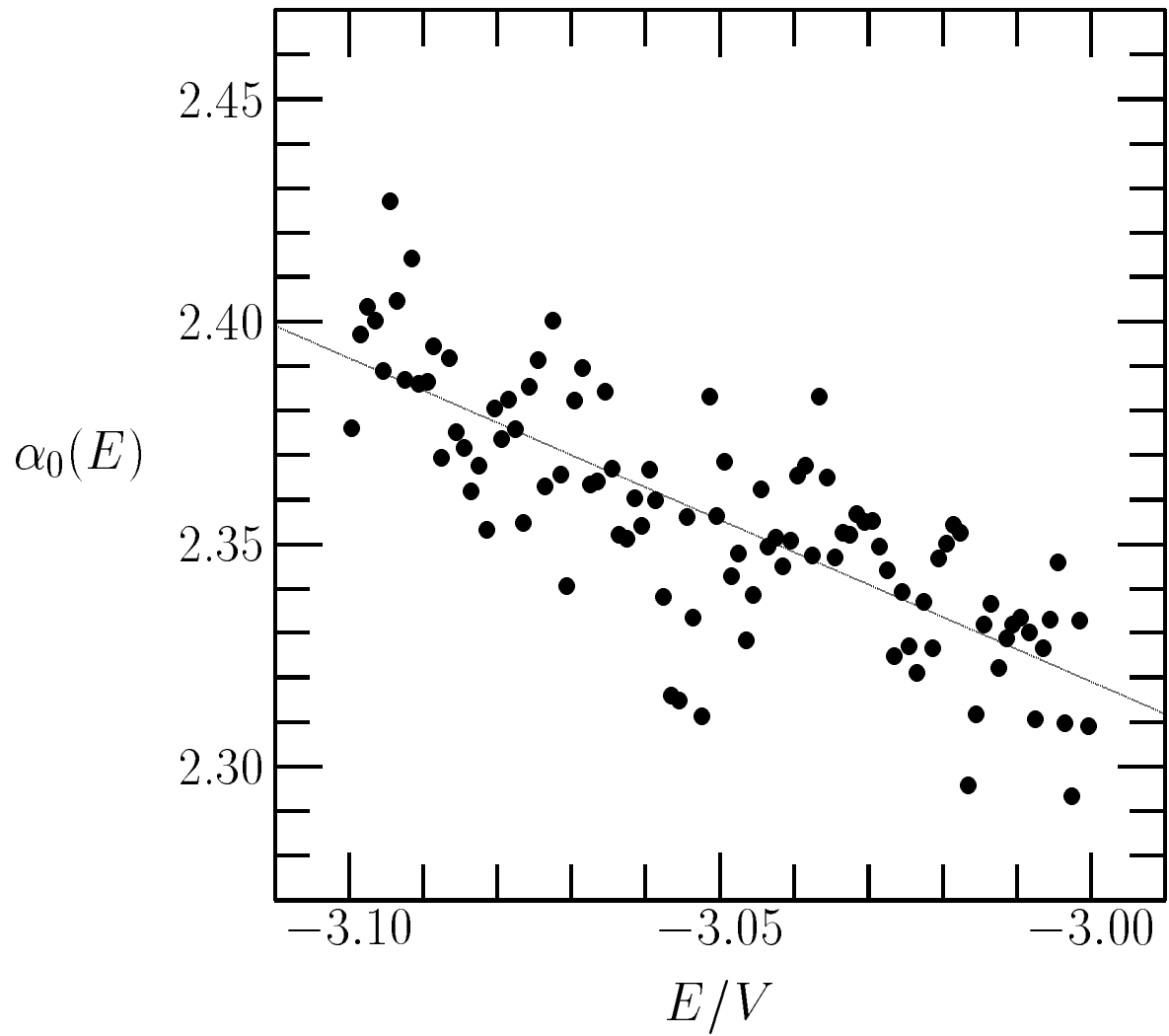
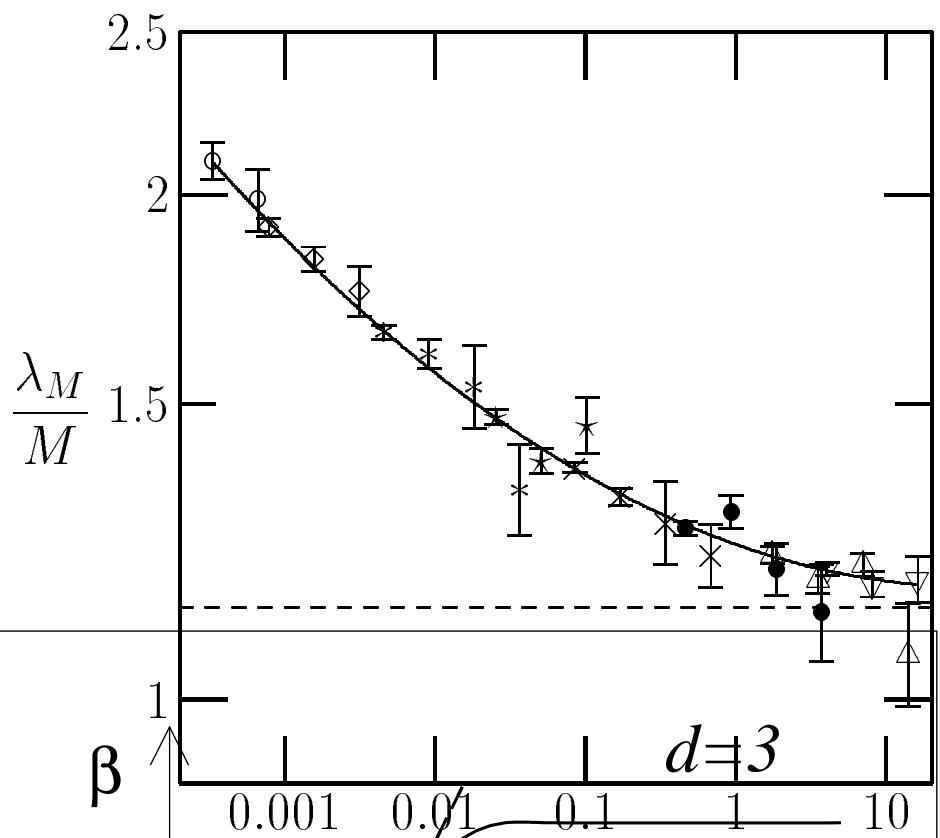
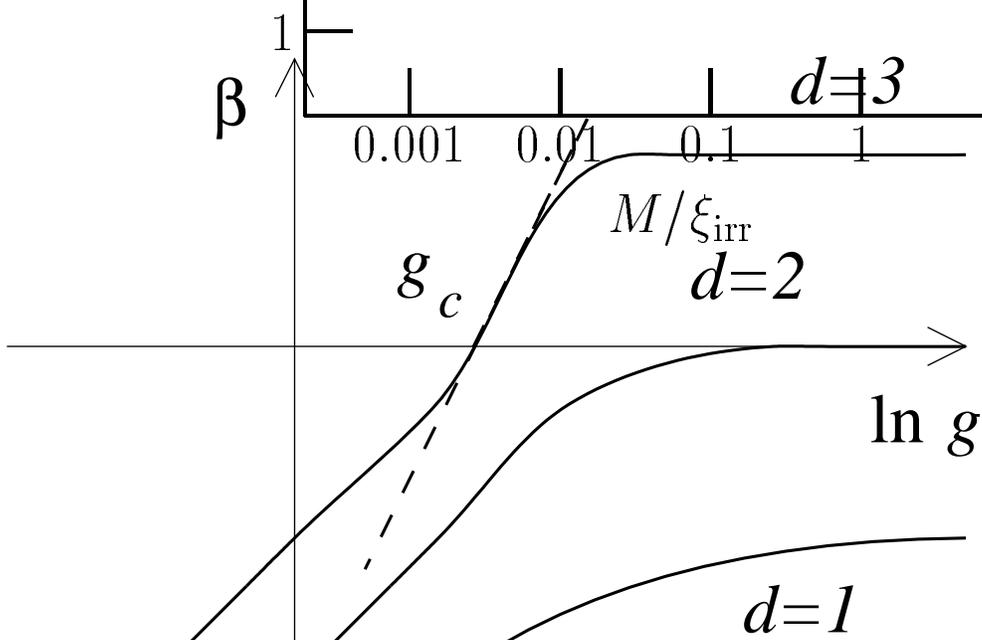
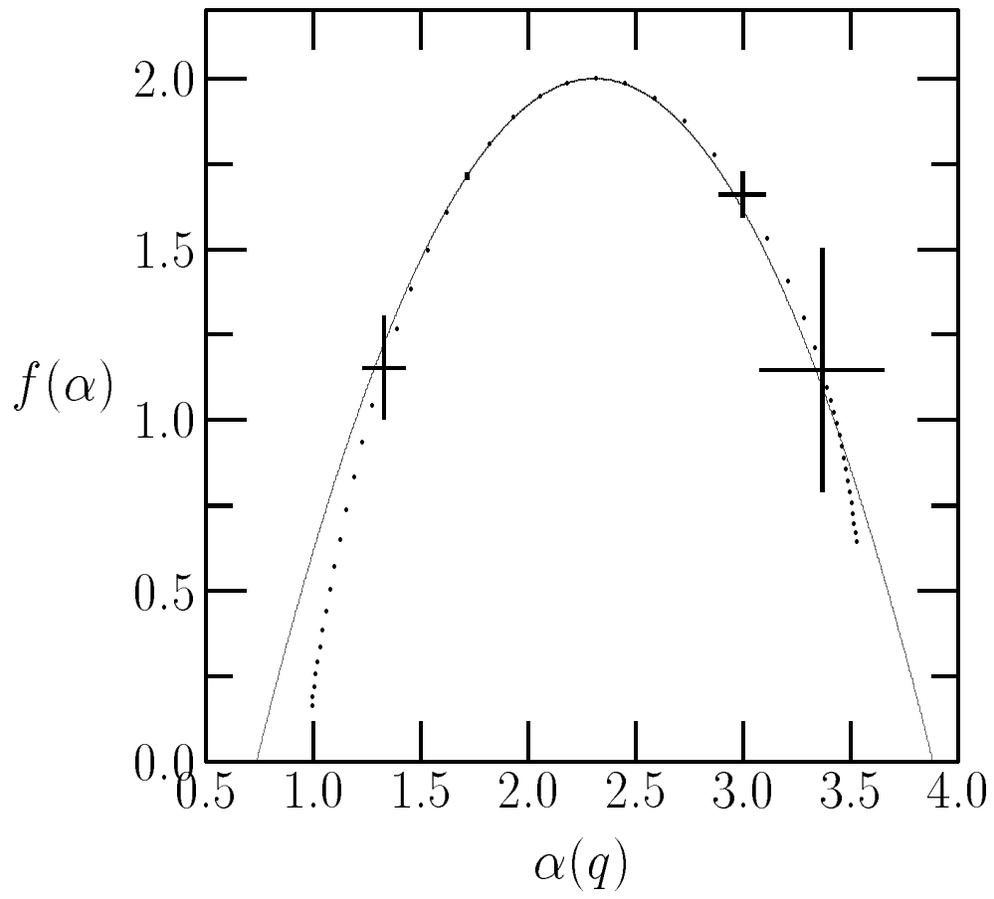
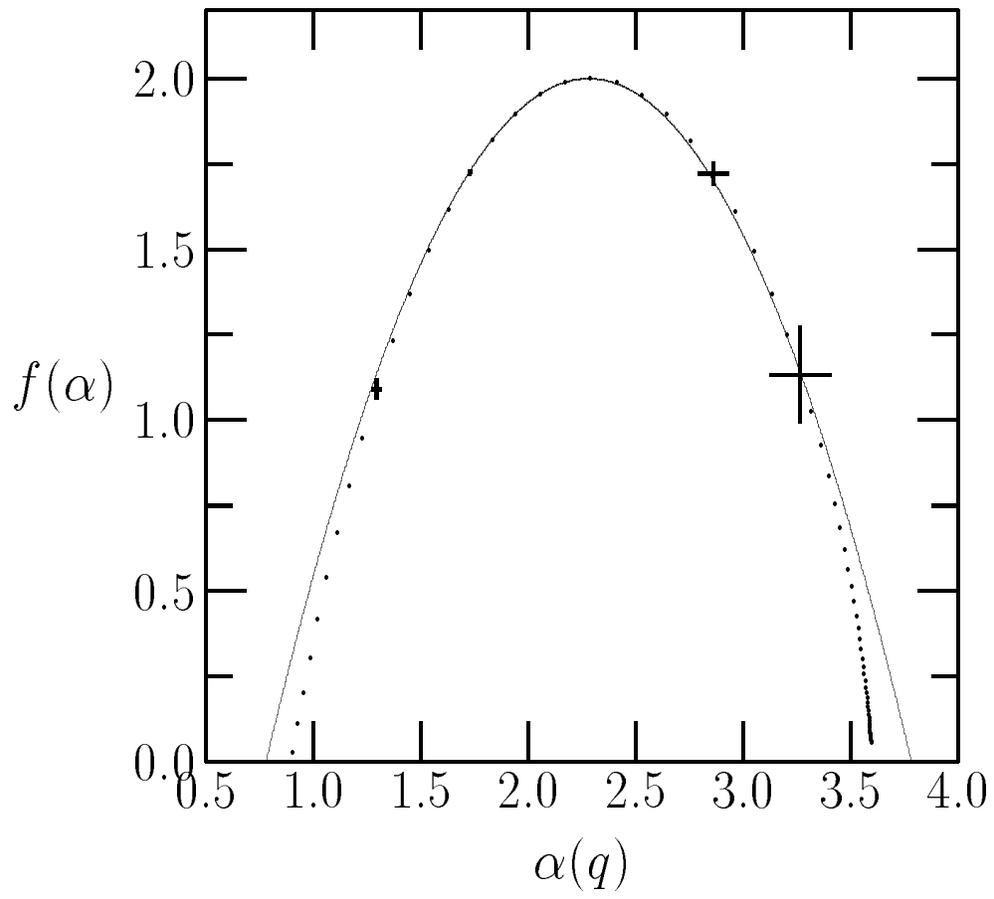
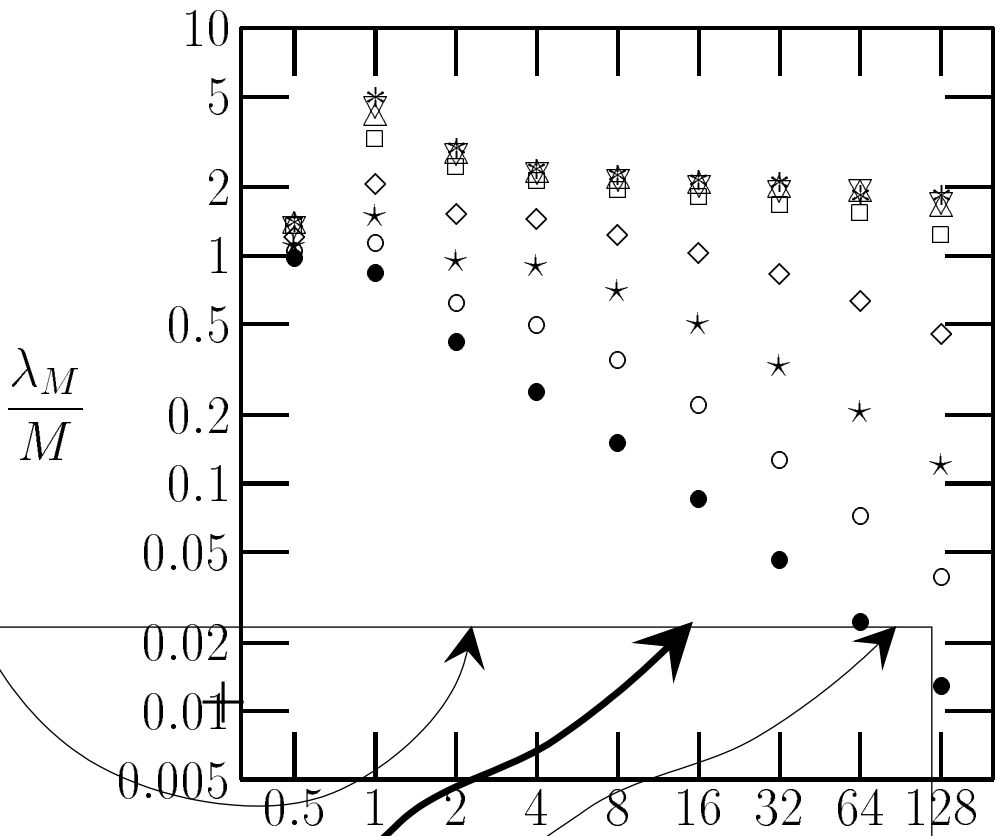
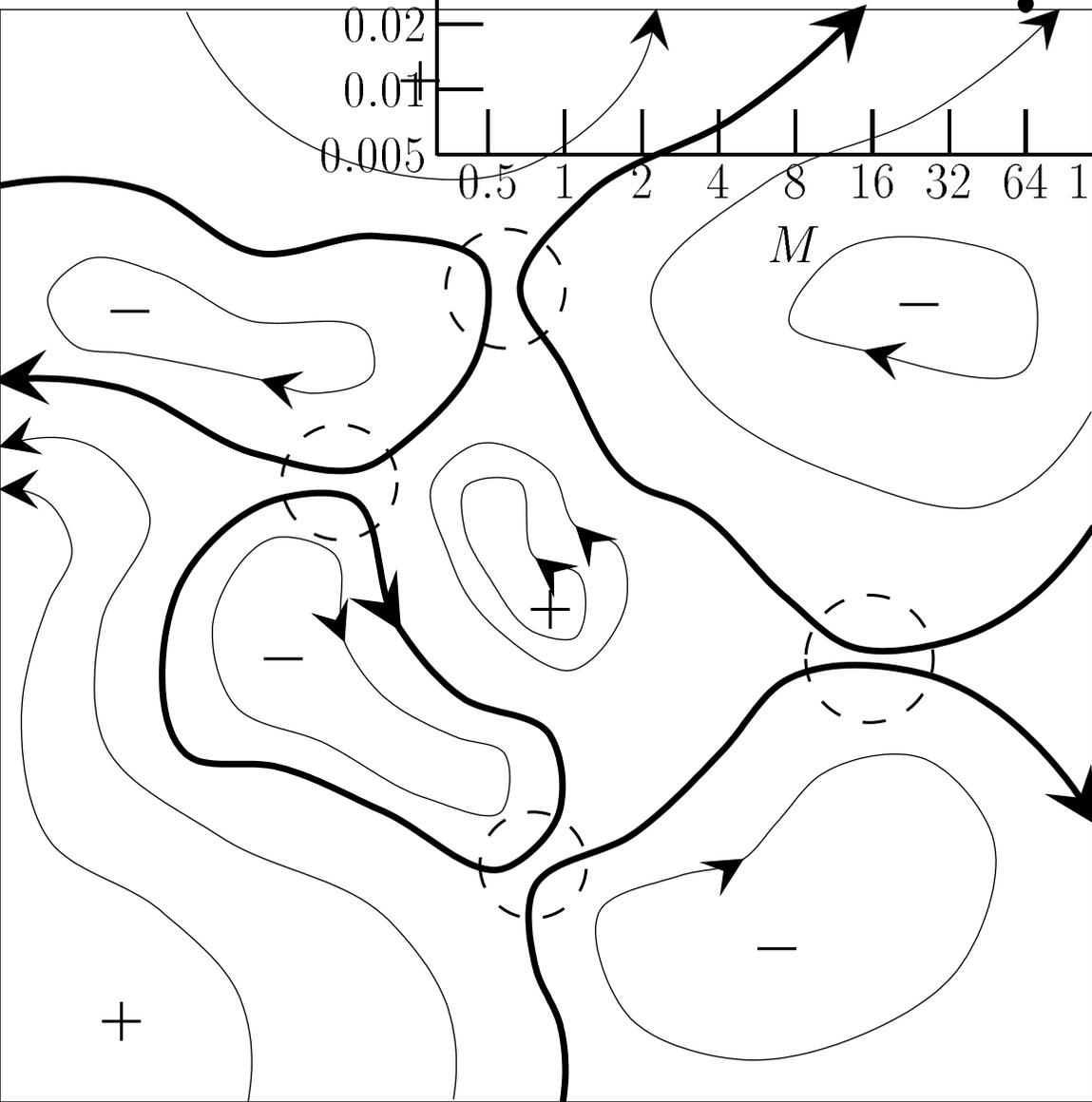
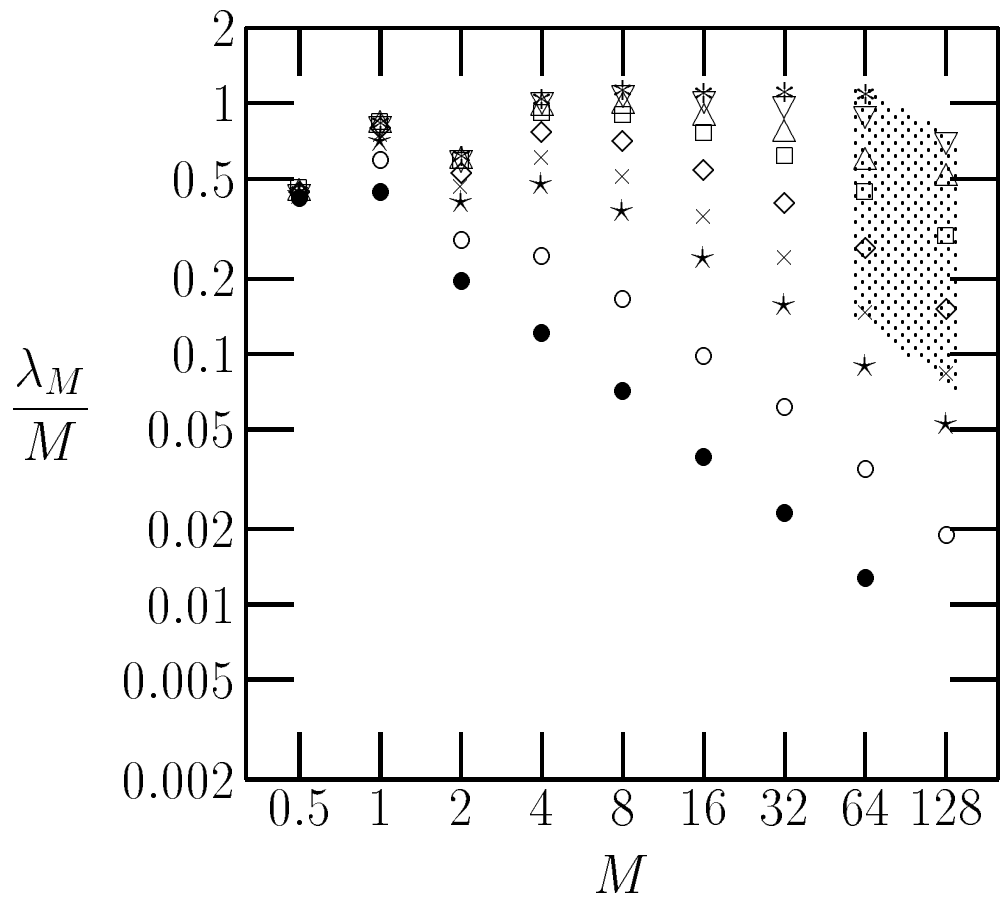
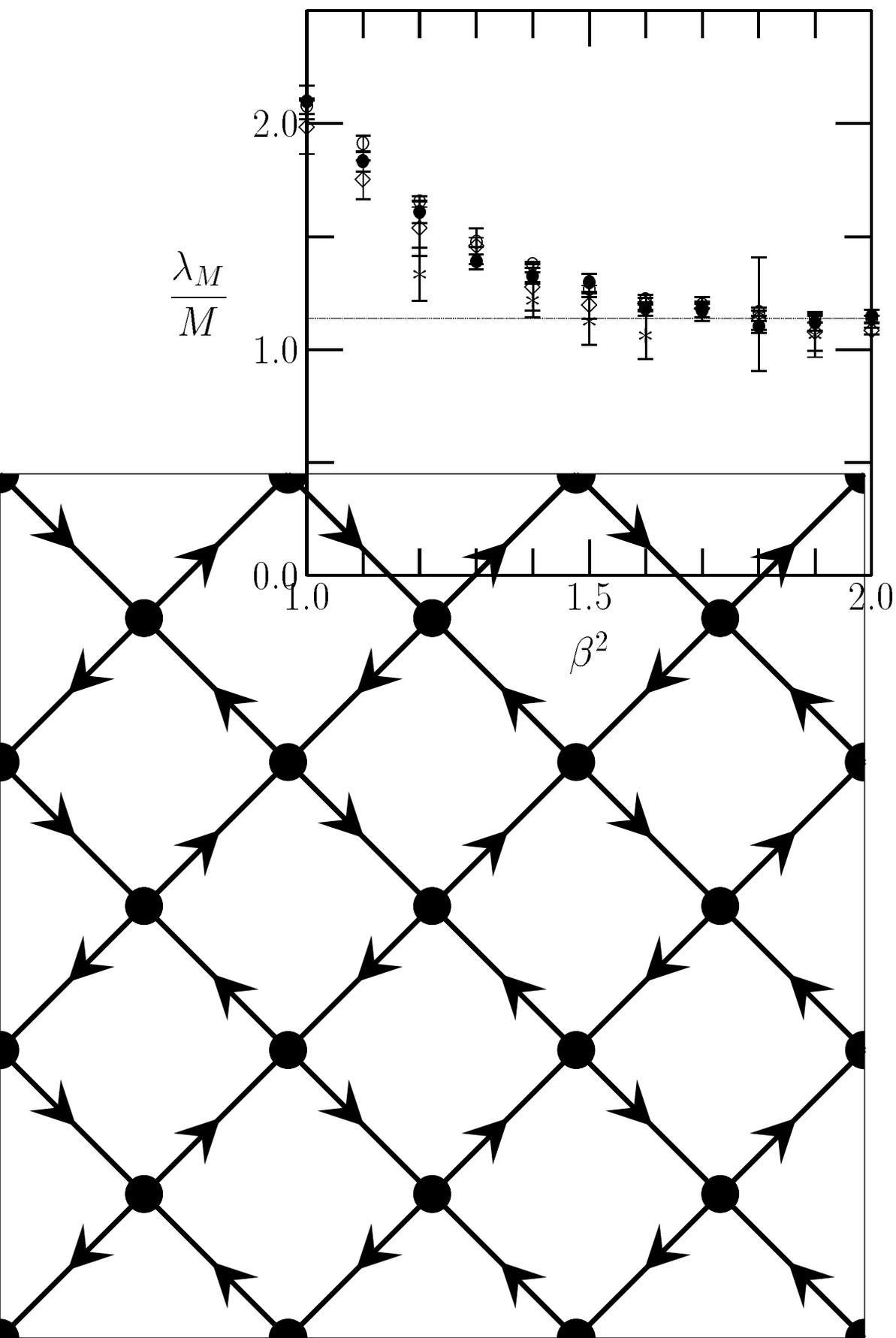
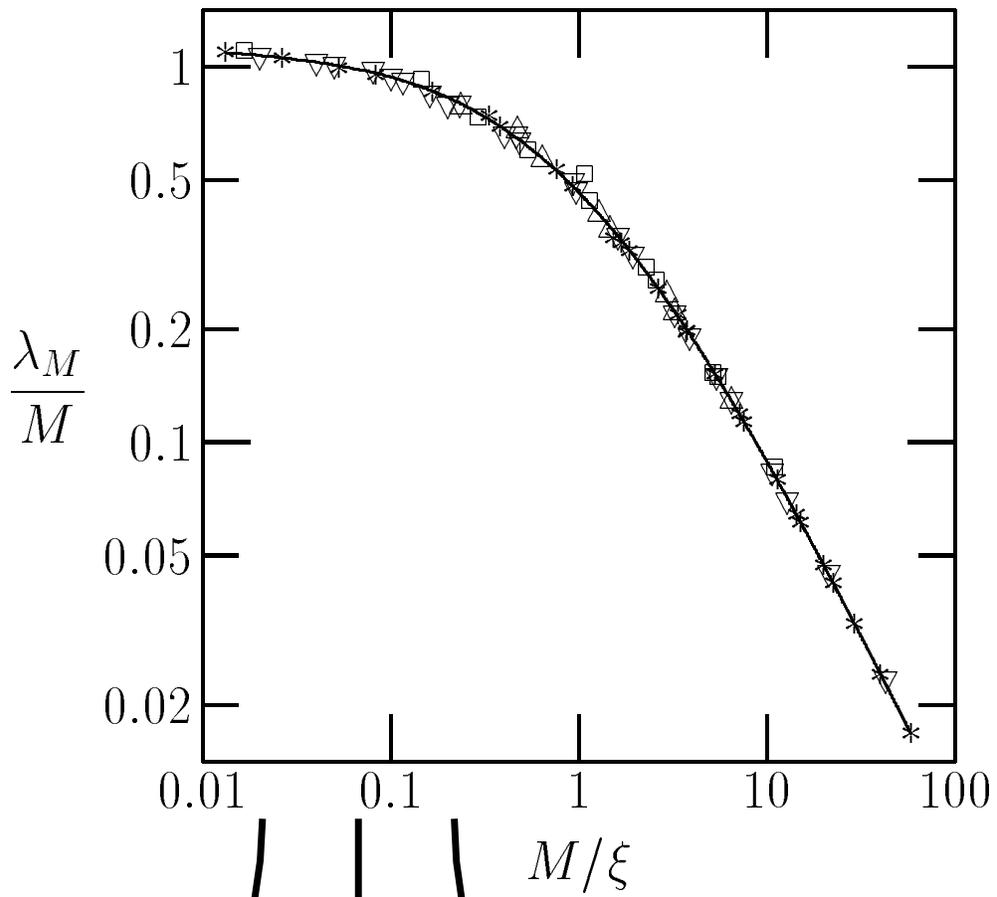
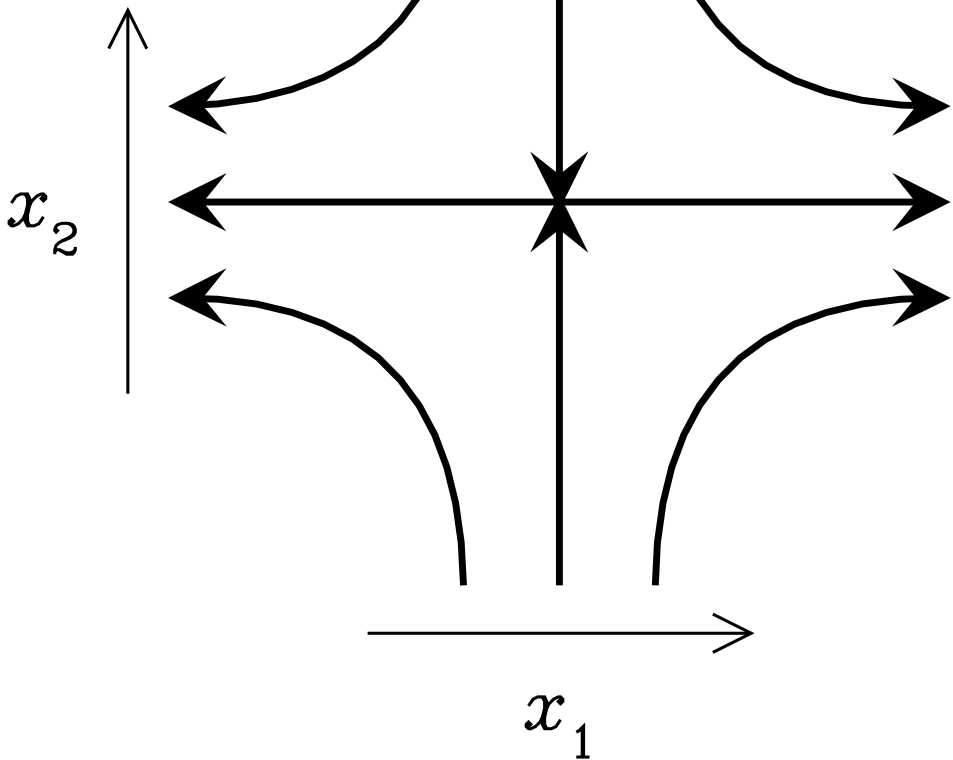
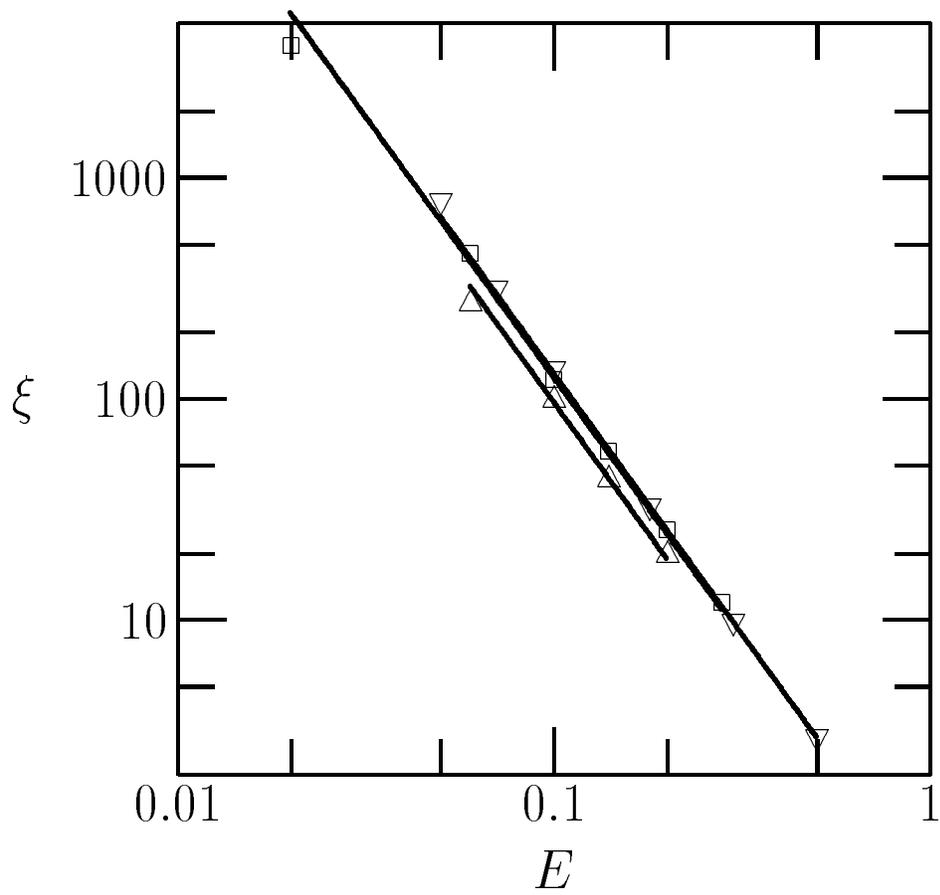
\end{document}